\documentclass[12pt]{article}
\usepackage{amsmath}
\usepackage[round]{natbib}
\usepackage{amssymb,xcolor}
\usepackage{appendix}
\usepackage{amsthm}
\usepackage{hyperref}
\usepackage{enumitem}
\usepackage{dsfont}
\usepackage{mathtools} 
\mathtoolsset{showonlyrefs=true} 

\usepackage{caption}

\usepackage[margin=0.7in]{geometry}

\numberwithin{equation}{section}
\usepackage{comment}

\usepackage{titling}
\newtheorem{thm}{Theorem}[section]
\newtheorem{cor}[thm]{Corollary}
\newtheorem{defi}{Definition}[section]
\newtheorem{assume}{Assumption}[section]
\newtheorem{prop}[thm]{Proposition}

\newtheorem{lemma}[thm]{Lemma}

\begin{document}
\author{
Jaehyun Kim\thanks{Department of Statistics and Data Science, The Chinese University of Hong Kong, Shatin, N.T. Hong Kong. EmaiL: jaehyunkim@cuhk.edu.hk}
\quad\textnormal{and}\quad
Hoi Ying Wong\thanks{Corresponding author; Department of Statistics and Data Science, The Chinese University of Hong Kong, Shatin, N.T. Hong Kong. Email: hywong@cuhk.edu.hk}
\\[1.5ex]
\normalsize Department of Statistics and Data Science
\\
\normalsize The Chinese University of Hong Kong
\\
\normalsize Shatin, N.T., Hong Kong
}

\title{Duality of Shortfall Aversion}

\maketitle	
\begin{abstract}
Shortfall aversion describes a form of consumption habit formation in which an individual's utility depends on consumption relative to the historical consumption peak. This study develops a duality framework for a broad class of utility functions, with and without discounting, avoiding the need for determining the local optimality required by the HJB approaches commonly adopted in the literature. We show that shortfall aversion can instead be incorporated directly into the dual problem through a singular-control mechanism characterized by a variational inequality. Furthermore, we establish that the resulting value function strictly dominates those obtained in the existing literature while remaining attainable under an admissible optimal control. We demonstrate that the suboptimality of existing solutions stems from their myopic treatment of updates to the consumption peak, whereas our duality approach enlarges the admissible class to treat the consumption peak as a separate nondecreasing control and determines its optimal evolution through duality, accounting for future consumption opportunities. These results reveal a previously unrecognized advantage of the dual formulation and provide a more general and flexible framework for consumption optimization under shortfall aversion.
\end{abstract}
\section{Introduction}
Consumption habit formation has long played an important role in finance by capturing the idea that individuals evaluate current consumption relative to a reference level shaped by past consumption \citep{sundaresan1989intertemporally, constantinides1990habit, detemple1992optimal}. Such preferences introduce intertemporal dependence into consumption decisions and have important implications for optimal consumption and portfolio choice. Habit formation has also been widely used in equilibrium asset pricing. In particular, it provides mechanisms for explaining the equity premium and other salient features of asset returns \citep{constantinides1990habit, Abel1990asset, Campbell1999by}, while empirical evidence on habit persistence in aggregate consumption is provided by \cite{Ferson1991habit}.

Habit formation essentially makes preferences path dependent. Increasing consumption today can increase the future reference level and thereby make future consumption requirements more demanding. At the extreme, habit persistence gives rise to consumption ratcheting, whereby individuals are unwilling to reduce their standard of living once a higher level has been attained. \cite{Dybvig1995} formalizes this idea as a dynamic consumption and investment problem in which consumption cannot decline over time and increases as wealth reaches new maxima. The corresponding robust setting is recently developed by \cite{ParkChenWong2025}.

The notion of shortfall aversion, introduced by \citet{guasoni2020shortfall} and further developed by \citet{deng2022optimal, Li2024change}, provides a flexible framework for modeling consumption relative to a historical reference level. Rather than imposing an irreversible-consumption constraint, shortfall aversion specifies utility as a function of current consumption relative to its historical peak. Raising consumption to a new peak therefore increases the reference level against which future consumption is evaluated, creating an intertemporal trade-off between the immediate benefit of higher consumption and the future cost associated with a higher consumption standard. The historical consumption peak thus becomes an endogenous state variable whose evolution is an integral part of the dynamic optimization problem.

The existing literature \citep{guasoni2020shortfall, deng2022optimal, Li2024change} addresses the resulting path-dependent optimization problem primarily through dynamic programming. A key step in this approach is to characterize the threshold at which the historical consumption peak should be raised. This characterization relies on a local optimality argument originating from \citet[Lemma A.1]{guasoni2020shortfall}, and subsequently used to other utility functions in related studies. We show, however, through a counterexample that the conditions stated in this lemma are not sufficient in general to ensure the asserted local optimality property within the HJB equation. This subtle issue is important because the location of the peak-adjustment threshold directly affects the characterization of the optimal consumption policy.

Motivated by this observation, we develop a direct duality approach that avoids reliance on a local characterization. By enlarging the admissible class to incorporate the consumption peak as a state and control variable, we recast the evolution of the historical consumption peak as a singular-control problem in the dual formulation, with its optimal dynamics characterized endogenously by a variational inequality. This formulation  provides a unified characterization of both the optimal timing and the magnitude of peak adjustments.

Beyond resolving the local-characterization issue, the duality approach considerably broadens the scope of the analysis. It accommodates a general class of utility functions under both discounted and undiscounted preferences, without relying on solution arguments tailored to specific utility forms. More importantly, the dual formulation decouples the singular control governing consumption-peak adjustments from the regular control governing optimal investment. The resulting dual problem reduces to a singular-control problem that can be analyzed through a variational inequality or an associated optimal stopping problem. This decomposition circumvents the difficulties created by the interaction of singular and regular controls in the primal problem.

Our contributions are threefold. First, we introduce a general duality framework for optimal consumption and investment under shortfall aversion, in which the historical consumption peak evolves as a singular control. Its optimal adjustment is characterized through a variational inequality, yielding a global characterization of the optimal policy. Second, the framework applies to a broad class of utility functions under both discounted and undiscounted preferences, without requiring solution arguments tailored to specific utility specifications. Third, we revisit the local optimality argument underlying the consumption-peak adjustment threshold in the existing literature and identify a subtle limitation. Specifically, we construct a counterexample showing that the stated conditions do not, in general, guarantee the asserted local optimality property. Our dual approach bypasses this issue altogether: by enlarging the admissible class, the consumption peak is determined endogenously as part of the global optimization problem rather than through a candidate local optimum. These contributions establish a more general and robust foundation for the analysis of optimal consumption and portfolio choice under shortfall aversion.

The rest of the paper is organized as follows. Section \ref{S2} presents the problem formulation and reviews relevant results from the existing literature. In particular, it motivates our duality approach and points to a counterexample provided in Appendix \ref{sec:benchmark-comparison}. Since the shortfall-aversion literature can be broadly divided into discounted and undiscounted settings, Section \ref{sec:dis} studies the discounted case using our duality approach. Our solution differs from existing results under exponential utility in that the resulting optimal consumption rate satisfies the budget constraint, whereas existing solutions may admit negative consumption rates. Auxiliary results for the discounted case are collected in Appendix \ref{sec:A}. Section \ref{sec:undis} then considers the undiscounted case and derives a new solution that provides a further contribution to the literature, with detailed proofs deferred to Appendix \ref{sec:B}. Finally, Section \ref{conclusion} concludes the paper.

\section{Preliminaries and Problem Formulation}\label{S2}
We consider a complete probability space $(\Omega,\mathcal{F},\mathbb{P})$ endowed with a filtration $\mathbb{F}=(\mathcal{F}_t)_{t\geq 0}$, given by the usual augmentation of the natural filtration generated by a standard one-dimensional Brownian motion $W=(W_t)_{t\geq 0}$. For ease of exposition, we restrict attention to a financial market with one risk-free asset and one risky asset, although our methodology extends straightforwardly to complete markets with multiple risky assets.

The riskless asset price $G = (G_t)_{t \ge 0}$ satisfies $dG_t = r G_t\,dt$ a positive constant $r$. The risky asset price $S=(S_t)_{t\ge0}$ satisfies
\begin{align}
    dS_t = \mu S_t\,dt + \sigma S_t\,dW_t\,,
\end{align}
where $\mu$ and $\sigma > 0$ are constants. The market price of risk is $\kappa \coloneqq \frac{\mu-r}{\sigma}$. Throughout the paper, we assume $\mu > r$, so that $\kappa > 0$.

Let $\pi = (\pi_t)_{t \ge 0}$ denote the amount invested in the risky asset and $c = (c_t)_{t \ge 0}$ the nonnegative consumption rate. For initial wealth $x \ge 0$, the associated wealth process $X = (X_t)_{t \ge 0}$ satisfies
\begin{align}
    dX_t = \big(rX_t + \pi_t(\mu-r) - c_t\big)\,dt + \pi_t\sigma\,dW_t\,,
    \quad X_0 = x\,.
\end{align}
For an initial habit level $h > 0$, define the habit process $H^{h,c} = (H_t^{h,c})_{t \ge 0}$ as the running maximum of consumption
\begin{align}
    H_t^{h,c} \coloneqq \max\Big(h, \sup_{0 \le s \le t} c_s\Big)\,.
\end{align}
Utility is evaluated at the relative consumption index:
\begin{align}
    D(H_t^{h,c})c_t-L(H_t^{h,c}),
\end{align}
where $D:(0,\infty)\to(0,\infty)$ and $L:(0,\infty)\to[0,\infty)$ are deterministic functions. For instance, $D(h) = 1$ and $L(h) = \alpha h$ for $0<\alpha<1$ are adopted in \cite{deng2022optimal, Li2024change} so that the individual evaluates consumption based on the difference between current consumption and a scaled consumption peak. In contrast, \cite{guasoni2020shortfall} use $D(h) = 1/h^\alpha$ and $L(h) = 0$ and the individual evaluates consumption based on the ratio of current consumption to a scaled consumption peak. The following assumptions are imposed throughout the paper.

\begin{assume}\label{assume:common}
The functions $u,D$, and $L$ satisfy the following conditions.
\begin{enumerate}[label=(\roman*)]
    \item
    $u\in C^2((0,\infty))$ is strictly increasing and strictly concave, with
    \begin{align}
        m_0:=\lim_{z\downarrow0}u'(z)\in(0,\infty],
        \qquad
        \lim_{z\to\infty}u'(z)=0.
    \end{align}
    \item
    $D,L\in C^2((0,\infty))$, $D(h)>0$, and
    \begin{align}\label{eq:common-DL}
        0\le \frac{L(h)}{D(h)}<h,
        \qquad
        D'(h)\le0,
        \qquad
        \left(\frac{L(h)}{D(h)}\right)'\ge0.
    \end{align}
    \item
    The following inequalities hold
    \begin{align}\label{eq:gprime-D}
        0<\big(D(h)h-L(h)\big)'<D(h),
        \qquad h>0.
    \end{align}
\end{enumerate}
\end{assume}
For notational convenience, write
\begin{align}
    \ell(h)=\frac{L(h)}{D(h)}\,,\qquad g(h)=D(h)h-L(h)\,.
\end{align}

We now define the admissible consumption and portfolio strategies.
\begin{defi}\label{def:ori_ad}
For initial wealth $x \ge 0$ and an initial habit level $h > 0$, a strategy $(c,\pi)$ is \textit{admissible}, written $(c,\pi)\in\mathcal A(x,h)$, if the following conditions hold.
\begin{enumerate}[label=(\roman*)]
    \item\label{def-i}
    The processes $c$ and $\pi$ are $\mathbb F$-progressively measurable, and $c$ has c\`adl\`ag paths.
    \item The consumption rate satisfies $c_t\ge\frac{L(H_t^{h,c})}{D(H_t^{h,c})}=\ell(H_t^{h,c})$ $\mathbb P$-a.s. for all $t\ge0$.
    \item\label{ad:2}
    The processes satisfy $\int_0^T(c_t+\pi_t^2)\,dt<\infty$ $\mathbb P$-a.s. for every $T>0$.
    \item\label{ad:3}
    The wealth process satisfies $X_t\ge0$ $\mathbb P$-a.s. for all $t\ge0$.
\end{enumerate}
\end{defi}

Since $D(h)c-L(h)=D(h)\bigl(c-\ell(h)\bigr)$,
 the constraint $c=\ell(h)$ corresponds to zero effective consumption. We extend $u$ to zero by
\begin{align}
    u(0):=\lim_{z\downarrow0}u(z)\in[-\infty,\infty).
\end{align}
In particular, if $m_0<\infty$, then $u(0)>-\infty$ by concavity.

Assumption~\ref{assume:common} has a natural interpretation.
Since $c_t\ge\ell(H_t^{h,c})$, the function $\ell(h)$ gives the minimum consumption level, which does not decrease with habit because $\ell'(h)\ge0$.
The coefficient $D(h)$ measures the effectiveness of consumption, and $D'(h)\le0$ means that this effectiveness does not increase with habit.
At $c=h$, effective consumption is $g(h)$, so $g'(h)>0$ means that effective consumption increases when consumption and habit increase together.
Since the marginal increase with habit fixed is $D(h)$, the condition $g'(h)<D(h)$ makes the marginal loss caused by increasing habit, $D(h)-g'(h)$, strictly positive.

We now introduce the objective functional and the value function.
Let $\rho\ge0$ denote the subjective discount rate.
For $(c,\pi)\in\mathcal A(x,h)$, define
\begin{align}\label{eqn:objective_functional}
    J_\rho(x,h;c,\pi)
    :=
    \mathbb E\bigg[
        \int_0^\infty
        e^{-\rho t}
        u\Big(
            D(H_t^{h,c})c_t-L(H_t^{h,c})
        \Big)\,dt
    \bigg]\,.
\end{align}
The value function is defined by
\begin{align}\label{eqn:value}
    v_\rho(x,h)
    :=
    \sup_{(c,\pi)\in\mathcal A(x,h)}
    J_\rho(x,h;c,\pi)\,.
\end{align}
At zero effective consumption, we use the convention for $u(0)$ stated above.

\citet{guasoni2020shortfall} derive a candidate solution from the HJB equation and the boundary condition in \cite[(22)--(23)]{guasoni2020shortfall}, which, in our notation, are \begin{align} \sup_{c,\pi} \Big\{ \frac{(c/h^\alpha)^{1-\gamma}}{1-\gamma} +(rx-c+\pi(\mu-r))V_x^G(x,h) +\frac12\sigma^2\pi^2V_{xx}^G(x,h) \Big\} &=0,\qquad c<h,\\ V_h^G(x,h)&=0,\qquad c=h. \end{align} Using a Legendre transform together with the homogeneity of power utility, they reduce the HJB equation to an ordinary differential equation. Their verification argument makes use of the upper bound in \cite[Lemma A.1]{guasoni2020shortfall}. We observe that this bound does not hold for all admissible consumption processes. In particular, for $u(\xi)=-\frac{1}{\xi}$, $D(h)=h^{-\frac{1}{2}}$, $L(h)=0$, $\rho=0$, $r=\frac{1}{2}$, and $\kappa=1$, we construct a consumption process satisfying the budget condition for which the inequality in \cite[(A.2)]{guasoni2020shortfall} is reversed at $h=1$, $x=\frac{41}{16}$, and $y=1$. The construction is provided in Appendix~\ref{sec:benchmark-comparison}. 

This observation motivates an alternative approach to the shortfall aversion problem. In particular, we explicitly incorporate the decision of when to update the consumption peak as part of the control. This leads us to enlarge the admissible class of consumption and investment strategies accordingly. To this end, we introduce an auxiliary value function. Let $\mathcal H(h)$ denote the class of adapted, finite-valued, c\`adl\`ag, nondecreasing processes with $H_{0-}=h$. Define the enlarged admissible class by \begin{align}\label{eq:Abar} \bar{\mathcal A}(x,h) :=\left\{(H,c,\pi): \begin{array}{l} H\in\mathcal H(h),\quad c,\pi\text{ satisfy conditions } \ref{def-i},\ \ref{ad:2},\ \ref{ad:3} \text{ in Definition~}\ref{def:ori_ad},\\ \ell(H_t)\le c_t\le H_t\quad \text{for every }t\ge0 \end{array} \right\}. \end{align} The corresponding auxiliary value function is \begin{align}\label{eq:vbar} \bar v_\rho(x,h) :=\sup_{(H,c,\pi)\in\bar{\mathcal A}(x,h)} \mathbb E\left[\int_0^\infty e^{-\rho t} u\big(D(H_t)c_t-L(H_t)\big)\,dt\right]. \end{align} The following lemma establishes that the original and auxiliary problems have the same value.
\begin{lemma}\label{lemma:exact-enlargement}
Suppose Assumption~\ref{assume:common} holds.
For every $x,h>0$ and $\rho\ge0$,
if $\bar v_\rho(x,h)$ is well defined in $[-\infty,\infty]$, then
\begin{align}\label{eq:v-equals-vbar}
    v_\rho(x,h)=\bar v_\rho(x,h)\,.
\end{align}
\end{lemma}

\begin{proof}
We first prove $v_\rho(x,h)\le\bar v_\rho(x,h)$.
Fix $(c,\pi)\in\mathcal A(x,h)$.
Since $c$ is adapted and has c\`adl\`ag paths,
$H^{h,c}\in\mathcal H(h)$.
Moreover, $\ell(H_t^{h,c})\le c_t\le H_t^{h,c}$ for all $t\ge0$. Thus $(H^{h,c},c,\pi)\in\bar{\mathcal A}(x,h)$, and hence
\begin{align}
    v_\rho(x,h)\le\bar v_\rho(x,h).
\end{align}

Next we prove $v_\rho(x,h)\ge\bar v_\rho(x,h)$.
Fix $(H,c,\pi)\in\bar{\mathcal A}(x,h)$.
Since $H$ is nondecreasing and $c_s\le H_s$, $H_t^{h,c}=h\vee\sup_{0\le s\le t}c_s\le H_t.$
The monotonicity of $\ell$ gives $c_t\ge\ell(H_t)\ge\ell(H_t^{h,c})$, Thus $(c,\pi)\in\mathcal A(x,h)$.
Moreover, $\frac{d(D(h)c-L(h))}{dh}
=D'(h)(c-\ell(h))-D(h)\ell'(h)\le0$
for $c\ge\ell(h)$.
Thus we have
\begin{align}
    u\big(D(H_t^{h,c})c_t-L(H_t^{h,c})\big)
    \ge
    u\big(D(H_t)c_t-L(H_t)\big).
\end{align}
This implies $\bar v_\rho\le v_\rho$ and completes the proof.
\end{proof}

Let $(K_t)_{t\ge0}$ satisfy the SDE: $dK_t=-rK_t\,dt-\kappa K_tdW_t$ with $K_0=1$.
For $(H,c,\pi)\in\bar{\mathcal A}(x,h)$, It\^{o}'s formula and Fatou's lemma yield the static budget constraint
\begin{align}
    \mathbb{E}\bigg[\int_0^\infty K_tc_t\,dt\bigg]\le x\,.
\end{align}
Since $H_t\ge h$ and $\ell$ is nondecreasing,
\begin{align}
    \mathbb{E}\Big[\int_0^\infty K_tc_t\,dt\Big]\ge\mathbb{E}\Big[\int_0^\infty K_t\ell(H_t)\,dt\Big]\ge \mathbb{E}\Big[\int_0^\infty K_t\ell(h)\,dt\Big]\ge \frac{1}{r}\ell(h)\,.
\end{align}
Consequently, every admissible strategy satisfies
\begin{align}
    x \ge \mathbb E\bigg[\int_0^\infty K_t c_t\,dt\bigg]
    \ge \frac{1}{r}\ell(h)\,.
\end{align}
Thus $\bar{\mathcal A}(x,h)$ is empty if $x<\frac{1}{r}\ell(h)$.
In what follows, we focus on the nontrivial case \begin{align} x > \frac{1}{r}\ell(h). \end{align} Using the static budget constraint and a Lagrange multiplier $y>0$, we derive an upper bound for the auxiliary value function. For fixed $h>0$, define
\begin{align}\label{eq:bar-u-definition}
    \bar u(y,h)
    :=\sup_{\ell(h)\le c\le h}
      \left\{u\big(D(h)c-L(h)\big)-yc\right\}.
\end{align}
Let $I=(u')^{-1}$ on $(0,m_0)$ and define
\begin{align}\label{eq:b-d-Gamma}
    b(h):=D(h)u'(g(h)),
    \qquad
    d(h):=D(h)m_0,
    \qquad
    \Gamma(h):=g'(h)u'(g(h)),
\end{align}
with $d(h)=\infty$ if $m_0=\infty$. The following proposition characterizes the optimizer in \eqref{eq:bar-u-definition}.
\begin{prop}\label{prop:pointwise}
Suppose Assumption~\ref{assume:common} holds. Then, for every $y,h>0$, the unique maximizer in \eqref{eq:bar-u-definition} is
\begin{align}\label{eq:c-pointwise-general}
    c^*(y,h)
    =\begin{cases}
       h,
       &0<y<b(h),\\[1mm]
       \displaystyle
       \ell(h)+\frac1{D(h)}I\left(\frac{y}{D(h)}\right),
       &b(h)\le y<d(h),\\[3mm]
       \ell(h),
       &d(h)\le y<\infty,
      \end{cases}
\end{align}
and 
\begin{align}\label{eq:bar-u-piecewise}
    \bar u(y,h)
    =\begin{cases}
       u(g(h))-yh,
       &0<y<b(h),\\[1mm]
       \displaystyle
       u\left(I\left(\frac{y}{D(h)}\right)\right)
       -y\left[
          \ell(h)+\frac1{D(h)}I\left(\frac{y}{D(h)}\right)
       \right],
       &b(h)\le y<d(h),\\[3mm]
       u(0)-y\ell(h),
       &d(h)\le y<\infty.
      \end{cases}
\end{align}
Moreover, $\bar u\in C^1((0,\infty)^2)$,  and $\bar u(\cdot,h)$ is convex. More precisely,
\begin{align}
    \bar u_y(y,h)&=-c^*(y,h),\\
    \bar u_h(y,h)
    &=\begin{cases}
       \Gamma(h)-y,
       &0<y<b(h),\\[1mm]
       \displaystyle
       \frac{y}{D(h)}\big(D'(h)c^*(y,h)-L'(h)\big),
       &b(h)\le y<d(h),\\[3mm]
       -y\ell'(h),
       &d(h)\le y<\infty.
      \end{cases}\label{eq:bar-u-h}
\end{align}
\end{prop}
\begin{proof}
The proof is a straightforward calculation and is omitted.
\end{proof}
Next, for $(H,c,\pi)\in\bar{\mathcal A}(x,h)$, the static budget constraint gives
\begin{align}
    &\mathbb{E}\left[\int_0^\infty e^{-\rho t}
       u\big(D(H_t)c_t-L(H_t)\big)\,d t\right]\\
    &\quad=\mathbb{E}\left[\int_0^\infty e^{-\rho t}
       \Big(u\big(D(H_t)c_t-L(H_t)\big)-Y_t^{\rho,y}c_t\Big)\,d t\right]
       +y\mathbb{E}\left[\int_0^\infty K_tc_t\,d t\right]\notag\\
    &\quad\le xy+
       \mathbb{E}\left[\int_0^\infty e^{-\rho t}
       \bar u(Y_t^{\rho,y},H_t)\,d t\right].
\end{align}
Here  $Y_t^{\rho,y}=ye^{\rho t}K_t$.
The process $Y^{\rho,y}$ satisfies
\begin{align}
    dY_t^{\rho,y}
    &=
    (\rho-r)Y_t^{\rho,y}\,dt
    -\kappa Y_t^{\rho,y}\,dW_t,
    \qquad
    Y_0^{\rho,y}=y.
    \label{eq:Y-rho-dynamics}
\end{align}
Taking the supremum over admissible strategies yields
\begin{align}\label{eqn:2-1}
    \bar{v}_\rho(x,h)&=\sup_{(H,c,\pi)\in\bar{\mathcal{A}}(x,h)}
      \mathbb{E}\left[\int_0^\infty e^{-\rho t}
      u\big(D(H_t)c_t-L(H_t)\big)\,d t\right]\\
      &\le xy+\sup_{H\in \mathcal{H}(h)}\mathbb{E}\left[\int_0^\infty e^{-\rho t}
       \bar u(Y_t^{\rho,y},H_t)\,d t\right].
\end{align}
We define the dual value function by
\begin{align}\label{eq:dual-J}
    \mathcal J_\rho(y,h)
    :=\sup_{H\in\mathcal{H}(h)}
       \mathbb{E}\left[\int_0^\infty e^{-\rho t}
       \bar u(Y_t^{\rho,y},H_t)\,d t\right]\,.
\end{align}
By \eqref{eqn:2-1} and Lemma~\ref{lemma:exact-enlargement},
\begin{align}\label{eq:weak-duality}
    v_\rho(x,h)=\bar v_\rho(x,h)
    \le xy+\mathcal J_\rho(y,h),\qquad y>0\,.
\end{align}
Thus the only control remaining in \eqref{eq:dual-J} is the nondecreasing process $H$.

\section{Discounted problem}\label{sec:dis}
This section is devoted to the case $\rho>0$, encompassing the discounted formulation considered by \cite{deng2022optimal}. The presence of a positive discount rate ensures the boundedness of the value function under mild conditions and, importantly, permits the use of optimal stopping times in the subsequent analysis. Exploiting this property, we prove the existence of an optimal strategy $(c^*,\pi^*)\in\mathcal A(x,h)$ attaining the value function:
\begin{align}
    v_\rho(x,h)=\sup_{(c,\pi)\in \mathcal{A}(x,h)}\mathbb{E}[\int_0^\infty e^{-\rho t}u(D(H_t^{h,c})c_t-L(H_t^{h,c}))\,dt]=\mathbb{E}[\int_0^\infty e^{-\rho t} u(D(H_t^{h,c^*})c^*_t-L(H_t^{h,c^*}))\,dt]\,.
\end{align}
In addition, we derive a semi-explicit representation of $v_\rho(x,h)$.

Let $\gamma_1>1>0>\gamma_2$ be the two roots of
\begin{align}\label{eq:discounted-roots}
    \frac{\kappa^2}{2}\gamma(\gamma-1)
    +(\rho-r)\gamma-\rho=0\,.
\end{align}
When $m_0<\infty$, extend the inverse marginal utility to all positive
arguments by
\begin{align}\label{eq:extended-inverse-marginal}
    \widehat I(q)
    :=\begin{cases}
       I(q),&0<q<m_0,\\
       0,&q\ge m_0.
      \end{cases}
\end{align}
We write $I$ for this extension.
\begin{assume}
\label{assume:discounted}
The functions $D,L,u$ satisfy the following conditions.
\begin{enumerate}[label=(\roman*)]
    \item\label{assume:discounted-IRRA}
    The relative risk aversion
    \begin{align}\label{eq:relative-risk-aversion}
        R_u(z):=-\frac{z u''(z)}{u'(z)},
        \qquad z>0,
    \end{align}
    is positive and nondecreasing on $(0,\infty)$.

    \item\label{assume:discounted-habit-shape}
    The two functions
    \begin{align}\label{eq:primitive-habit-shape}
        h\longmapsto \ell'(h),
        \qquad
        h\longmapsto
        -\frac{D'(h)}{D(h)}\bigl(h-\ell(h)\bigr)
    \end{align}
    are nondecreasing on $(0,\infty)$.

    \item\label{assume:discounted-direct-tail}
    There exists $h_0>0$ such that
    \begin{align}\label{eq:discounted-direct-tail}
        \int_{h_0}^{\infty}
        \left(\Gamma(a)\right)^{1-\gamma_2}\,d a
        <\infty.
    \end{align}
\end{enumerate}
\end{assume}
For $H\in\mathcal H(h)$ and $a>h$, define
\begin{align}\label{eq:tau-H-detailed}
    \tau_a^H:=\inf\{t\ge0:H_t\ge a\},
    \qquad
    \mathcal T(\mathcal H(h))
    :=\left\{
       (\tau_a^H)_{a>h}:H\in\mathcal H(h)
    \right\}.
\end{align}
Since $H$ is adapted, nondecreasing, and right-continuous, for $t\ge0$,
\begin{align}\label{eq:hitting-event}
    \{\tau_a^H\le t\}&=\{H_t\ge a\}\,,\quad 
    H_t=h\vee\sup\{a>h:\tau_a^H\le t\}\,. 
\end{align}
The following theorem gives a representation of the dual value function in terms of these families of stopping times.
\begin{thm}
\label{thm:ordered-stopping}
Suppose Assumptions~\ref{assume:common} and~\ref{assume:discounted} hold.
Then, for every $y,h>0$,
\begin{align}\label{eq:ordered-stopping-representation}
    \mathcal J_\rho(y,h)
    =\mathbb{E}\left[\int_0^\infty e^{-\rho t}
      \bar u(Y_t^{\rho,y},h)\,d t\right]+\sup_{(\tau_a)_{a>h}\in\mathcal{T}(\mathcal{H}(h))}
      \int_h^\infty
      \mathbb{E}\left[\int_{\tau_a}^\infty e^{-\rho t}
      \bar u_h(Y_t^{\rho,y},a)\,d t\right]\,da\,.
\end{align}

\end{thm}
\begin{proof}
First, we prove that, for every $H\in\mathcal H(h)$,
\begin{align}
    \mathbb{E}\Big[\int_0^\infty e^{-\rho t}\bar u(Y_t^{\rho,y},H_t)\,dt\Big]=\mathbb{E}\Big[\int_0^\infty e^{-\rho t}\bar u (Y_t^{\rho,y},h)\,dt\Big]+\int_h^\infty \mathbb{E}\Big[\int_{\tau_a^H}^\infty e^{-\rho t}\bar u_h(Y_t^{\rho,y},a)\,dt\Big]\,da\,.\label{eqn:3-5}
\end{align}
By the fundamental theorem of calculus and \eqref{eq:hitting-event},
\begin{align}
    \bar u(Y_t^{\rho,y},H_t)=\bar u (Y_t^{\rho,y},h)+\int_h^\infty \mathds{1}_{a\le H_t}\bar u_h(Y_t^{\rho,y},a)\,da=\bar u (Y_t^{\rho,y},h)+\int_h^\infty \mathds{1}_{\tau_a^H\le t}\bar u_h(Y_t^{\rho,y},a)\,da.
\end{align}
By Lemma~\ref{lemma:primitive-shape-discounted}, we have
\begin{align}
    \mathbb{E}\Big[\int_0^\infty e^{-\rho t}|\bar u(Y_t^{\rho,y},h)|\,dt\Big]<\infty\,.\label{eqn:3-6}
\end{align}
Observe that $(\bar u_h(y,a))^+=(\Gamma(a)-y)^+$. Applying  \cite[Proposition 4.1 ]{knudsen1998valuation} for $y\ge\Gamma(a)$, we obtain 
\begin{align}
    \mathbb{E}\Big[\int_0^\infty e^{-\rho t} (\bar u_h(Y_t^{\rho,y},a))^+\,dt\Big]=
    \frac{2y^{\gamma_2}}
         {\kappa^2(\gamma_1-\gamma_2)}
    \int_0^{\Gamma(a)}
       t^{-\gamma_2-1}
       \big(\Gamma(a)-t\big)\,dt\le Cy^{\gamma_2}\Gamma(a)^{1-\gamma_2},\label{eqn:3-7}
\end{align}
for some constant $C>0$ independent of $y$ and $a$.
For $0<y<\Gamma(a)$, we have
\begin{align}
    \mathbb{E}\Big[\int_0^\infty e^{-\rho t} \mathds{1}_{a\le H_t}(\bar u_h(Y_t^{\rho,y},a))^+\,dt\Big]
    &\le \mathbb{E}\Big[\int_0^\infty e^{-\rho t} (\bar u_h(Y_t^{\rho,y},a))^+\,dt\Big]\\
    &\le \frac{\Gamma(a)}{\rho}\le \frac{y^{\gamma_2}\Gamma(a)^{1-\gamma_2}}{\rho}<\infty\,.\label{eqn:3-8}
\end{align}
Next, \eqref{eqn:3-7}, \eqref{eqn:3-8}, and \eqref{eq:discounted-direct-tail} imply
\begin{align}
    \mathbb{E}\Big[\int_0^\infty \int_h^\infty e^{-\rho t}\mathds{1}_{a\le H_t}(\bar u_h(Y_t^{\rho,y},a))^+\,da\,dt\Big]<\infty\,.\label{eqn:3-9}
\end{align}
By \eqref{eqn:3-6} and \eqref{eqn:3-9}, Tonelli's theorem applied separately to the positive and negative parts yields
\begin{align}
\mathbb{E}\Big[\int_0^\infty e^{-\rho t}\bar u(Y_t^{\rho,y},H_t)\,dt\Big]&=\mathbb{E}\Big[\int_0^\infty e^{-\rho t}\bar u(Y_t^{\rho,y},h)\,dt+\int_0^\infty \int_h^\infty e^{-\rho t}\mathds{1}_{a\le H_t}\bar u_h(Y_t^{\rho,y},a)\,da\,dt\Big] \\
&=\mathbb{E}\Big[\int_0^\infty e^{-\rho t}\bar u(Y_t^{\rho,y},h)\,dt+\int_h^\infty \int_0^\infty  e^{-\rho t}\mathds{1}_{\tau_a^H\le t}\bar u_h(Y_t^{\rho,y},a)\,dt\,da\Big]\,.
\end{align}
This proves \eqref{eqn:3-5}. Finally, taking the supremum over $H\in\mathcal H(h)$ yields \eqref{eq:ordered-stopping-representation}.

\end{proof}

\begin{prop}
\label{prop:discounted-boundary-scale}
Suppose Assumptions~\ref{assume:common} and~\ref{assume:discounted} hold. For each $a>0$, the equation
\begin{align}\label{eq:z-rho-general-equation}
    \int_z^\infty s^{-\gamma_1-1}\bar u_h(s,a)\,d s=0
\end{align}
has a unique solution $z_\rho(a)$ such that $(0,\Gamma(a))$.
Define $\zeta_\rho(a):=\frac{z_\rho(a)}{\Gamma(a)}\in(0,1)$.
Then $\zeta_\rho$ is continuous and nonincreasing, while $z_\rho$ is
continuous and strictly decreasing, with
\begin{align}
    \lim_{a\to\infty}z_\rho(a)=0.
\end{align}
Moreover, define
\begin{align}\label{eq:A-rho-definition}
    A_\rho(a)
    :=\frac{2}{\kappa^2(\gamma_1-\gamma_2)}
      \int_0^{z_\rho(a)}s^{-\gamma_2-1}\bar u_h(s,a)\,d s.
\end{align}
Then
\begin{align}\label{eq:A-rho-Gamma-form}
    A_\rho(a)
    &=\frac{2}{\kappa^2(\gamma_1-\gamma_2)}
      \Gamma(a)^{1-\gamma_2}
      \int_0^{\zeta_\rho(a)}s^{-\gamma_2-1}(1-s)\,d s,\\
    0<A_\rho(a)
    &\le C\Gamma(a)^{1-\gamma_2}\,,\label{eq:A-rho-Gamma-bound}
\end{align}
for some constant $C>0$.
Consequently,
\begin{align}\label{eq:A-rho-tail-finite}
    \int_h^\infty A_\rho(a)\,d a<\infty,
    \qquad h>0.
\end{align}
\end{prop}

\begin{proof}
First, we prove the existence and uniqueness of $z_\rho(a)$ in $(0,\Gamma(a))$. Define
\begin{align}
    F_a(z)=\int_z^\infty s^{-\gamma_1-1}\bar u_h(s,a)\,ds\,.
\end{align}
Observe that $\bar{u}_h(s,a)=\Gamma(a)-s$ for $0<s<b(a)$ and 
\begin{align}
\lim_{z\to 0}F_a(z)= \lim_{z\to 0} \int_z^{b(a)}s^{-\gamma_1-1} (\Gamma(a)-s)\,ds+ \int_{b(a)}^\infty s^{-\gamma_1-1}\bar u_h(s,a)\,ds\,.
\end{align}
By Lemma~\ref{lemma:primitive-shape-discounted} and direct calculation, we have
\begin{align}
    \lim_{z\to 0} F_a(z)=\infty\,.\label{eqn:3-2}
\end{align}
By \eqref{eq:bar-u-h} and Proposition~\ref{prop:normalized-marginal-order}, we obtain
\begin{align}
    F_a(\Gamma(a))=\int_{\Gamma(a)}^{b(a)} s^{-\gamma_1-1}\bar u_h(s,a)\,ds+\int_{b(a)}^\infty s^{-\gamma_1-1}\bar u_h(s,a)\,ds<0\,.\label{eqn:3-4}
\end{align}
Here we used $\bar u_h(s,a)<0$ for $\Gamma(a)<s\le b(a)$ and $\bar u_h(s,a)\le0$ for $s\ge b(a)$. Moreover, $F'_a(z)=-z^{-\gamma_1-1}(\Gamma(a)-z)<0$ for $0<z<\Gamma(a)\,.$ Thus \eqref{eqn:3-2}, \eqref{eqn:3-4}, and the intermediate value theorem imply the existence and uniqueness of $z_\rho(a)$ in $(0,\Gamma(a))$. 
Since $0<z_\rho(a)<\Gamma(a)$, we have $\lim_{a\to\infty}z_\rho(a)=0$ by Lemma~\ref{lemma:primitive-shape-discounted}.

Next, we prove that $\zeta_\rho$ is nonincreasing and continuous. Define
\begin{align}
    \bar F_a(z)
    :=
    \int_z^\infty
       s^{-\gamma_1-1}\psi_a(s)\,d s\,,\label{eqn:6-1}
\end{align}
where $\psi_a(s)=\frac{\bar u_h(\Gamma(a)s,a)}{\Gamma(a)}$.
By a change of variables, we have $\bar F_a(\zeta_\rho(a))=0$.
For $a_1<a_2$, Proposition~\ref{prop:normalized-marginal-order} implies $\bar F_{a_1}(z)\ge\bar F_{a_2}(z)$. Thus $0=\bar F_{a_2}(\zeta_\rho(a_2))\le \bar F_{a_1}(\zeta_\rho(a_2))$.
By Proposition~\ref{prop:normalized-marginal-order}, $\bar F_{a_1}'(z)=-z^{-\gamma_1-1}(1-z)<0$ for $0<z<1$. Thus,
\begin{align}
    \zeta_\rho(a_2)\le \zeta_\rho(a_1)\,.\label{eqn:3-3}
\end{align}
This proves that $\zeta_\rho$ is nonincreasing.
To prove continuity, let $(a_n)_{n\ge1}$ satisfy $a_n\to a>0$.
By Proposition~\ref{prop:normalized-marginal-order} and the dominated convergence theorem, $\bar F_a(z)$ is continuous in $a$ for every fixed $z>0$.
Since $\bar F_a$ is strictly decreasing on $(0,1)$, for $0<\varepsilon<\min\{\zeta_\rho(a),1-\zeta_\rho(a)\}$, 
\begin{align}
 \bar F_a(\zeta_\rho(a)-\varepsilon)>0\,,\quad \bar F_a(\zeta_\rho(a)+\varepsilon)<0\,.   
\end{align}
By continuity, there exists $N$ such that, for $n\ge N$,
\begin{align}
    \bar F_{a_n}(\zeta_\rho(a)-\varepsilon)>0\,,\quad \bar F_{a_n}(\zeta_\rho(a)+\varepsilon)<0\,.
\end{align}
Since $\zeta_\rho(a_n)$ is the unique zero of $\bar F_{a_n}$ in $(0,1)$, we have
\begin{align}
    \zeta_\rho(a)-\varepsilon<\zeta_\rho(a_n)<\zeta_\rho(a)+\varepsilon\,\quad \text{ for } n\ge N\,.
\end{align}
Thus $\zeta_\rho$ is continuous. By Lemma~\ref{lemma:primitive-shape-discounted}, $z_\rho(a)=\Gamma(a)\zeta_\rho(a)$ is continuous and strictly decreasing. Finally, a change of variables in \eqref{eq:A-rho-definition} implies \eqref{eq:A-rho-Gamma-form}. Since $0<\zeta_\rho(a)<1$ and $\gamma_2<0$, we obtain \eqref{eq:A-rho-Gamma-bound}, and \eqref{eq:discounted-direct-tail} implies \eqref{eq:A-rho-tail-finite}.

\end{proof}

For a fixed $a>0$, define
\begin{align}\label{eq:G-discounted-definition}
    G_\rho(y,a)
    :=\mathbb{E}_y\left[\int_0^\infty e^{-\rho t}
       \bar u_h(Y_t^{\rho,y},a)\,d t\right]\,,\quad R_\rho(y,a)
    :=\mathbb{E}\left[\int_0^\infty e^{-\rho t}
      \bar u(Y_t^{\rho,y},a)\,d t\right]\,.
\end{align}

By Theorem \ref{thm:ordered-stopping}, we have
\begin{align}
\mathcal J_\rho(y,h)
    &=R_\rho(y,h)+\sup_{\mathcal{T}(\mathcal{H}(h))}
      \int_h^\infty
      \mathbb{E}\left[\int_{\tau_a}^\infty e^{-\rho t}
      \bar u_h(Y_t^{\rho,y},a)\,d t\right]\,da\,\\
      &\le R_\rho(y,h)+
      \int_h^\infty
      \sup_{\tau \in \mathcal{T}}\mathbb{E}\left[\int_{\tau}^\infty e^{-\rho t}
      \bar u_h(Y_t^{\rho,y},a)\,d t\right]\,da\,,
\end{align}
where $\mathcal T$ denotes the class of $[0,\infty]$-valued stopping times.
Proposition \ref{prop:Green-discounted} provides an explicit representation
of $R_\rho$.  
We therefore turn to the optimal stopping problem
\begin{align}
    \sup_{\tau\in\mathcal T}
    \mathbb E_y\left[
       \int_\tau^\infty e^{-\rho t}
       \bar u_h(Y_t^{\rho,y},a)\,d t
    \right],
    \qquad a>0,
\end{align}
where the integral is defined to be zero on $\{\tau=\infty\}$.

By the strong Markov property, for any $\tau\in\mathcal T$,
\begin{align}\label{eq:strong-Markov-G}
    \mathbb E_y\left[
       \int_\tau^\infty e^{-\rho t}
       \bar u_h(Y_t^{\rho,y},a)\,d t
    \right]
    =
    \mathbb E_y\left[
       e^{-\rho\tau}
       G_\rho(Y_\tau^{\rho,y},a)
       \mathds{1}_{\{\tau<\infty\}}
    \right].
\end{align}
We therefore define
\begin{align}\label{eq:single-level-value}
    S_\rho(y,a)
    :=
    \sup_{\tau\in\mathcal T}
    \mathbb E_y\left[
       e^{-\rho\tau}
       G_\rho(Y_\tau^{\rho,y},a)
       \mathds{1}_{\{\tau<\infty\}}
    \right]\,.
\end{align}
The following theorem provides a representation of $\mathcal J_\rho$ and an optimal process.
\begin{thm}
\label{thm:discounted-dual-value}
Suppose Assumptions \ref{assume:common} and
\ref{assume:discounted} hold. Then, for every $y,h>0$,
\begin{align}\label{eq:J-R-plus-S}
    \mathcal J_\rho(y,h)
    =R_\rho(y,h)+\int_h^\infty S_\rho(y,a)\,d a.
\end{align}
Moreover,
\begin{align}\label{eq:J-discounted-piecewise}
    \mathcal J_\rho(y,h)
    =\begin{cases}
       \displaystyle
       R_\rho(y,h)+y^{\gamma_2}
       \int_h^\infty A_\rho(a)\,d a,
       &y>z_\rho(h),\\[4mm]
       \displaystyle
       R_\rho\big(y,z_\rho^{-1}(y)\big)
       +y^{\gamma_2}
       \int_{z_\rho^{-1}(y)}^\infty A_\rho(a)\,d a,
       &0<y\le z_\rho(h).
      \end{cases}
\end{align}
An optimal process is
\begin{align}\label{eq:H-star-discounted}
    H_t^{\rho,y,h}
    =z_\rho^{-1}\left(
       \min\left\{z_\rho(h),\inf_{0\le s\le t}Y_s^{\rho,y}\right\}
      \right).
\end{align}
\end{thm}

\begin{proof}
First, we prove that
\begin{align}
    \int_h^\infty S_\rho(y,a)\,da<\infty\,.\label{eqn:6-3}
\end{align}
For $y\ge z_\rho(h)$, Proposition~\ref{prop:discounted-boundary-scale} implies $z_\rho(a)\le z_\rho(h)\le y$ for every $a\ge h$. By Theorem~\ref{thm:fixed-level-discounted}, we have
\begin{align}
    \int_h^\infty S_\rho(y,a)\,d a
    =y^{\gamma_2}\int_h^\infty A_\rho(a)\,d a<\infty.\label{eqn:6-2}
\end{align}
Next, let $0<y\le z_\rho(h)$. By Proposition~\ref{prop:discounted-boundary-scale}, $z_\rho^{-1}(y)$ is well defined, and 
\begin{align}
    \int_h^\infty S_\rho(y,a)\,da=\int_h^{z_\rho^{-1}(y)}S_\rho(y,a)\,da+\int_{z_\rho^{-1}(y)}^\infty S_\rho(y,a)\,da\,.
\end{align}
Since $y<z_\rho(a)$ on $a\in [h,z_\rho^{-1}(y))$ and $y\ge z_\rho(a)$ on $a\in [z_\rho^{-1}(y),\infty)$, we have
\begin{align}\label{eqn:3-23}
    \int_h^\infty S_\rho(y,a)\,da=\int_h^{z_\rho^{-1}(y)}G_\rho(y,a)\,da+\int_{z_\rho^{-1}(y)}^\infty A_\rho(a)y^{\gamma_2}\,da<\infty\,.
\end{align}
The finiteness in \eqref{eqn:3-23} follows from the continuity of $G_\rho(y,\cdot)$ and \eqref{eq:A-rho-tail-finite}. Together with \eqref{eqn:6-2}, this proves \eqref{eqn:6-3}. Hence the right-hand side of \eqref{eq:J-R-plus-S} is finite.

We now prove \eqref{eq:J-R-plus-S}. Let $(\tau_a^*)_{a>h}$ be the family of optimal stopping times given by Theorem~\ref{thm:fixed-level-discounted}. It suffices to show that
\begin{align}
    (\tau_a^*)_{a>h}\in\mathcal{T}(\mathcal{H}(h))\,.
\end{align}
Indeed, from the definition of $S_\rho$, it follows that
\begin{align}
    \mathcal J_\rho(y,h)
    \le
    R_\rho(y,h)+\int_h^\infty S_\rho(y,a)\,da\,.
\end{align}
If $(\tau_a^*)_{a>h}\in \mathcal{T}(\mathcal{H}(h))$, then, since $\tau_a^*$ is optimal
for $S_\rho(y,a)$ for every $a\ge h$,
\begin{align}
    \mathcal J_\rho(y,h)
    \ge
    R_\rho(y,h)+\int_h^\infty
      \mathbb{E}\left[\int_{\tau^*_a}^\infty e^{-\rho t}
      \bar u_h(Y_t^{\rho,y},a)\,d t\right]\,da=R_\rho(y,h)+\int_h^\infty S_\rho(y,a)\,d a\,.    
\end{align}
We now define the process
\begin{align}
      H_t^{\rho,y,h}
    :=
    z_\rho^{-1}\left(
        \min\left\{
            z_\rho(h),
            \inf_{0\le s\le t}Y_s^{\rho,y}
        \right\}
    \right)\,,
    \qquad t\ge0\,.
\end{align}
Since $z_\rho$ is continuous and strictly decreasing and $Y^{\rho,y}$ has positive continuous paths, we have $H^{\rho,y,h}\in\mathcal H(h)$ with the convention $H_{0-}^{\rho,y,h}=h$. Moreover, for every $a>h$,

\begin{align*}
    \inf\left\{t\ge0:H_t^{\rho,y,h}\ge a\right\}
    &=
    \inf\left\{t\ge0:
        \inf_{0\le s\le t}Y_s^{\rho,y}\le z_\rho(a)
    \right\} \\
    &=
    \inf\left\{t\ge0:Y_t^{\rho,y}\le z_\rho(a)\right\}\,.
\end{align*}
By Theorem~\ref{thm:fixed-level-discounted}, $H^{\rho,y,h}$ induces the family $(\tau_a^*)_{a>h}$. Thus $(\tau_a^*)_{a>h}\in\mathcal T(\mathcal H(h))$, which proves \eqref{eq:J-R-plus-S}. Since each $\tau_a^*$ attains
$S_\rho(y,a)$, the process $H^{\rho,y,h}$   is optimal for $\mathcal J_\rho(y,h)$.
Finally, for $0<y\le z_\rho(h)$, we have
\begin{align}
    \int_h^{z_\rho^{-1}(y)}G_\rho(y,a)\,da=\int_h^{z_\rho^{-1}(y)}\frac{\partial R_\rho}{\partial a}(y,a)\,da= R_\rho(y,z_\rho^{-1}(y))-R_\rho(y,h)\,.\label{eqn:3-22}
\end{align}
Here we used Proposition~\ref{prop:Green-discounted}. Combining this identity with \eqref{eq:J-R-plus-S}, \eqref{eqn:6-2}, and \eqref{eqn:3-23} proves \eqref{eq:J-discounted-piecewise}.
\end{proof}

We now construct the optimal consumption process. For $y,h>0$, define
\begin{align}\label{eq:c-rho-y}
    c_t^{\rho,y,h}
    :=c^*\big(Y_t^{\rho,y},H_t^{\rho,y,h}\big).
\end{align}
The following lemma verifies that $H^{\rho,y,h}$ is the habit process associated with $c^{\rho,y,h}$.
\begin{lemma}
\label{lemma:discounted-running-maximum}

Suppose Assumptions \ref{assume:common} and
\ref{assume:discounted} hold. Then, for every $y,h>0$ and $t\ge0$,
\begin{align}\label{eq:H-is-running-rho}
    H_t^{\rho,y,h}
    =h\vee\sup_{0\le s\le t}c_s^{\rho,y,h}.
\end{align}
\end{lemma}

\begin{proof}
Since $c_s^{\rho,y,h}\le H_s^{\rho,y,h}$, $h\le H_s^{\rho,y,h}$, and $H^{\rho,y,h}$ is nondecreasing, we have
\begin{align}
    h \vee \sup_{0\le s\le t}c_s^{\rho,y,h}\le H_t^{\rho,y,h}\,.\label{eqn:3-33}
\end{align}
Next, if $H_t^{\rho,y,h}=h$, \eqref{eq:H-is-running-rho} follows from \eqref{eqn:3-33}. If $H_t^{\rho,y,h}>h$, then by \eqref{eq:H-star-discounted}, we have
\begin{align}\label{eq:H-at-running-minimum}
    H_t^{\rho,y,h}
    =z_\rho^{-1}\left(\inf_{0\le s\le t}Y_s^{\rho,y}\right).
\end{align}
By continuity of $Y^{\rho,y}$, choose $s_t\in[0,t]$ such that $Y_{s_t}^{\rho,y}=\inf_{0\le s\le t}Y_s^{\rho,y}$.
By \eqref{eq:H-star-discounted}, we have $ H_{s_t}^{\rho,y,h}=z_\rho^{-1}(Y_{s_t}^{\rho,y})=H_t^{\rho,y,h}$ and $Y_{s_t}^{\rho,y}=z_\rho(H_{s_t}^{\rho,y,h})\,$. Since $z_\rho(a)<\Gamma(a)<b(a)$, we have $Y_{s_t}^{\rho,y}<b(H_{s_t}^{\rho,y,h})$. By \eqref{eq:c-pointwise-general}, we obtain $c_{s_t}^{\rho,y,h}=H_{s_t}^{\rho,y,h}=H_t^{\rho,y,h}$.
Therefore,
\begin{align}
    h\vee\sup_{0\le s\le t}c_s^{\rho,y,h}
    \ge c_{s_t}^{\rho,y,h}
    =H_t^{\rho,y,h}.\label{eqn:3-34}
\end{align}
Combining \eqref{eqn:3-33} and \eqref{eqn:3-34}, we have
\begin{align}    H_t^{\rho,y,h}
    =h\vee\sup_{0\le s\le t}c_s^{\rho,y,h}.
\end{align}

\end{proof}

Now we state the main theorem of this section.
\begin{thm}
\label{thm:discounted-primal}

Suppose Assumptions \ref{assume:common} and
\ref{assume:discounted} hold. Then, for every $h>0$ and $x>\frac{\ell(h)}r$, there is a unique $y^*=y^*(x,h)>0$ satisfying
\begin{align}\label{eq:y-star-discounted}
    -\mathcal J_{\rho,y}(y^*,h)=x.
\end{align}
Moreover,
\begin{align}\label{eq:discounted-primal-value}
    v_\rho(x,h)=\bar v_\rho(x,h)
    =\mathcal J_\rho(y^*,h)+xy^*
    =\min_{y>0}\{\mathcal J_\rho(y,h)+xy\}\,,
\end{align}
and the optimal consumption and portfolio are 
\begin{align}
    c_t^*&:=c^*(Y_t^{\rho,y^*},H_t^{\rho,y^*,h})\,,\quad 
    \pi_t^*
    =\frac{\mu-r}{\sigma^2}
      Y_t^{\rho,y^*}\mathcal J_{\rho,yy}(Y_t^{\rho,y^*},H_t^{\rho,y^*,h})\label{eq:pi-star-discounted}\,.
\end{align}

\end{thm}

\begin{proof}
First, Proposition~\ref{prop:discounted-convexity} and Lemma~\ref{lemma:discounted-budget-derivative} imply the existence and uniqueness of $y^*$ in \eqref{eq:y-star-discounted}. By strict convexity, $y^*$ minimizes $\mathcal J_\rho(y,h)+xy$ over $y>0$.

Now we prove \eqref{eq:discounted-primal-value} and \eqref{eq:pi-star-discounted}.
By Lemma \ref{lemma:discounted-budget-derivative} and the Markov property, we have
\begin{align}
    \mathbb{E}\left[\int_0^\infty K_tc_t^*\,d t\right]=x\,,\quad X_t^*:=-\mathcal J_{\rho,y}(Y_t^{\rho,y^*},H_t^{\rho,y^*,h})=\frac1{K_t}
      \mathbb{E}_t\left[\int_t^\infty K_sc_s^*\,d s\right]\,.
    \label{eq:budget-equality-discounted}
\end{align}
Observe that $K_tX_t^*+\int_0^tK_sc_s^*\,d s=\mathbb{E}_t\left[\int_0^\infty K_sc_s^*\,d s\right]$
is a martingale. By the martingale representation theorem, there exists a locally square-integrable portfolio $\hat\pi$ such that
\begin{align}
    d X_t^*
    =
    \left(
        rX_t^*+(\mu-r)\hat \pi_t-c_t^*
    \right)d t
    +\sigma\hat \pi_t\,d W_t.
    \label{eq:X-star-financing-dynamics}
\end{align}
Since $\mathcal J_{\rho,y}\in C^1((0,\infty)^2)$
by Proposition \ref{prop:discounted-convexity}
and $H^{\rho,y^*,h}$ has continuous paths of finite variation,
we have
\begin{align}
    d\langle X^*,W\rangle_t
    =
    -\mathcal J_{\rho,yy}(Y_t^{\rho,y^*},H_t^{\rho,y^*,h})
      \,d\langle Y^{\rho,y^*},W\rangle_t=
    \kappa Y_t^{\rho,y^*}
    \mathcal J_{\rho,yy}(Y_t^{\rho,y^*},H_t^{\rho,y^*,h})\,dt\,.\label{eqn:3-39}
\end{align}
Thus, by \eqref{eq:X-star-financing-dynamics} and \eqref{eqn:3-39}, we obtain 
\begin{align}
    \hat \pi_t
    =\frac{\mu-r}{\sigma^2}
    Y_t^{\rho,y^*}\mathcal J_{\rho,yy}(Y_t^{\rho,y^*},H_t^{\rho,y^*,h})=\pi_t^*\,.
\end{align}
Thus $(X^*,c^*,\pi^*)$ satisfies the wealth equation.
By the continuity of $c^*$
and the local square integrability of $\pi^*$,
$(H^{\rho,y^*,h},c^*,\pi^*)\in\bar{\mathcal A}(x,h)$. Moreover,  $(c^*,\pi^*)\in\mathcal A(x,h)$ by
Lemma~\ref{lemma:discounted-running-maximum}.
By the pointwise optimality of $c^*$,
Theorem~\ref{thm:discounted-dual-value},
and \eqref{eq:budget-equality-discounted}, we obtain
$v_\rho(x,h)\ge J_\rho(x,h;c^*,\pi^*)
=\mathcal J_\rho(y^*,h)+xy^*$.
Together with \eqref{eq:weak-duality},
this proves \eqref{eq:discounted-primal-value}.

\end{proof}

\begin{cor}
The consideration of \cite{deng2022optimal} with an exponential utility corresponds to $\rho=r>0$, $\beta>0$, $0<\lambda<1$,
$u(\xi)=-\frac{1}{\beta}e^{-\beta\xi}$,  $D(h)=1$, and $L(h)=\lambda h$ under our setting.
Let $k_r\in(0,1-\lambda)$ be the unique root of
\begin{align}
 (1-\lambda)k_r^{\gamma_1}-\gamma_1k_r
 +(1-\lambda)(\gamma_1-1)=0,
\end{align}
where $\gamma_1:=\frac{1+\sqrt{1+\frac{8r}{\kappa^2}}}{2}$.
The corresponding dual value function is given by
\begin{align}
 \mathcal J_r(y,h)=
 \begin{cases}
 -\frac{1}{\beta r}-\frac{\lambda h}{r}y
 +\big[(C_{2,r}+C_{3,r})e^{-\beta(1-\lambda)\gamma_1h}-C_{2,r}\big]y^{\gamma_2},
 &y\ge1,\\[2mm]
 \begin{aligned}
 &\frac{y}{\beta r}
 \left(\log y-\beta\lambda h+\frac{\kappa^2}{2r}-1\right)
 -C_{1,r}y^{\gamma_1}\\
 &\quad+(C_{2,r}+C_{3,r})e^{-\beta(1-\lambda)\gamma_1h}y^{\gamma_2}
 \end{aligned},
 &e^{-\beta(1-\lambda)h}\le y<1,\\[2mm]
 \begin{aligned}
 &-\frac{e^{-\beta(1-\lambda)h}}{\beta r}-\frac{h}{r}y\\
 &\quad+C_{1,r}\big(e^{-\beta(1-\lambda)\gamma_2h}-1\big)y^{\gamma_1}
 +C_{3,r}e^{-\beta(1-\lambda)\gamma_1h}y^{\gamma_2}
 \end{aligned},
 &k_re^{-\beta(1-\lambda)h}\le y<e^{-\beta(1-\lambda)h},\\[2mm]
 \frac{y}{\beta r(1-\lambda)}
 \left(\log\frac{y}{k_r}+\frac{\kappa^2}{2r}\right)
 -\frac{y}{\beta r k_r}-C_{1,r}y^{\gamma_1},
 &0<y<k_re^{-\beta(1-\lambda)h},
 \end{cases}
\end{align}
where
\begin{align}
 \gamma_2&=1-\gamma_1,&C_{1,r}&:=\frac{1}{\beta r(\gamma_1-\gamma_2)(\gamma_1-1)},\\
 C_{2,r}&:=-\frac{1}{\beta r(\gamma_1-\gamma_2)\gamma_1},&C_{3,r}&:=\frac{k_r^{\gamma_1-1}}{\beta r(\gamma_1-\gamma_2)}
 \left(1-\frac{(\gamma_1-1)k_r}{(1-\lambda)\gamma_1}\right).
\end{align}
Moreover, for every $h>0$ and $x>\frac{\lambda h}{r}$, there exists
a unique $y^*=y^*(x,h)>0$ satisfying
\begin{align}
 -\mathcal J_{r,y}(y^*,h)=x,
\end{align}
and
\begin{align}
 v_r(x,h)=\bar v_r(x,h)
 =\mathcal J_r(y^*,h)+xy^*
 =\min_{y>0}\{\mathcal J_r(y,h)+xy\}.
\end{align}
The optimal consumption and portfolio are given by
\begin{align}
 c_t^*&=\min\left\{H_t^{r,y^*,h},\lambda H_t^{r,y^*,h}
 +\frac{1}{\beta}(-\log Y_t^{r,y^*})^+\right\},\\
 \pi_t^*&=\frac{\mu-r}{\sigma^2}Y_t^{r,y^*}
 \mathcal J_{r,yy}(Y_t^{r,y^*},H_t^{r,y^*,h}).
\end{align}
\end{cor}
\begin{proof}
The functions $u$, $D$, and $L$ satisfy
Assumptions~\ref{assume:common} and \ref{assume:discounted}.
The result follows by evaluating \eqref{eq:J-R-plus-S}
with $\rho=r$ and applying Theorem~\ref{thm:discounted-primal}.
\end{proof}

Note that our solution is not directly comparable to that in \cite{deng2022optimal}. Their analysis is based on dynamic programming and the associated HJB equation, under which the optimal consumption rate may reach zero. By contrast, our duality approach imposes a strictly positive lower bound on the consumption rate, which can be interpreted as a subsistence consumption constraint. If zero consumption is viewed as economically unrealistic, our approach provides a complementary solution and a novel contribution to the literature.

\section{Undiscounted Problem}\label{sec:undis}
This section is devoted to the case $\rho=0$, consistent with the setting considered by \cite{guasoni2020shortfall}. In the absence of discounting, the value function may be unbounded, precluding the use of the optimal-stopping arguments employed in the case $\rho>0$. We therefore adopt a variational inequality approach. Using this approach, we establish the existence of an optimal strategy $(c^*,\pi^*)\in\mathcal{A}(x,h)$ such that
\begin{align}
    v_0(x,h)=\sup_{(c,\pi)\in \mathcal{A}(x,h)}\mathbb{E}\Big[\int_0^\infty u(D(H_t^{h,c})c_t-L(H_t^{h,c}))\,dt\Big]
    =\mathbb{E}\Big[\int_0^\infty  u(D(H_t^{h,c^*})c^*_t-L(H_t^{h,c^*}))\,dt\Big]\,.
\end{align}
Moreover, we derive a semi-closed-form representation of $v_0(x,h)$.

When $m_0<\infty$, we extend the inverse marginal utility by
\begin{align}
    \widehat I(q)
    :=\begin{cases}
       I(q),&0<q<m_0,\\
       0,&q\ge m_0,
      \end{cases}
    \label{eq:Ihat-zero-revised}
\end{align}
and set $\widehat I=I$ when $m_0=\infty$.
Recall that $\ell(h)=\frac{L(h)}{D(h)}$ and $g(h)=D(h)h-L(h)$.

\begin{assume}
\label{assume:zero-general-new}
The functions $D$, $L$, and $u$ satisfy the following conditions.
\begin{enumerate}[label=(\roman*)]
    \item\label{assume:zero-utility-tail-weak}The utility function $u$ satisfies
    \begin{align}
        \lim_{z\to\infty}u(z)=0.
        \label{eq:zero-utility-tail-weak}
    \end{align}
    The relative risk aversion
    \begin{align}
        R_u(z):=-\frac{z u''(z)}{u'(z)},
        \qquad z>0,
        \label{eq:zero-RRA-weak}
    \end{align}
    is nondecreasing.

    \item\label{assume:zero-habit-order-weak}
    The two functions
    \begin{align}
        h\longmapsto \ell'(h),
        \qquad
        h\longmapsto
        -\frac{D'(h)}{D(h)}\bigl(h-\ell(h)\bigr)
        \label{eq:zero-habit-order-weak}
    \end{align}
    are nondecreasing.

    \item\label{assume:zero-crossing-weak}
    For every $h>0$ and every $0<z<g(h)$,
    \begin{align}
        {\ell'(h)}
        +\left(-\frac{D'(h)}{D(h)}\right)
          \frac{z}{D(h)}
          \left(1-\frac1{R_u(z)}\right)
        \ge0.
        \label{eq:zero-single-crossing-weak}
    \end{align}

    \item\label{assume:zero-trace-weak}
    There exists $h_0>0$ such that
    \begin{align}
        \int_{h_0}^\infty
        |u(g(a))|
        \left(-\frac{g''(a)}{g'(a)}\right)d a<\infty.
        \label{eq:zero-integrated-curvature-weak}
    \end{align}
\end{enumerate}
\end{assume}

For notational convenience, set
$\delta_0:=1+\frac{2r}{\kappa^2}>1$
and write $Y_t^y$ for $Y_t^{0,y}$. The following lemma will be used to construct the undiscounted dual value function.
\begin{lemma}
\label{lemma:zero-boundary-new}
 Suppose Assumptions~\ref{assume:common} and \ref{assume:zero-general-new} hold. For $h,s>0$, define  $d_h(s)=\frac{\bar u_h(\Gamma(h)s,h)}{\Gamma(h)}$. Then, for every $h>0$, the equation 
\begin{align}
    \int_{z}^\infty s^{-\delta_0-1}d_h(s)d s=0
\end{align}
has a unique solution $\zeta_0(h)$ such that $0<\zeta_0(h)<1$. The function $\zeta_0$ is continuously differentiable and nonincreasing. Moreover,
there is a constant $\underline\zeta>0$ such that
\begin{align}
    \underline\zeta\le\zeta_0(h)<1,
    \qquad h>0.
    \label{eq:zeta-uniform-bounds-revised}
\end{align}
Define $z_0(h):=\Gamma(h)\zeta_0(h)$. Then $z_0$ is continuously differentiable, strictly decreasing, and
converges to zero as $h\to\infty$. Moreover, $\int_{z_0(h)}^\infty
       s^{-\delta_0-1}\bar u_h(s,h)d s=0$.
\end{lemma}

\begin{proof}
First, arguing as in Proposition \ref{prop:discounted-boundary-scale},
using Lemma \ref{lemma:zero-primitive-tail-revised}
and Proposition \ref{prop:dh-ordering-revised},
we obtain a unique solution $\zeta_0(h)\in(0,1)$
that is nonincreasing in $h$.
Now we prove that $\zeta_0$ is continuously differentiable. By a simple calculation, we have
\begin{align}
    \frac{\partial}{\partial z}\int_z^\infty s^{-\delta_0-1}d_h(s)\,ds=-z^{-\delta_0-1}d_h(z)<0\,,\quad (h,z)\in(0,\infty)\times (0,1)\label{eqn:6-4}
\end{align}
and this derivative is continuous in $(h,z)$. Fix $h_0>0$. By Proposition~\ref{prop:dh-ordering-revised}, for $0<|\varepsilon|<h_0/2$ and $s>0$, 
\begin{align}
    \Big|\frac{d_{h_0+\varepsilon}(s)-d_{h_0}(s)}{\varepsilon}\Big|\le C(1+s)\,. 
\end{align}
By Proposition~\ref{prop:dh-ordering-revised},
for each fixed $h>0$, $d_h(s)$ is $C^1$ near $(h,s)$
for all but finitely many $s>0$.
The preceding bound and the dominated convergence theorem imply
\begin{align}
    \frac{\partial}{\partial h}\int_z^\infty s^{-\delta_0-1}d_h(s)\,ds=
    \int_z^\infty s^{-\delta_0-1}\frac{\partial}{\partial h}d_h(s)\,ds\,.\label{eqn:6-5}
\end{align}
By Proposition~\ref{prop:dh-ordering-revised}
and the dominated convergence theorem,
$\int_z^\infty s^{-\delta_0-1}
\frac{\partial}{\partial h}d_h(s)\,ds$
is continuous in $(h,z)$.
By \eqref{eqn:6-4}, \eqref{eqn:6-5}, the implicit function theorem, and uniqueness, $\zeta_0$ is continuously differentiable.

Next, we prove the uniform lower bound \eqref{eq:zeta-uniform-bounds-revised}.
Observe that 
\begin{align}
    \int_{\zeta_0(h)}^1s^{-\delta_0-1}d_h(s)\,ds=-\int_1^\infty s^{-\delta_0-1}d_h(s)\,ds\le C\,,
\end{align}
for some constant $C>0$.
Here we used \ref{B-2-3} in Proposition \ref{prop:dh-ordering-revised}.
Since $\lim_{\zeta\to0}\int_\zeta^1s^{-\delta_0-1}d_h(s)\,ds=\infty$, choose $\underline \zeta\in (0,1)$ such that
\begin{align}
    \int_{\underline \zeta}^1s^{-\delta_0-1}d_h(s)\,ds=\int_{\underline \zeta}^1s^{-\delta_0-1}(1-s)\,ds>C\,.
\end{align}
Since $\int_z^1s^{-\delta_0-1}d_h(s)\,ds$ is strictly decreasing in $z\in(0,1)$, we obtain $\underline\zeta\le\zeta_0(h)$ for every $h>0$.

Finally, $\Gamma$ is continuously differentiable by Assumption~\ref{assume:common}, and Lemma~\ref{lemma:zero-primitive-tail-revised} implies that it is strictly decreasing and converges to zero. 
Thus $z_0=\Gamma\zeta_0$ is continuously differentiable
and strictly decreasing to zero.
The final identity follows by a change of variables.
This completes the proof.
\end{proof}

For $y,h>0$, define
\begin{align}
    \Phi_0(y,h)
    :=\frac{2}{\kappa^2\delta_0}
      \left[
         \int_1^y\frac{\bar u(s,h)}sd s
         +y^{\delta_0}\int_y^\infty
            s^{-\delta_0-1}\bar u(s,h)d s
      \right].
    \label{eq:Phi-zero-new}
\end{align}
Then
\begin{align}
    \Phi_{0,y}(y,h)
    &=\frac{2}{\kappa^2}y^{\delta_0-1}
      \int_y^\infty s^{-\delta_0-1}\bar u(s,h)d s,
      \label{eq:Phi-y-zero-new}\\
    \Phi_{0,yh}(y,h)
    &=\frac{2}{\kappa^2}y^{\delta_0-1}
      \int_y^\infty s^{-\delta_0-1}\bar u_h(s,h)d s,
      \label{eq:Phi-yh-zero-new}
\end{align}
and $\frac{\kappa^2}{2}y^2\Phi_{0,yy}(y,h)-ry\Phi_{0,y}(y,h)+\bar u(y,h)=0$.
Now we define the candidate dual value function. Let 
\begin{align}
    B_0(h):=-\int_h^\infty
       \Phi_{0,y}(z_0(a),a)z_0'(a)d a,
    \label{eq:B-zero-new}
\end{align}
and 
\begin{align}
    V_0(y,h)
    :=\begin{cases}
       \displaystyle
       B_0(h)+\Phi_0(y,h)-\Phi_0( z_0(h),h),
       &y> z_0(h),\\[2mm]
       \displaystyle
       B_0\bigl( z_0^{-1}(y)\bigr),
       &0<y\le z_0(h).
      \end{cases}
    \label{eq:V-zero-general-new}
\end{align}
By Lemma~\ref{lemma:B-finite-zero-new}, $V_0(y,h)$ is finite for all $y,h>0$.
The following proposition shows that $V_0$ satisfies
the associated variational inequality.
\begin{prop}
\label{prop:VI-zero-general-new}

Suppose Assumptions~\ref{assume:common} and \ref{assume:zero-general-new} hold. Then $V_0\in C^2((0,\infty)^2)$. For each $h>0$, $V_0(\cdot,h)$ is strictly decreasing and strictly convex. Moreover, $V_0$ satisfies
\begin{align}
    \max\left\{
       \frac{\kappa^2}{2}y^2V_{0,yy}(y,h)-ryV_{0,y}(y,h)+\bar u(y,h),
       \ V_{0,h}(y,h)
    \right\}=0\,.
    \label{eq:VI-zero-general-new}
\end{align}
More precisely,
\begin{align}
   \frac{\kappa^2}{2}y^2V_{0,yy}-ryV_{0,y}+\bar u&=0,
   &V_{0,h}&<0,
   &&y> z_0(h),\\
   \frac{\kappa^2}{2}y^2V_{0,yy}-ryV_{0,y}+\bar u&\le0,
   &V_{0,h}&=0,
   &&0<y\le z_0(h)\,.
\end{align}
Moreover,
\begin{align}
    V_{0,yh}( z_0(h),h)=0.
    \label{eq:Vyh-boundary-zero-revised}
\end{align}
\end{prop}

\begin{proof}
We first verify \eqref{eq:VI-zero-general-new} for $y\ne z_0(h)$. By Lemma~\ref{lemma:B-finite-zero-new}, for $y>z_0(h)$, we have
\begin{align}
    V_{0,h}(y,h)=\Phi_{0,h}(y,h)-\Phi_{0,h}(z_0(h),h)=\int_{z_0(h)}^y\Phi_{0,yh}(s,h)\,ds\,.\label{eqn:4-21}
\end{align}
By Proposition \ref{prop:dh-ordering-revised} and Lemma \ref{lemma:zero-boundary-new}, 
\begin{align}\label{eqn:6-6}
    \int_y^\infty s^{-\delta_0-1}\bar u_h(s,h)\,ds=
    \begin{cases}
        -\int_{z_0(h)}^ys^{-\delta_0-1}\bar u_h(s,h)\,ds<0\,,\text{ for }z_0(h)<y\le \Gamma(h)\,,\\
    \int_y^\infty s^{-\delta_0-1}\bar u_h(s,h)\,ds\le 0\text{ for }y\ge\Gamma(h)\,.   
    \end{cases}
\end{align}
By \eqref{eq:Phi-yh-zero-new} and \eqref{eqn:6-6}, we have  $\Phi_{0,yh}(y,h)\le 0$ for $y>z_0(h)$ and $\Phi_{0,yh}(y,h)< 0$ for $z_0(h)<y\le \Gamma(h)$. Thus $V_{0,h}(y,h)<0$, and  $V_0(y,h)$ satisfies \eqref{eq:VI-zero-general-new} for $y>z_0(h)$. Next, we verify \eqref{eq:VI-zero-general-new} for $0<y<z_0(h)$.
By Assumption~\ref{assume:common}
and Lemma~\ref{lemma:zero-primitive-tail-revised},
$g''(h)\le0$, $g'(h)>0$, and $u''(g(h))<0$.
Thus $\Gamma'(h)=g''(h)u'(g(h))+(g'(h))^2u''(g(h))<0$.
Hence $z_0'(h)=\Gamma'(h)\zeta_0(h)+\Gamma(h)\zeta_0'(h)<0$.
By Lemmas~\ref{lemma:zero-boundary-new} and \ref{lemma:B-finite-zero-new}, we obtain
\begin{align}
    V_{0,y}(y,h)&=\frac{B_0'(z_0^{-1}(y))}{z_0'(z_0^{-1}(y))}=\Phi_{0,y}(y,z_0^{-1}(y))\\
    V_{0,yy}(y,h)&=\Phi_{0,yy}(y,z_0^{-1}(y))+\frac{\Phi_{0,yh}(y,z_0^{-1}(y))}{z_0'(z_0^{-1}(y))}=\Phi_{0,yy}(y,z_0^{-1}(y))\label{eqn:4-20}
\end{align}
Here we used $\Phi_{0,yh}(y,z_0^{-1}(y))=0$, which follows from \eqref{eq:Phi-yh-zero-new} and Lemma~\ref{lemma:zero-boundary-new}.
Consequently, 
\begin{align*}
    &\quad\frac{\kappa^2}{2}y^2V_{0,yy}(y,h)
      -ryV_{0,y}(y,h)+\bar u(y,h)=-\bar u(y,z_0^{-1}(y))+\bar u(y,h)\\
      &=-\int_h^{z_0^{-1}(y)}\bar u_h(y,a)\,da=-\int_h^{z_0^{-1}(y)}\bigl(\Gamma(a)-y\bigr)\,da \le 0\,, \text{ for }0<y<z_0(h)
\end{align*}
Here we used that $y\le z_0(a)\le \Gamma(a)$ for $a\le z_0^{-1}(y)$. Moreover, $V_{0,h}(y,h)=0$ for $0<y<z_0(h)$. Thus \eqref{eq:VI-zero-general-new} holds for $0<y<z_0(h)$.

Next, we prove that $V_0\in C^2((0,\infty)^2)$ and statisfy \eqref{eq:VI-zero-general-new}. By \eqref{eq:V-zero-general-new} and \eqref{eqn:4-20}, $V_0$, $V_{0,y}$, and $V_{0,yy}$ are continuous. 
For $0<y<z_0(h)$,  we have
\begin{align}
    V_{0,h}(y,h)=V_{0,yh}(y,h)=V_{0,hh}(y,h)=0\,.\label{eqn:4-22}
\end{align}
By \eqref{eq:Phi-yh-zero-new}, \eqref{eqn:4-21},  \eqref{eqn:4-22},  $V_{0,h}$ and $V_{0,yh}$ are continuous. Now we prove $V_{0,hh}$ is continuous.For $h$ in a compact interval $K\subset(0,\infty)$, Proposition~\ref{prop:dh-ordering-revised} implies
\begin{align}
    |\bar u_h(s,h)|&=|\Gamma(h)d_h(q)|\le C(1+s)\,,\\
    |\bar u_{hh}(s,h)|
    &=|\Gamma'(h)\bigl[d_h(q)-q\partial_q d_h(q)\bigr]
      +\Gamma(h)\partial_h d_h(q)|\le C(1+s)\,\label{eqn:4-56}
\end{align}
where $q=\frac{s}{\Gamma(h)}$.
By \eqref{eq:bar-u-h}, for each fixed $h>0$,
$\bar u_{hh}$ exists and is continuous near $(s,h)$
for almost every $s>0$.
By Proposition~\ref{prop:dh-ordering-revised},
\eqref{eqn:4-56}, and the dominated convergence theorem, we obtain
\begin{align}
    \Phi_{0,hh}(y,h)
    =\frac{2}{\kappa^2\delta_0}
      \left[
       \int_1^y\frac{\bar u_{hh}(s,h)}s\,ds
       +y^{\delta_0}\int_y^\infty
        s^{-\delta_0-1}\bar u_{hh}(s,h)\,ds
      \right]\,.
\end{align}By \eqref{eqn:4-56} and the dominated convergence theorem, $\Phi_{0,hh}$ is jointly continuous. Hence $\Phi_0\in C^2((0,\infty)^2)$. For $y>z_0(h)$, differentiating \eqref{eqn:4-21}, we obtain
\begin{align} V_{0,hh}(y,h)=\Phi_{0,hh}(y,h)-\Phi_{0,hh}(z_0(h),h)\,.\label{eqn:4-23}
\end{align}
Here we used
$\Phi_{0,hy}(z_0(h),h)=\Phi_{0,yh}(z_0(h),h)=0$. By \eqref{eqn:4-22}, \eqref{eqn:4-23} and $\Phi_0\in C^2((0,\infty)^2)$, $V_{0,hh}(y,h)$ is continuous. Thus $V_0\in C^2((0,\infty)^2)$. The equality \eqref{eq:VI-zero-general-new} at $y=z_0(h)$ and \eqref{eq:Vyh-boundary-zero-revised} follow by continuity.

Now we prove that $V_0$ is strictly decreasing and strictly convex.
For $y\ge z_0(h)$,
\begin{align}
    V_{0,y}(y,h)= \frac{2}{\kappa^2}y^{\delta_0-1}
      \int_y^\infty s^{-\delta_0-1}\bar u(s,h)d s<0\,.
\end{align}
By \eqref{eqn:4-20}, $V_{0,y}(y,h)<0$ holds for $0<y<z_0(h)$. Thus $V_0(\cdot,h)$ is strictly decreasing. Next, we prove strict convexity. By \eqref{eq:Phi-y-zero-new}, we have
\begin{align}
    \Phi_{0,yy}(y,h)
    &=\frac{2}{\kappa^2}
      \Big(
       (\delta_0-1)y^{\delta_0-2}
       \int_y^\infty s^{-\delta_0-1}\bar u(s,h)\,ds
       -\frac{\bar u(y,h)}{y^2}
      \Big)\,.
\end{align}
Observe that
\begin{align}
    \int_y^\infty s^{-\delta_0-1}\bar u(s,h)\,ds&=\int_y^\infty s^{-\delta_0-1}\big(\bar u(y,h)-\int_y^sc^*(q,h)\,dq\big)\,ds\\
    &=\frac{\bar u(y,h)y^{-\delta_0}}{\delta_0}
      -\frac1{\delta_0}
       \int_y^\infty q^{-\delta_0}c^*(q,h)\,dq\,.
\end{align}
Since $c^*(\cdot,h)$ is nonincreasing, we have
\begin{align}
    \Phi_{0,yy}(y,h)
    &=\frac{2}{\kappa^2\delta_0y^2}
      \Big(
       -\bar u(y,h)
       -(\delta_0-1)y^{\delta_0}
       \int_y^\infty q^{-\delta_0}c^*(q,h)\,dq
      \Big)\\
      &\ge \frac{2}{\kappa^2\delta_0y^2}
      \Big(
       -\bar u(y,h)
       -(\delta_0-1)y^{\delta_0}c^*(y,h)
       \int_y^\infty q^{-\delta_0}\,dq
      \Big)=\frac{2}{\kappa^2\delta_0y^2}(-\bar u(y,h)-yc^*(y,h))\\
      &\ge -\frac{2}{\kappa^2\delta_0y^2} u(D(h)c^*(y,h)-L(h))>0\,.\label{eqn:4-57}
\end{align}
By \eqref{eqn:4-20} and \eqref{eqn:4-57}, $V_{0,yy}>0$ for all $y,h>0$. This proves strict convexity and completes the proof.

\end{proof}

For $y,h>0$, define $H_{0-}^{y,h}:=h$ and
\begin{align}
    H_t^{y,h}
    := z_0^{-1}\left(
       \min\left\{
           z_0(h),
          \inf_{0\le s\le t}Y_s^y
       \right\}
      \right),
    \qquad t\ge0.
    \label{eq:H-star-zero-general-new}
\end{align}
Since $z_0^{-1}$ is continuous and strictly decreasing, $H^{y,h}$ is adapted, nondecreasing, and continuous for $t\ge0$. Thus $H^{y,h}\in\mathcal H(h)$.
Let
\begin{align}
    c_t^{y,h}
    :=c^*(Y_t^y,H_t^{y,h}).
    \label{eq:c-zero-general-new}
\end{align}
The following lemma verifies that $H^{y,h}$ is the habit process associated with $c^{y,h}$.
\begin{lemma}
 Suppose Assumptions~\ref{assume:common} and \ref{assume:zero-general-new} hold. Then, for every $y,h>0$ and $t\ge0$,
\begin{align}
    H_t^{y,h}
    =h\vee\sup_{0\le s\le t}c_s^{y,h}\,.
    \label{eq:running-maximum-zero-general-new}
\end{align}
\end{lemma}
\begin{proof}
   The proof is similar to that of
Lemma~\ref{lemma:discounted-running-maximum} and is omitted. 
\end{proof}

The following theorem shows that $V_0$ is the undiscounted dual value function.
\begin{thm}
\label{thm:dual-value-zero-general-new}
Suppose Assumptions~\ref{assume:common} and \ref{assume:zero-general-new} hold. Then, for every $y,h>0$,
\begin{align}
    \mathcal J_0(y,h)=V_0(y,h)\,,\label{eq:dual-value-zero-general-new}
\end{align}
and the process $H^{y,h}$ in
\eqref{eq:H-star-zero-general-new} is optimal for $\mathcal J_0(y,h)$.
\end{thm}

\begin{proof}
 Recall the auxiliary functions $\mathfrak a$, $\Psi$, and $w$
defined in \eqref{eq:a-transform-zero-new}, \eqref{eq:Psi-transform-zero-new} and \eqref{eq:w-general-revised}, respectively.
We first prove that, for every $H\in\mathcal H(h)$,
\begin{align}
    \mathbb E\left[\int_0^\infty
       \bar u(Y_t^y,H_t)\,dt\right]
    \le V_0(y,h)\,.
    \label{eq:dual-comparison-goal-explained}
\end{align}
Let $A_t=\mathfrak a(H_t)$, $A_t^*=\mathfrak a(H_t^{y,h})$ and $\Delta A_t=A_t-A_t^*$.
By It\^{o}'s formula and Lemma~\ref{lemma:adjoint-zero-revised}, we have
\begin{align}
    \Delta A_Tw(Y_T^y, A^*_T)
    &=\Delta A_0w(Y_0^y, A^*_0)
      +\int_0^T\Delta A_t\Psi_A(Y_t^y,A_t^*)\,dt
      +\int_0^T w(Y_t^y, A^*_t)\,dA_t\\
      &\quad
      -\int_0^T w(Y_t^y, A^*_t)\,dA_t^*
      -\kappa\int_0^T
       \Delta A_{t-}Y_t^yw_y(Y^y_t,A_t^*)\,dW_t\,.
    \label{eq:comparison-product-expanded-explained}
\end{align}
By the definitions of $w$ and $A^*$ and \eqref{eqn:3-43}, we have 
\begin{align}
  w(Y_t^y, A_t^*)=\frac{V_{0,h}(Y_t^y,H_t^{y,h})}{\Gamma(H_t^{y,h})}\le0\,,\quad \int_0^Tw(Y_t^y,A_t^*)\,dA_t^*=0\,,\label{eqn:4-46}
\end{align}
For $y\ge z_0(h)$, $A_0^*=\mathfrak a(h)$. Since $H_0\ge h$ and $\mathfrak a$ is decreasing, \eqref{eqn:4-46} implies
\begin{align}
    \Delta A_0w(Y^y_0,A_0^*)\ge0\,.\label{eqn:4-47}
\end{align}
For $y\le z_0(h)$, we have $y=z_0(H_0^{y,h})$. Thus,
\begin{align}
    \Delta A_0w(Y^y_0, A_0^*)=\Delta A_0\frac{V_{0,h}(y,H_0^{y,h})}{\Gamma(H_0^{y,h})}=\Delta A_0\frac{V_{0,h}(z_0(H_0^{y,h}),H_0^{y,h})}{\Gamma(H_0^{y,h})}=0\,.\label{eqn:4-48}
\end{align}
Since $A$ is nonincreasing, \eqref{eqn:4-46}, \eqref{eqn:4-47}, and \eqref{eqn:4-48} imply
\begin{align}
    \Delta A_Tw(Y_T^y, A^*_T)
    &\ge\int_0^T\Delta A_t\Psi_A(Y_t^y,A_t^*)\,dt
      -\kappa\int_0^T
       \Delta A_{t-}Y_t^yw_y(Y^y_t,A_t^*)\,dW_t\,,
\end{align}
By Proposition~\ref{prop:dh-ordering-revised} and Lemmas~\ref{lemma:Psi-concavity-revised} and \ref{lemma:adjoint-zero-revised}, we have 
\begin{align}
    \Big|\Psi_A(Y^y_t,A_t^*)\Big|&=\Big|\frac{\bar u_h(Y^y_t,H_t^{y,h})}{\Gamma(H_t^{y,h})}\Big|= \Big|d_{H_t^{y,h}}(\frac{Y^y_t}{\Gamma(H_t^{y,h})})\Big|\le 1+\frac{Y^y_t}{\Gamma(H^{y,h}_t)}\le 1+\frac{Y^y_t}{z_0(H^{y,h}_t)}\,,\\
    \Big|Y_t^yw_y(Y^y_t,A_t^*)\Big|&\le C(h)(1+\frac{Y^y_t}{z_0(H_t^{y,h})})\,,\quad|\Delta A_{t-}|\le \mathfrak a(h)\,.
\end{align}
By \eqref{eqn:5-1} and Lemma~\ref{lemma:reflected-moments-zero-new}, the stochastic integral is a martingale. Taking expectations, we obtain

\begin{align}
    \mathbb{E}\Big[\int_0^T\Delta A_t\Psi_A(Y_t^y,A_t^*)\,dt\Big]\le \mathbb{E}[\Delta A_Tw(Y_T^y, A_T^*)]\label{eqn:4-49}
\end{align}
By Lemma \ref{lemma:Psi-concavity-revised}, 
\begin{align}
    \bar u(Y^y_t,H_t)-\bar u(Y_t^y,H_t^{y,h})=\Psi(Y_t^y,A_t)-\Psi(Y_t^y,A_t^*)\le \Psi_A(Y_t^y,A_t^*)\Delta A_t\,.
\end{align}
Thus, for $H$ satisfying \eqref{eq:finite-payoff-H-zero-new}, we obtain 
\begin{align}
    \lim_{T\to \infty }\mathbb{E}[\int_0^T \bar u (Y_t^y,H_t)\,dt]&\le \lim_{T\to \infty}\mathbb{E}[\int_0^T\bar u (Y_t^y,H_t^{y,h})+\Delta A_t\Psi_A(Y^y_t,A_t^*)\,dt]\\
    &\le \lim_{T\to \infty }\mathbb{E}[\int_0^T\bar u (Y_t^y,H_t^{y,h})\,dt+\Delta A_Tw(Y_T^y,A_T^*)]\,.\label{eqn:4-50}
\end{align}
By Lemma~\ref{lemma:competitor-terminal-zero-new}
and Proposition~\ref{prop:candidate-payoff-zero-general-new},
\eqref{eq:dual-comparison-goal-explained} holds.
For $H\in\mathcal H(h)$ the inequality
\eqref{eq:dual-comparison-goal-explained} holds trivially if
\begin{align*}
    \mathbb E\left[\int_0^\infty
    \bar u(Y_t^y,H_t)\,dt\right]=-\infty.
\end{align*}
Taking the supremum over $H\in\mathcal H(h)$, we obtain $\mathcal J_0(y,h)\le V_0(y,h)$. Since $H^{y,h}\in\mathcal H(h)$, Proposition~\ref{prop:candidate-payoff-zero-general-new} implies the reverse inequality and the optimality of $H^{y,h}$. This proves \eqref{eq:dual-value-zero-general-new} and completes the proof.

\end{proof}

\begin{lemma}
\label{lemma:budget-zero-general-new}
Suppose Assumptions~\ref{assume:common} and \ref{assume:zero-general-new} hold. Then, for every $y,h>0$,
\begin{align}
    -V_{0,y}(y,h)=\mathbb{E}\left[\int_0^\infty K_t c_t^{y,h}d t\right]\,.
    \label{eq:budget-derivative-zero-general-new}
\end{align}
Moreover,
\begin{align}
    \lim_{y\to\infty}\bigl(-V_{0,y}(y,h)\bigr)
    &=\frac{\ell(h)}r\,,\quad \lim_{y\to 0}\bigl(-V_{0,y}(y,h)\bigr)
    =\infty.
    \label{eq:budget-limit-zero-zero-new}
\end{align}
\end{lemma}

\begin{proof}
First, \eqref{eq:budget-derivative-zero-general-new} follows as in the proof of Lemma~\ref{lemma:discounted-budget-derivative}, using Theorem~\ref{thm:dual-value-zero-general-new}. Next, we prove \eqref{eq:budget-limit-zero-zero-new}.
If $m_0<\infty$, then, for sufficiently large $y$,
\begin{align}
    V_{0,y}(y,h)=\Phi_{0,y}(y,h)
    &=\frac{2}{\kappa^2}y^{\delta_0-1}
      \int_y^\infty s^{-\delta_0-1}\bar u(s,h)d s=\frac{2}{\kappa^2}y^{\delta_0-1}
      \int_y^\infty s^{-\delta_0-1}\bigl(u(0)-s\ell(h)\bigr)\,d s\,.
\end{align}
By direct calculation, we obtain $\lim_{y\to\infty}(-V_{0,y}(y,h))=\frac{\ell(h)}r$. 
Suppose now that $m_0=\infty$. For all sufficiently large $y$,
\begin{align}
    \frac{\bar u(y,h)}{y}
    =-\ell(h)
      +\frac{u\!\left(I(y/D(h))\right)}{y}
      -\frac{I(y/D(h))}{D(h)}\,.\label{eqn:4-51}
\end{align}
For $\eta>0$, by concavity of $u$, we have $u(\eta)\le u\!\left(I(y/D(h))\right)+\frac{y}{D(h)}\left(\eta-I(y/D(h))\right)$.
Since $u<0$, this implies
\begin{align*}
    \frac{u(\eta)}{y}
    -\frac{\eta}{D(h)}
    +\frac{I(y/D(h))}{D(h)}
    \le
    \frac{u\!\left(I(y/D(h))\right)}{y}
    \le0.
\end{align*}
Since $\lim_{y\to \infty}I(y/D(h))=0$ as $y\to\infty$, letting $y\to\infty$ and then $\eta\downarrow0$, we obtain
\begin{align}
   \lim_{y\to \infty} \frac{u\!\left(I\!\left(\frac{y}{D(h)}\right)\right)}{y}=0\,.\label{eq:utility-inverse-tail-zero-review}
\end{align}
By \eqref{eqn:4-51}, \eqref{eq:utility-inverse-tail-zero-review}, and l'H\^opital's rule, we have
\begin{align}
    \lim_{y\to \infty}V_{0,y}(y,h)=\lim_{y\to \infty} \frac{2}{\kappa^2}\frac{
      \int_y^\infty s^{-\delta_0-1}\bar u(s,h)d s}{y^{1-\delta_0}}=\lim_{y\to\infty}\frac{2}{\kappa^2(\delta_0-1)}\frac{\bar u(y,h)}{y}\,.
\end{align}
Thus the first limit in \eqref{eq:budget-limit-zero-zero-new} follows from $\delta_0-1=2r/\kappa^2$. The second limit follows as in Lemma~\ref{lemma:discounted-budget-derivative}, using the properties of $z_0$ and $\zeta_0$ in Lemma~\ref{lemma:zero-boundary-new}.

\end{proof}
Now we state the main theorem of this section.
\begin{thm}
\label{thm:primal-zero-general-new}
Suppose Assumptions~\ref{assume:common} and \ref{assume:zero-general-new} hold. Then, for every $h>0$ and $x>\frac{\ell(h)}r$, there is a unique $y^*=y^*(x,h)>0$ satisfying
\begin{align}
    -V_{0,y}(y^*,h)=x.
\end{align}
Moreover,
\begin{align}
   v_0(x,h)=\bar v_0(x,h)
    =V_0(y^*,h)+xy^*
    =\min_{y>0}\{V_0(y,h)+xy\}\,,
\end{align}
and the optimal consumption and portfolio are
\begin{align}
    c_t^*&:=c^*(Y_t^{y^*},H_t^{y^*,h})\,,\quad 
    \pi_t^*
    =\frac{\mu-r}{\sigma^2}
      Y_t^{y^*}V_{0,yy}(Y_t^{y^*},H_t^{y^*,h})\,.
\end{align}

\end{thm}

\begin{proof}
The proof follows the same argument as in
Theorem~\ref{thm:discounted-primal} and is omitted.
\end{proof}

\begin{cor}
The consideration of \cite{guasoni2020shortfall} corresponds to $\rho=0$, $\gamma>1$, $0<\alpha<1$,
$u(\xi)=\frac{\xi^{1-\gamma}}{1-\gamma}$, $D(h)=h^{-\alpha}$, and $L(h)=0$ in our setting. Let $k\in(0,1-\alpha)$ be the unique root of
\begin{align}
 (1-\alpha)(\delta-1)-\delta k
 +\frac{\delta+p(\delta-1)}{\gamma\delta-\gamma+1}k^\delta=0,
\end{align}
where $p:=(1-\alpha)(\gamma-1)$ and
$\delta:=1+\frac{2r}{\kappa^2}$.
The corresponding dual value function is given by
\begin{align}
 V_0(y,h)=
 \begin{cases}
 C_3h^{-p}-\frac{\gamma C_1}{\gamma-1}
 h^{\frac{\alpha(\gamma-1)}{\gamma}}
 y^{\frac{\gamma-1}{\gamma}},
 &y\ge h^{-1-p},\\[2mm]
 h^{-p}\left\{
 \begin{aligned}
 &-\frac{(1+p)kC_2}{p}
 -\frac{2}{\kappa^2}\left[
 \frac{\log\frac{yh^{1+p}}{k}}{(\gamma-1)\delta}
 +\frac{yh^{1+p}-k}{\delta-1}\right]\\
 &\quad+\frac{2\big[(yh^{1+p})^\delta-k^\delta\big]}
 {\kappa^2\delta^2(\delta-1)(\gamma\delta-\gamma+1)}
 \end{aligned}
 \right\},
 &kh^{-1-p}\le y<h^{-1-p},\\[3mm]
 -\frac{(1+p)kC_2}{p}
 \left(\frac{y}{k}\right)^{\frac{p}{1+p}},
 &0<y<kh^{-1-p},
 \end{cases}
\end{align}
where
\begin{align}
 C_1&=\frac{2\gamma^2}
 {\kappa^2(\gamma-1)(\gamma\delta-\gamma+1)},\quad C_2=\frac{2}{\kappa^2}\left[
 \frac{1}{(\gamma-1)\delta k}
 +\frac{1}{\delta-1}
 -\frac{k^{\delta-1}}
 {\delta(\delta-1)(\gamma\delta-\gamma+1)}\right],\\
 C_3&=-\frac{(1+p)kC_2}{p}+\frac{\gamma C_1}{\gamma-1}-\frac{2}{\kappa^2}\left[
 \frac{\log\frac{1}{k}}{(\gamma-1)\delta}
 +\frac{1-k}{\delta-1}
 -\frac{1-k^\delta}
 {\delta^2(\delta-1)(\gamma\delta-\gamma+1)}\right].
\end{align}
Moreover, for every $x,h>0$, there exists a unique
$y^*=y^*(x,h)>0$ satisfying
\begin{align}
 -V_{0,y}(y^*,h)=x,
\end{align}
and
\begin{align}
 v_0(x,h)=\bar v_0(x,h)
 =V_0(y^*,h)+xy^*
 =\min_{y>0}\{V_0(y,h)+xy\}.
\end{align}
The optimal consumption and portfolio are given by
\begin{align}
 c_t^*&=\min\left\{H_t^{y^*,h},
 (H_t^{y^*,h})^{\frac{\alpha(\gamma-1)}{\gamma}}
 (Y_t^{y^*})^{-\frac{1}{\gamma}}\right\},\\
 \pi_t^*&=\frac{\mu-r}{\sigma^2}Y_t^{y^*}
 V_{0,yy}(Y_t^{y^*},H_t^{y^*,h}).
\end{align}
\end{cor}
\begin{proof}
The functions $u$, $D$, and $L$ satisfy
Assumptions~\ref{assume:common} and \ref{assume:zero-general-new}.
The result follows by evaluating \eqref{eq:V-zero-general-new}
and applying Theorems~\ref{thm:dual-value-zero-general-new}
and \ref{thm:primal-zero-general-new}.
\end{proof}

Under the power utility specification, both the dynamic programming and duality approaches yield a positive optimal consumption rate. Our solution, however, differs from that of \cite{guasoni2020shortfall}, as our derivation does not rely on \cite[Lemma A.1]{guasoni2020shortfall}. How the two solutions can be reconciled remains an interesting question that warrants further investigation. As an initial observation, straightforward, albeit tedious, calculations show that our value function dominates that obtained in \cite{guasoni2020shortfall}.

\section{Conclusion}\label{conclusion}
This paper revisits the shortfall-aversion problem through a duality approach, enlarging the admissible control space to treat the consumption peak as both a control and a state variable. The dual formulation reveals a useful structural separation between the regular consumption-investment decisions and the singular control governing increases in the consumption peak. As a result, the singular component can be characterized through an optimal stopping formulation in the discounted setting or a variational inequality in the undiscounted setting. This separation not only simplifies the analysis but also clarifies the distinct economic roles of regular portfolio-consumption choices and adjustments to the consumption benchmark. More importantly, the duality approach accommodates a broader class of utility functions and thereby provides a unified framework for shortfall-aversion problems beyond the specific formulations considered in the existing literature.\\
\bigskip

\noindent {\bf Funding:} H.Y. Wong received funding from the RGC-NSFC Joint Research Scheme 2024/25.\\

\noindent {\bf Ethics Statement:} Ethics approval was not required for this study.\\

\noindent {\bf Conflicts of Interest:} The authors declare no conflicts of interest.\\

\noindent {\bf Data Availability Statement:} Data sharing is not applicable to this article, as no datasets were generated or analysed.

\bibliographystyle{apalike}
\bibliography{references}

\newpage
\appendix
\section{ Some auxiliary results in Section \ref{sec:dis}}\label{sec:A}

For later calculations, let
\begin{align}\label{eq:vartheta-chi}
    \vartheta(h)
    :=\frac{g'(h)}{D(h)},
    \qquad
    \chi(h)
    :=-\frac{D'(h)}{D(h)}\bigl(h-\ell(h)\bigr)\,.
\end{align}
Since $g(h)=D(h)(h-\ell(h))$, we have
\begin{align}\label{eq:vartheta-identity}
    \vartheta(h)=1-\ell'(h)-\chi(h).
\end{align}
By Assumption~\ref{assume:common}, $0<\vartheta(h)<1$.

\begin{lemma}\label{lemma:primitive-shape-discounted}

Suppose Assumptions~\ref{assume:common} and
\ref{assume:discounted} hold. Then the following properties hold.
\begin{enumerate}[label=(\roman*)]
    \item\label{3-i}
    The functions $\vartheta$ and $g'$ are nonincreasing.

    \item\label{3-ii}
    The function $\Gamma(h)$ is continuous and strictly decreasing, and satisfies
    \begin{align}\label{eq:Gamma-endpoints}
        \lim_{h\to\infty}\Gamma(h)=0.
    \end{align}

    \item\label{3-iii}
    For every fixed $t>1$, the map
    \begin{align}\label{eq:inverse-marginal-ratio}
        z\longmapsto
        \frac{ I(tu'(z))}{z}
    \end{align}
    is nondecreasing on $(0,\infty)$. Moreover, for every fixed $z>0$,
    it is nonincreasing as a function of $t$.
    \item\label{lemma:local-linear-growth}
 There is a continuous function $C:(0,\infty)\to (0,\infty)$ such that
\begin{align}\label{eq:local-linear-growth}
    |\bar u(y,h)|+|\bar u_h(y,h)|
    \le C(h)(1+y),
    \qquad y>0\,.
\end{align}
Moreover, for every $y,h>0$,
\begin{align}\label{eqn:3-1}
   \int_0^y s^{-\gamma_2-1}
     \big(|\bar u(s,h)|+|\bar u_h(s,h)|\big)\,ds+
     \int_y^\infty s^{-\gamma_1-1}
     \big(|\bar u(s,h)|+|\bar u_h(s,h)|\big)\,d s<\infty.
\end{align}
\end{enumerate}
\end{lemma}

\begin{proof}
First, we prove \ref{3-i}. By Assumption~\ref{assume:discounted} and \eqref{eq:vartheta-identity}, $\vartheta$ is nonincreasing. Since $D$ and $\vartheta$ are positive and nonincreasing, $g'=D\vartheta$ is nonincreasing.

Next, we prove \ref{3-ii}. Since $g$ is strictly increasing
and $u'$ is strictly decreasing, \ref{3-i} implies that
$\Gamma(h)=g'(h)u'(g(h))$ is positive, continuous, and strictly decreasing.
Hence $\lim_{h\to\infty}\Gamma(h)$ exists. If $\lim_{h\to\infty}\Gamma(h)>0$, $1-\gamma_2>0$
implies that the integral in \eqref{eq:discounted-direct-tail} diverges,
a contradiction. Thus 
\begin{align}
    \lim_{h\to \infty }\Gamma(h)=0\,.
\end{align}

Now, we prove \ref{3-iii}. For $t>1$, let $\xi_t(z):=I(tu'(z))$ for $z>0$. If $tu'(z)\ge m_0$, then $\xi_t(z)=0$. For $tu'(z)<m_0$, we have
\begin{align} 
    \frac{d}{dz}(\log\xi_t(z)-\log z)&=\frac{\xi'_t(z)}{\xi_t(z)}-\frac{1}{z}=\frac{1}{z}\Big(\frac{R_u(z)}{R_u(\xi_t(z))}-1\Big)\,,
\end{align}
and $0<\xi_t(z)<z$. By Assumption~\ref{assume:discounted}, $\log\frac{\xi_t(z)}{z}$ is nondecreasing wherever $\xi_t(z)>0$. Together with the zero extension and continuity of $I$, this proves that $\frac{I(tu'(z))}{z}$ is nondecreasing on $(0,\infty)$. For fixed $z>0$, it is nonincreasing in $t$ because $I$ is nonincreasing.

Finally, we prove \ref{lemma:local-linear-growth}.
Observe that
\begin{align}
    u(g(h))-yh\le  \bar u(y,h)
    =\sup_{\ell(h)\le c\le h}
      \left\{u\big(D(h)c-L(h)\big)-yc\right\}\le u(g(h))\,.
\end{align}
Then 
\begin{align}
    |\bar u(y,h)|\le |u(g(h))|+yh\le C(h)(1+y)\,, \label{eqn:6-7} 
\end{align}
where $C(h):=1+|u(g(h))|+h+\Gamma(h)$ is positive and continuous. In addition, For $y\ge b(h)$, \eqref{eq:bar-u-h},
$c^*(y,h)\le h$, and \eqref{eq:vartheta-identity} imply
\begin{align*}
    |\bar u_h(y,h)|
    &=y\left[\ell'(h)-\frac{D'(h)}{D(h)}
       \bigl(c^*(y,h)-\ell(h)\bigr)\right]\\
    &\le y\bigl(\ell'(h)+\chi(h)\bigr)
     =y\bigl(1-\vartheta(h)\bigr)<y.
\end{align*}
For $0<y<b(h)$, we have
$|\bar u_h(y,h)|=|\Gamma(h)-y|\le\Gamma(h)+y$.
Thus
\begin{align}
    |\bar u_h(y,h)|
    \le \Gamma(h)+y\le C(h)(1+y).
    \label{eqn:6-8}
\end{align}
Thus \eqref{eqn:6-7} and \eqref{eqn:6-8} imply \eqref{eq:local-linear-growth}. Since $\gamma_1>1$ and $\gamma_2<0$, we obtain \eqref{eqn:3-1}. This completes the proof.
\end{proof}

\begin{prop}
\label{prop:normalized-marginal-order}
Suppose Assumptions~\ref{assume:common} and \ref{assume:discounted} hold. For $q,h>0$, define $\psi_h(q):=\frac{\bar u_h(\Gamma(h)q,h)}{\Gamma(h)}$.
Then
\begin{align}\label{eq:psi-discounted}
    \psi_h(q)
    =\begin{cases}
       1-q,
       &0<q\le\frac{1}{\vartheta(h)},\\[1mm]
       \displaystyle
       q\left(
          -\chi(h)
          \frac{ I(\vartheta(h)q\,u'(g(h)))}{g(h)}
          -\ell'(h)
       \right),
       &q>\frac{1}{\vartheta(h)}.
      \end{cases}
\end{align}
For every fixed $q>0$, the map $h\mapsto\psi_h(q)$ is nonincreasing.
Moreover,
\begin{align}\label{eq:psi-sign-growth}
    \psi_h(q)&>0\,, \text{ for } 0<q<1\,,
    \quad \psi_h(1)=0\,,\quad
    \psi_h(q)\le0,\quad \text{ for } q>1,\\
    |\psi_h(q)|&\le C(h)(1+q),\quad q>0\,,
\end{align}
where $C:(0,\infty)\to (0,\infty)$ is a continuous function.
\end{prop}

\begin{proof}
First, we prove \eqref{eq:psi-discounted}. Recall from \eqref{eq:bar-u-h} that
\begin{align}
    \bar u_h(y,h)
    &=\begin{cases}
       \Gamma(h)-y,
       &0<y<b(h),\\[1mm]
       \displaystyle
       \frac{y}{D(h)}\big(D'(h)c^*(y,h)-L'(h)\big),
       &b(h)\le y<d(h),\\[3mm]
       -y\ell'(h),
       &d(h)\le y<\infty.
      \end{cases}
\end{align}
Since $\Gamma(h)=\vartheta(h)D(h)u'(g(h))$, $\psi_h(q)=1-q$ for $0<q\le \frac{1}{\vartheta(h)}$. 
For $b(h)<\Gamma(h)q<d(h)$, we have
\begin{align}
    q\Big(\frac{D'(h)c^*(\Gamma(h)q,h)-L'(h)}{D(h)}\Big)
    &=q\Big(\frac{D'(h)}{D(h)^2}I(\vartheta(h)q\,u'(g(h)))-\ell'(h)\Big)\\
    &=q\Big(-\chi(h)
      \frac{ I(\vartheta(h)q\,u'(g(h)))}{g(h)}-\ell'(h)\Big)\,.
\end{align}
For $m_0\le\vartheta(h)q\,u'(g(h))$, the extension of $I$ implies
\begin{align}
    q\Big(-\chi(h)
      \frac{ I(\vartheta(h)q\,u'(g(h)))}{g(h)}-\ell'(h)\Big)=-q\ell'(h)\,.
\end{align}
This proves \eqref{eq:psi-discounted}.

Next, we prove that $\psi_h(q)$ is nonincreasing in $h$.
Fix $q>0$ and let $0<h_1<h_2$. By Lemma~\ref{lemma:primitive-shape-discounted}, $\frac{1}{\vartheta(h_1)}\le \frac{1}{\vartheta(h_2)}$.
For $q\le \frac{1}{\vartheta(h_1)}$, $\psi_{h_1}(q)=\psi_{h_2}(q)=1-q$. For $q>\frac{1}{\vartheta(h_2)}$, Assumption~\ref{assume:discounted} and Lemma~\ref{lemma:primitive-shape-discounted} imply
\begin{align}
    \ell'(h_1)&\le \ell'(h_2)\,,\quad \chi(h_1)\le \chi(h_2)\,,\\
    \frac{I(\vartheta(h_1)qu'(g(h_1)))}{g(h_1)}&\le \frac{I(\vartheta(h_2)qu'(g(h_1)))}{g(h_1)}\le \frac{I(\vartheta(h_2)qu'(g(h_2)))}{g(h_2)}\,.
\end{align}
This implies that $\psi_{h_1}(q)\ge \psi_{h_2}(q)$.
For $\frac{1}{\vartheta(h_1)}<q\le\frac{1}{\vartheta(h_2)}$, we have
\begin{align}
    \psi_{h_1}(q)-(1-q)
    &=q\vartheta(h_1)-1
      +q\chi(h_1)
       \left(1-\frac{ I(\vartheta(h_1)q u'(g(h_1)))}{g(h_1)}\right)>0\,.
\end{align}
Here we used $\vartheta(h_1)q>1$ and $0\le I(\vartheta(h_1)q u'(g(h_1)))\le g(h_1)$. Thus $\psi_{h_1}(q)\ge\psi_{h_2}(q)$.

Finally, we prove \eqref{eq:psi-sign-growth}. Since $\chi(h),\ell'(h)\ge0$ and $\frac{1}{\vartheta(h)}>1$, \eqref{eq:psi-discounted} implies
\begin{align}
    &\psi_h(q)>0\,, \text{ for } 0<q<1\,,
    \quad \psi_h(1)=0\,,\quad
    \psi_h(q)\le0,\quad \text{ for } q>1\,,\\
    &|\psi_h(q)|\le 1+q \text{ for }0<q\le \frac{1}{\vartheta(h)}\,.
\end{align} 
Moreover, for $q>\frac{1}{\vartheta(h)}$, observe that
\begin{align}
    0\le I(\vartheta(h)qu'(g(h)))\le g(h)\,.
\end{align} 
By \eqref{eq:vartheta-identity}, we obtain $|\psi_h(q)|\le q(\chi(h)+\ell'(h))=q(1-\vartheta(h))<q$. Thus \eqref{eq:psi-sign-growth} holds with $C(h)=1$. This completes the proof.
\end{proof}

For $\varphi\in C^2((0,\infty))$, define
\begin{align}
    \mathcal L_\rho\varphi(y)
    :=\frac{\kappa^2}{2}y^2\varphi''(y)
      +(\rho-r)y\varphi'(y).
\end{align}

\begin{prop}
\label{prop:Green-discounted}

Suppose Assumptions~\ref{assume:common} and \ref{assume:discounted} hold. Then, for every $y,a>0$,
\begin{equation}
\begin{aligned}
    R_\rho(y,a)
    &=\frac{2}{\kappa^2(\gamma_1-\gamma_2)}
      \left[
      y^{\gamma_2}\int_0^y s^{-\gamma_2-1}\bar u(s,a)\,d s
      +y^{\gamma_1}\int_y^\infty s^{-\gamma_1-1}\bar u(s,a)\,d s
      \right],\\
    G_\rho(y,a)
    &=\frac{2}{\kappa^2(\gamma_1-\gamma_2)}
      \left[
      y^{\gamma_2}\int_0^y s^{-\gamma_2-1}\bar u_h(s,a)\,d s
      +y^{\gamma_1}\int_y^\infty s^{-\gamma_1-1}\bar u_h(s,a)\,d s
      \right].\label{eq:G-Green}
\end{aligned}
\end{equation}
Moreover, $R_\rho$ and $G_\rho$ satisfy
    \begin{equation}
\begin{aligned}
    (\mathcal L_\rho-\rho)R_\rho(y,a)+\bar u(y,a)&=0\,,\\
    (\mathcal L_\rho-\rho)G_\rho(y,a)+\bar u_h(y,a)&=0\,,
\end{aligned}
\end{equation}
and 
\begin{align}
    \frac{\partial R_\rho}{\partial a}(y,a)=G_\rho(y,a)\label{eq:R-h-equals-G}\,.
\end{align}
\end{prop}

\begin{proof}

The representations and differential equations follow from
\eqref{eqn:3-1} and  \cite[Proposition 4.1]{knudsen1998valuation}.
For a compact interval $K_a\subset(0,\infty)$ containing $a$ in its interior,
Lemma~\ref{lemma:primitive-shape-discounted} implies
\begin{align}
    \sup_{\alpha\in K_a}
    \bigl(|\bar u(s,\alpha)|+|\bar u_h(s,\alpha)|\bigr)
    \le C(1+s)
\end{align}
for some $C>0$.
Thus $R_{\rho,a}=G_\rho$ by the dominated convergence theorem.
\end{proof}

\begin{thm}
\label{thm:fixed-level-discounted}
Suppose Assumptions~\ref{assume:common} and \ref{assume:discounted} hold, and fix $a>0$.
Then
\begin{align}\label{eq:S-discounted}
    S_\rho(y,a)
    =\begin{cases}
       G_\rho(y,a),&0<y\le z_\rho(a),\\[1mm]
       A_\rho(a)y^{\gamma_2},&y>z_\rho(a),
      \end{cases}
\end{align}
where $z_\rho$ and $A_\rho$ are given by
\eqref{eq:z-rho-general-equation} and \eqref{eq:A-rho-definition}. An optimal stopping time is
\begin{align}\label{eq:tau-star-discounted}
    \tau_a^*:=\inf\{t\ge0:Y_t^{\rho,y}\le z_\rho(a)\}.
\end{align}
Moreover, $S_\rho(\cdot,a)$ is a bounded function in 
$C^1((0,\infty))\cap C^2((0,\infty)\setminus\{z_\rho(a)\})$ satisfying
\begin{align}\label{eq:obstacle-discounted}
    \max\left\{
      (\mathcal L_\rho-\rho)S_\rho(y,a),
      G_\rho(y,a)-S_\rho(y,a)
    \right\}=0\,,\quad  y\ne z_\rho(a)\,,
\end{align}
and the boundary condition
$\lim_{y\to\infty}S_\rho(y,a)=0$.
\end{thm}

\begin{proof}
For fixed $a>0$, define
\begin{align}
    \hat S_\rho(y,a)
    =\begin{cases}
       G_\rho(y,a),&0<y\le z_\rho(a),\\[1mm]
       A_\rho(a)y^{\gamma_2},&y>z_\rho(a)\,.
      \end{cases}
\end{align}
We first prove that
$\hat S_\rho(\cdot,a)\in C^1((0,\infty))
\cap C^2((0,\infty)\setminus\{z_\rho(a)\})$
and is bounded. Recall that
\begin{align}
    G_\rho(y,a)
    &=\frac{2}{\kappa^2(\gamma_1-\gamma_2)}
      \left[
      y^{\gamma_2}\int_0^y s^{-\gamma_2-1}\bar u_h(s,a)\,d s
      +y^{\gamma_1}\int_y^\infty s^{-\gamma_1-1}\bar u_h(s,a)\,d s
      \right]\,,\\
      A_\rho(a)
    &=\frac{2}{\kappa^2(\gamma_1-\gamma_2)}
      \int_0^{z_\rho(a)}s^{-\gamma_2-1}\bar u_h(s,a)\,d s,
      \end{align} and
      \begin{align}
      &\int_{z_\rho(a)}^\infty s^{-\gamma_1-1}\bar u_h(s,a)\,d s=0\,.
\end{align}
By a simple calculation, we have
\begin{align}
    G_\rho(z_\rho(a),a)=A_\rho(a)(z_\rho(a))^{\gamma_2}\,,\quad 
    G_{\rho,y}(z_\rho(a),a)
    =
    \gamma_2A_\rho(a)
    z_\rho(a)^{\gamma_2-1}\,.\label{eq:smooth-fit-discounted}
\end{align}
Thus $\hat S_\rho(\cdot,a)$ is continuously differentiable. By Proposition~\ref{prop:Green-discounted}, it is twice continuously differentiable for $y\ne z_\rho(a)$.
Let $\varepsilon\in(0,b(a))$. For $0<y<\varepsilon$,
\begin{align}
    G_\rho(y,a)
    =
    \frac{2}{\kappa^2(\gamma_1-\gamma_2)}
    \Bigg[
       &y^{\gamma_2}
       \int_0^y
          s^{-\gamma_2-1}(\Gamma(a)-s)\,d s\\
       &+
       y^{\gamma_1}
       \int_y^\varepsilon
          s^{-\gamma_1-1}(\Gamma(a)-s)\,d s+
       y^{\gamma_1}
       \int_\varepsilon^\infty
          s^{-\gamma_1-1}\bar u_h(s,a)\,d s
    \Bigg].
    \label{eq:G-zero-split}
\end{align}
Here we used \eqref{eq:bar-u-h}. By Lemma~\ref{lemma:primitive-shape-discounted} and \eqref{eq:discounted-roots}, we obtain
\begin{align}
    \lim_{y\to 0} G_\rho(y,a)=
    \frac{2\Gamma(a)}
         {\kappa^2(\gamma_1-\gamma_2)}
    \left(
       \frac1{-\gamma_2}
       +
       \frac1{\gamma_1}
    \right)=\frac{\Gamma(a)}{\rho}\,.\label{eqn:3-10}
\end{align}
By \eqref{eqn:3-10} and $\gamma_2<0$, $\hat S_\rho(\cdot,a)$ is bounded and satisfies $\lim_{y\to\infty}\hat S_\rho(y,a)=0$.

Next, we prove that $\hat S_\rho$ satisfies \eqref{eq:obstacle-discounted}.
First, we prove that $\hat S_\rho(y,a)\ge G_\rho(y,a)$ for $y>z_\rho(a)$.
By a simple calculation, we have
\begin{align}
    \frac{\partial}{\partial y}
    \left(
       \frac{G_\rho(y,a)}{y^{\gamma_2}}
    \right)
    =
    \frac{2}{\kappa^2}
    y^{\gamma_1-\gamma_2-1}
    \int_y^\infty
       s^{-\gamma_1-1}\bar u_h(s,a)\,d s\,.
    \label{eqn:3-11}
\end{align}
By Proposition~\ref{prop:normalized-marginal-order}, we have
\begin{align}\label{eqn:3-12}
    \bar u_h(y,a)>0 \text{ for }0<y<\Gamma(a)\,,\quad \bar u_h(y,a)\le0 \text{ for }y\ge\Gamma(a)\,.
\end{align}
Observe that 
\begin{align}\label{eqn:3-13}
    \int_y^\infty
       s^{-\gamma_1-1}\bar u_h(s,a)\,d s=\int_y^{z_\rho(a)}s^{-\gamma_1-1}\bar u_h(s,a)\,d s+\int_{z_\rho(a)}^\infty s^{-\gamma_1-1}\bar u_h(s,a)\,d s\,.
\end{align}
By \eqref{eq:z-rho-general-equation} and \eqref{eqn:3-11}--\eqref{eqn:3-13}, $\frac{G_\rho(y,a)}{y^{\gamma_2}}$ is strictly increasing on $(0,z_\rho(a))$, strictly decreasing on $(z_\rho(a),\Gamma(a)]$, and nonincreasing on $[\Gamma(a),\infty)$. Thus, for $y>z_\rho(a)$,
\begin{align}
    G_\rho(y,a)< \frac{G_\rho(z_\rho(a),a)}{(z_\rho(a))^{\gamma_2}}y^{\gamma_2}=A_\rho(a)y^{\gamma_2}\,.\label{eqn:3-14}
\end{align}
Here we used \eqref{eq:smooth-fit-discounted}. By Proposition~\ref{prop:Green-discounted} and $z_\rho(a)<\Gamma(a)$, for $0<y<z_\rho(a)$,
\begin{align}
    (\mathcal L_\rho-\rho)\hat S_\rho(y,a)=-\bar u_h(y,a)\le 0\,,\label{eqn:3-15}
\end{align}
and by \eqref{eq:discounted-roots}, for $y>z_\rho(a)$
\begin{align}
    (\mathcal L_\rho-\rho)\hat S_\rho(y,a)= 0\,.\label{eqn:3-16}
\end{align}
Thus \eqref{eqn:3-14}--\eqref{eqn:3-16} prove that $\hat S_\rho$ satisfies \eqref{eq:obstacle-discounted}.

Now we prove that $\hat S_\rho(y,a)=S_\rho(y,a)$ and that $\tau_a^*$ is optimal.
Let $\tau\in\mathcal T$ and $\eta_n:=\inf\{t\ge0:Y_t^{\rho,y}\notin(1/n,n)\}$.
By \cite[Theorem 2.1]{peskir2005change},  applying
It\^{o}'s formula to
$e^{-\rho t}\hat S_\rho(Y_t^{\rho,y},a)$, we obtain

\begin{align}
    \mathbb{E}\left[
       e^{-\rho(\tau\wedge\eta_n\wedge T)}
       \hat S_\rho
          (Y_{\tau\wedge\eta_n\wedge T}^{\rho,y},a)
    \right]=
    \hat S_\rho(y,a)
    +
    \mathbb{E}_y\left[
       \int_0^{\tau\wedge\eta_n\wedge T}
       e^{-\rho t}
       (\mathcal L_\rho-\rho)
       \hat S_\rho(Y_t^{\rho,y},a)\,d t
    \right]\le
    \hat S_\rho(y,a).
\end{align}
Here we used \eqref{eq:obstacle-discounted}. Since $\hat S_\rho$ is bounded and $\rho>0$, we apply
the dominated convergence theorem.
Letting $n\to\infty$ and then $T\to\infty$, we have
\begin{align}
    \mathbb{E}_y\left[
       e^{-\rho\tau}
       \hat S_\rho(Y_\tau^{\rho,y},a)
       \mathds{1}_{\{\tau<\infty\}}
    \right]
    \le
    \hat S_\rho(y,a)\,.
\end{align}
Since $\hat S_\rho\ge G_\rho$ and $\tau\in\mathcal T$ is arbitrary, we obtain
\begin{align}
    S_\rho(y,a)\le \hat S_\rho(y,a)\,.\label{eqn:3-17}
\end{align}
Next, we verify the optimality of $\tau_a^*$. By definition, $\tau_a^*=0$ for $y\le z_\rho(a)$.
Thus
\begin{align}\label{eqn:3-18}
    S_\rho(y,a)\ge \hat S_\rho(y,a)=G_\rho(y,a) \text{ for }y\le z_\rho(a)\,.
\end{align}
For $y>z_\rho(a)$, applying It\^{o}'s formula yields
\begin{align}
 e^{-\rho(T \wedge \tau_a^*)}\hat S_\rho(Y^{\rho,y}_{T\wedge\tau_a^*},a)=e^{-\rho(T\wedge\tau_a^*)}
    A_\rho(a)
    \bigl(Y_{T\wedge\tau_a^*}^{\rho,y}\bigr)^{\gamma_2}=
    A_\rho(a)y^{\gamma_2}
    -\kappa\gamma_2A_\rho(a)
    \int_0^{T\wedge\tau_a^*}
       e^{-\rho t}
       \bigl(Y_t^{\rho,y}\bigr)^{\gamma_2}\,dW_t\,.
\end{align} 
Since $\gamma_2<0$ and $Y_s^{\rho,y}\ge z_\rho(a)$
for $s\le\tau_a^*$, the stochastic integral is a martingale.
Thus, we have
\begin{align}
    \hat S_\rho(y,a)
    =
    \mathbb E_y\left[
       e^{-\rho(T\wedge\tau_a^*)}
       \hat S_\rho
       (Y_{T\wedge\tau_a^*}^{\rho,y},a)
    \right].
\end{align}
Since $\hat S_\rho$ is bounded and $\hat S_\rho(Y_{\tau_a^*}^{\rho,y},a)=G_\rho(Y_{\tau_a^*}^{\rho,y},a)$ on $\{\tau_a^*<\infty\}$, letting $T\to\infty$  yields
\begin{align}
    \widehat S_\rho(y,a)
    =
    \mathbb E_y\left[
       e^{-\rho\tau_a^*}
       G_\rho(Y_{\tau_a^*}^{\rho,y},a)
       \mathds{1}_{\{\tau_a^*<\infty\}}
    \right].\label{eqn:3-19}
\end{align}
By \eqref{eqn:3-17}--\eqref{eqn:3-19}, we have 
\begin{align}
\hat S_\rho(y,a)= S_\rho(y,a)\,,   
\end{align}
and $\tau_a^*$ is optimal.
This completes the proof.
\end{proof}

\begin{prop}
\label{prop:discounted-convexity}
Suppose Assumptions~\ref{assume:common} and
\ref{assume:discounted} hold.
For each $h>0$, $\mathcal J_\rho(\cdot,h)$ is twice continuously differentiable
and strictly convex. Moreover, $\mathcal J_{\rho,y}\in C^1((0,\infty)^2)$,
\begin{align}
    \mathcal J_{\rho,h}(y,h)
    &=\begin{cases}
       G_\rho(y,h)-A_\rho(h)y^{\gamma_2}<0,
       &y>z_\rho(h),\\
       0,&0<y\le z_\rho(h)\,,
      \end{cases}\label{eq:J-h-gradient}
\end{align}
and
\begin{align}
    \mathcal J_{\rho,yh}(y,h)&=0,
    \qquad 0<y\le z_\rho(h)\,.\label{eq:J-yh-zero-action}
    \end{align}
\end{prop}

\begin{proof}
First, we prove that $\mathcal J_\rho(\cdot,h)\in C^2((0,\infty))$
for fixed $h>0$.
Fix $y>0$ and choose $0<\underline y<y<\bar y$.
By Proposition~\ref{prop:discounted-boundary-scale}, we can choose $\underline a>h$
such that, for every $a\ge\underline a$,
\begin{align}
    z_\rho(a)<\underline y<y\,.
\end{align}
Thus, for $(y,a)\in[\underline y,\bar y]\times[\underline a,\infty)$,
Theorem~\ref{thm:fixed-level-discounted} implies
\begin{align}
    S_{\rho,y}(y,a)&=\gamma_2A_\rho(a)y^{\gamma_2-1},\\
    S_{\rho,yy}(y,a)
    &=\gamma_2(\gamma_2-1)A_\rho(a)y^{\gamma_2-2}\,.\label{eqn:3-30}
\end{align}
For $a\in [h,\underline a]$ and $\underline y<y<y'<\bar y$, observe that
\begin{align}
    S_\rho(y,a)-S_\rho(y',a)&=
    \begin{cases}
    G_{\rho}(y,a)-G_\rho(y',a)& \text{ if }y,y'\le z_\rho(a)    \,,\\
    \begin{aligned}[t]
       &G_\rho(y,a)-G_\rho(z_\rho(a),a)\\
       &\quad+A_\rho(a)(z_\rho(a))^{\gamma_2}-A_\rho(a)(y')^{\gamma_2}
    \end{aligned}
    &\text{ if }y<z_\rho(a)<y'\,,\\
    A_\rho(a)(y)^{\gamma_2}-A_\rho(a)(y')^{\gamma_2}&\text{ if }z_\rho(a)\le y<y'\,,
    \end{cases}\\
    S_{\rho,y}(y,a)-S_{\rho,y}(y',a)&=
    \begin{cases}
    G_{\rho,y}(y,a)-G_{\rho,y}(y',a)& \text{ if }y,y'\le z_\rho(a)    \,,\\
    \begin{aligned}[t]
       &G_{\rho,y}(y,a)-G_{\rho,y}(z_\rho(a),a)\\
       &\quad+\gamma_2 A_\rho(a)(z_\rho(a))^{\gamma_2-1}
          -\gamma_2 A_\rho(a)(y')^{\gamma_2-1}
    \end{aligned}
    &\text{ if }y<z_\rho(a)<y'\,,\\
    \gamma_2A_\rho(a)(y)^{\gamma_2-1}-\gamma_2A_\rho(a)(y')^{\gamma_2-1}&\text{ if }z_\rho(a)\le y<y'\,.
    \end{cases}
\end{align}
By \eqref{eq:G-Green} and a simple calculation, we have
\begin{align}
    G_{\rho,y}(y,a)
    &=\frac{2}{\kappa^2(\gamma_1-\gamma_2)}
      \left[
      \gamma_2y^{\gamma_2-1}\int_0^y s^{-\gamma_2-1}\bar u_h(s,a)\,d s
      +\gamma_1y^{\gamma_1-1}\int_y^\infty s^{-\gamma_1-1}\bar u_h(s,a)\,d s
      \right]\,,\\
      G_{\rho,yy}(y,a)
    &=\frac{2}{\kappa^2(\gamma_1-\gamma_2)}
      \Big[
      \gamma_2(\gamma_2-1)y^{\gamma_2-2}\int_0^y s^{-\gamma_2-1}\bar u_h(s,a)\,d s+\gamma_2y^{-2}\bar u_h(y,a)\\&\quad
      +\gamma_1(\gamma_1-1)y^{\gamma_1-2}\int_y^\infty s^{-\gamma_1-1}\bar u_h(s,a)\,d s-\gamma_1y^{-2}\bar u_h(y,a)
      \Big]\,.
\label{eq:G-rho-yy-green}
\end{align}
By Lemma~\ref{lemma:primitive-shape-discounted} and
Proposition~\ref{prop:discounted-boundary-scale}, the functions
$G_{\rho,y}(y,a)$, $G_{\rho,yy}(y,a)$,
$\gamma_2 A_\rho(a)y^{\gamma_2-1}$, and
$\gamma_2(\gamma_2-1)A_\rho(a)y^{\gamma_2-2}$ are bounded on
$[\underline y,\bar y]\times[h,\underline a]$.
Together with \eqref{eq:smooth-fit-discounted}, these bounds imply that,
for $\underline y<y<y'<\bar y$ and $a\in[h,\underline a]$,
\begin{equation}
\begin{aligned}
    |S_\rho(y,a)-S_\rho(y',a)|&\le C|y-y'|\,,\\
    |S_{\rho,y}(y,a)-S_{\rho,y}(y',a)|&\le C|y-y'| \,.\label{eqn:3-31}
\end{aligned}
\end{equation}
Since $z_\rho$ is strictly decreasing, $y=z_\rho(a)$ holds for
at most one $a$ for each fixed $y$.
By \eqref{eqn:3-30}, \eqref{eqn:3-31}, and \eqref{eq:A-rho-tail-finite},
the dominated convergence theorem applies separately on $[h,\underline a]$
and $[\underline a,\infty)$.
Differentiating \eqref{eq:J-R-plus-S} under the integral, we obtain
\begin{align}
    \mathcal{J}_{\rho,y}(y,h)&=R_{\rho,y}(y,h)+\int_h^\infty S_{\rho,y}(y,a)\,da\,,\\
    \mathcal{J}_{\rho,yy}(y,h)&=R_{\rho,yy}(y,h)+\int_h^\infty \hat S_{\rho,yy}(y,a)\,da\,,
\end{align}
where 
\begin{align}
    \hat S_{\rho,yy}(y,a)
    =\begin{cases}
       G_{\rho,yy}(y,a),&y<z_\rho(a),\\[1mm]
       \gamma_2(\gamma_2-1)A_\rho(a)y^{\gamma_2-2},
       &y>z_\rho(a),\\[1mm]
       0,&y=z_\rho(a).
      \end{cases}
\end{align}
The same bounds and dominated convergence imply continuity of both
derivatives in $y$, since $\hat S_{\rho,yy}(\cdot,a)$ is continuous at $y$
for all but at most one $a$.
Thus $\mathcal J_\rho(\cdot,h)\in C^2((0,\infty))$.

Next, we prove that $\mathcal J_{\rho,y}$ and $\mathcal J_{\rho,yy}$
are jointly continuous and that $\mathcal J_\rho(\cdot,h)$ is strictly convex.
By Lemma~\ref{lemma:primitive-shape-discounted}, we have
\begin{align}
        |\bar u(s,a)|+|\bar u_h(s,a)|
    \le C(a)(1+s),
    \qquad s>0,
\end{align}
where $C$ is positive and continuous.
By \eqref{eq:G-Green}, \eqref{eq:G-rho-yy-green},
and the dominated convergence theorem,
$R_{\rho,y}$, $R_{\rho,yy}$, $G_{\rho,y}$, and $G_{\rho,yy}$
are jointly continuous.
For $\bar h>h$, \eqref{eq:R-h-equals-G} implies
\begin{align}
    R_\rho(y,\bar h)-R_\rho(y,h)
    =
    \int_h^{\bar h} G_\rho(y,a)\,d a.
    \label{eq:R-difference-finite-level}
\end{align}
Since $G_{\rho,y}$ and $G_{\rho,yy}$ are jointly continuous,
we differentiate twice in $y$ under the integral to obtain 
\begin{align}
    R_{\rho,y}(y,\bar h)-R_{\rho,y}(y,h)
    &=
    \int_h^{\bar h} G_{\rho,y}(y,a)\,d a,
    \label{eq:R-y-difference-finite-level}\\
    R_{\rho,yy}(y,\bar h)-R_{\rho,yy}(y,h)
    &=
    \int_h^{\bar h} G_{\rho,yy}(y,a)\,d a.
    \label{eq:R-yy-difference-finite-level}
\end{align}
By \eqref{eq:R-y-difference-finite-level}
and \eqref{eq:R-yy-difference-finite-level}, we obtain
\begin{align}\label{eqn:6-9}
    \mathcal J_{\rho,y}(y,h)&=\begin{cases}
      R_{\rho,y}(y,h)
      +\gamma_2y^{\gamma_2-1}
       \int_h^\infty A_\rho(a)d a\,,&\text{ if }y>z_\rho(h)\,,\\
      R_{\rho,y}\bigl(y,z_\rho^{-1}(y)\bigr)
      +\gamma_2y^{\gamma_2-1}
       \int_{z_\rho^{-1}(y)}^\infty A_\rho(a)d a\,,&\text{ if }0<y\le z_\rho(h)\,, 
    \end{cases}\\
    \mathcal J_{\rho,yy}(y,h)&=\begin{cases}
      R_{\rho,yy}(y,h)
      +\gamma_2(\gamma_2-1)y^{\gamma_2-2}
       \int_h^\infty A_\rho(a)d a\,,&\text{ if }y>z_\rho(h)\,,\\
      R_{\rho,yy}\bigl(y,z_\rho^{-1}(y)\bigr)
      +\gamma_2(\gamma_2-1)y^{\gamma_2-2}
       \int_{z_\rho^{-1}(y)}^\infty A_\rho(a)d a\,,&\text{ if }0<y\le z_\rho(h) \,.
    \end{cases}
\end{align}
By \eqref{eqn:6-9}, $\mathcal J_{\rho,y}$ and
$\mathcal J_{\rho,yy}$ are jointly continuous
Next, we prove that $\mathcal J_\rho(\cdot,h)$ is strictly convex.
Since $\bar u(\cdot,h)$ is convex and $Y_t^{\rho,y}=yY_t^{\rho,1}$,
we have
\begin{align}
    R_{\rho,yy}(y,h)\ge0\,.\label{eqn:3-32}
\end{align}
Observe that $\gamma_2(\gamma_2-1)>0$, and $ A_\rho(a)>0$. Thus, by \eqref{eqn:6-9} and \eqref{eqn:3-32},
$\mathcal J_{\rho,yy}>0$ and $\mathcal J_\rho(\cdot,h)$ is strictly convex.

Finally, we prove \eqref{eq:J-h-gradient},
\eqref{eq:J-yh-zero-action}, and $\mathcal J_{\rho,y}\in C^1((0,\infty)^2)$.
Differentiating \eqref{eq:J-R-plus-S} and \eqref{eqn:6-9} with respect to $h$, we have
$\mathcal J_{\rho,h}=G_\rho-S_\rho$ and
$\mathcal J_{\rho,yh}=G_{\rho,y}-S_{\rho,y}$.
Thus, for $0<y\le z_\rho(h)$,
\begin{align}
    \mathcal J_{\rho,h}(y,h)=0,
    \qquad
    \mathcal J_{\rho,yh}(y,h)=0\,,\label{eqn:6-10}
\end{align}
whereas, for $y>z_\rho(h)$, \eqref{eqn:3-14} implies
\begin{align}
    \mathcal J_{\rho,h}(y,h)
 =R_{\rho,h}(y,h)-A_\rho(h)y^{\gamma_2}
 =G_\rho(y,h)-A_\rho(h)y^{\gamma_2}<0\,.
\end{align}
This proves \eqref{eq:J-h-gradient} and \eqref{eq:J-yh-zero-action}.
By \eqref{eq:smooth-fit-discounted}, $S_{\rho,y}$ is jointly continuous.
Thus $\mathcal J_{\rho,yh}=G_{\rho,y}-S_{\rho,y}$ is jointly continuous.
Together with the joint continuity of $\mathcal J_{\rho,yy}$, this proves
$\mathcal J_{\rho,y}\in C^1((0,\infty)^2)$.
This completes the proof.

\end{proof}

\begin{lemma}
\label{lemma:discounted-budget-derivative}
Suppose Assumptions~\ref{assume:common} and
\ref{assume:discounted} hold. Then, for every $y,h>0$,
\begin{align}\label{eq:J-y-budget}
    -\mathcal J_{\rho,y}(y,h)
    =\mathbb{E}\left[\int_0^\infty K_t
      c_t^{\rho,y,h}\,d t\right].
\end{align}
Moreover, for every $h>0$,
\begin{equation}\label{eq:budget-limits-discounted}
\begin{aligned}
    \lim_{y\to\infty}\big(-\mathcal J_{\rho,y}(y,h)\big)
    &=\frac{\ell(h)}r,\\
    \lim_{y\to0}\big(-\mathcal J_{\rho,y}(y,h)\big)
    &=\infty.
\end{aligned}
\end{equation}
\end{lemma}

\begin{proof}
First, we prove \eqref{eq:J-y-budget}.
By Proposition~\ref{prop:pointwise}, for $0<y_1<y_2$ and
$H\in\mathcal H(h)$, we have
\begin{align}
    \bar u(Y_t^{\rho,y_2},H_t)
     -\bar u(Y_t^{\rho,y_1},H_t)=
    -\int_{y_1}^{y_2}
       e^{\rho t}K_t
       c^*(Y_t^{\rho,s},H_t)\,d s\,,
    \label{eq:bar-u-secant-integral}
\end{align}
Since $c^*(\cdot,h)$ is nonincreasing, we obtain
    \begin{align}
    -K_t c^*(Y_t^{\rho,y_1},H_t)
    \le
    \frac{e^{-\rho t}
    \left[
       \bar u(Y_t^{\rho,y_2},H_t)
       -\bar u(Y_t^{\rho,y_1},H_t)
    \right]}{y_2-y_1}\le
    -K_t c^*(Y_t^{\rho,y_2},H_t).
\end{align}
For $0<\varepsilon<y$, the optimality of
$H^{\rho,y,h}$ in Theorem~\ref{thm:discounted-dual-value} implies
\begin{align}
    &-\frac{\mathcal J_\rho(y+\varepsilon,h)-\mathcal J_\rho(y,h)}{\varepsilon}\\
    &\quad\le -\frac{1}{\varepsilon}\mathbb E\Big[\int_0^\infty e^{-\rho t}
       \big(\bar u(Y_t^{\rho,y+\varepsilon},H_t^{\rho,y,h})
       -\bar u(Y_t^{\rho,y},H_t^{\rho,y,h})\big)\,dt\Big]\\
    &\quad\le \mathbb{E}\Big[\int_0^\infty K_tc^*(Y_t^{\rho,y}, H_t^{\rho,y,h})\,dt\Big]\,,\label{eqn:3-36}
\end{align}
Similarly,
\begin{align}
    -\frac{\mathcal{J}_\rho(y,h)-\mathcal{J}_\rho(y-\varepsilon,h)}{\varepsilon}\ge \mathbb{E}\Big[\int_0^\infty K_tc^*(Y_t^{\rho,y}, H_t^{\rho,y,h})\,dt\Big]\,.\label{eqn:3-35}
\end{align}
The expectation in \eqref{eqn:3-35} is therefore finite.
By Proposition~\ref{prop:discounted-convexity}, $\mathcal J_\rho(\cdot,h)$ is differentiable.
Letting $\varepsilon\downarrow0$ in \eqref{eqn:3-36} and \eqref{eqn:3-35},
we obtain \eqref{eq:J-y-budget}.

Next, we prove the first limit in \eqref{eq:budget-limits-discounted}.
By \eqref{eq:J-discounted-piecewise}, for $y>z_\rho(h)$, we have
\begin{align}
    -\mathcal J_{\rho,y}(y,h)
    &=-R_{\rho,y}(y,h)
      -\gamma_2y^{\gamma_2-1}
       \int_h^\infty A_\rho(a)\,d a\\
       &=-\frac{\partial}{\partial y}\mathbb{E}\left[\int_0^\infty e^{-\rho t}
       \bar u(Y_t^{\rho,y},h)\,d t\right]-\gamma_2y^{\gamma_2-1}
       \int_h^\infty A_\rho(a)\,d a.
    \label{eq:large-y-budget-decomposition}
\end{align}
Observe that
\begin{equation}\label{eqn:3-37}
\begin{aligned}
    \Big| e^{-\rho t}\frac{\partial \bar u(Y_t^{\rho,y},h)}{\partial y}\Big|&=\Big|e^{-\rho t} \bar u_y(Y_t^{\rho,y},h)e^{\rho t}K_t\Big|=\Big|c^*(Y_t^{\rho,y},h)K_t\Big|\le K_th\,,\\
    \lim_{y\to \infty} c^*(Y_t^{\rho,y},h)&=\ell(h)\,.
\end{aligned}
\end{equation}
Here we used $\bar u_y(y,h)=-c^*(y,h)$.
Since $\mathbb E[\int_0^\infty K_t\,dt]=1/r$ and $\gamma_2<0$,
\eqref{eqn:3-37}, \eqref{eq:A-rho-tail-finite}, and the dominated convergence theorem imply
\begin{align}
    \lim_{y\to \infty }-\mathcal{J}_{\rho,y}(y,h)&=\lim_{y\to \infty}-\frac{\partial}{\partial y}\mathbb{E}\left[\int_0^\infty e^{-\rho t}
       \bar u(Y_t^{\rho,y},h)\,d t\right]-\lim_{y\to \infty}\gamma_2y^{\gamma_2-1}
       \int_h^\infty A_\rho(a)\,d a\\
       &=\frac{\ell(h)}{r}\,.
\end{align}

Finally, we prove
\begin{align}
    \lim_{y\to0}\big(-\mathcal J_{\rho,y}(y,h)\big)
    &=\infty.
\end{align}
For $0<y<z_\rho(h)$, \eqref{eq:H-star-discounted} implies
\begin{align}\label{eqn:3-38}
    z_\rho^{-1}(y)\le H_t^{\rho,y,h}=z_\rho^{-1}(\inf_{0\le s\le t}Y_s^{\rho,y})\,.
\end{align}
Define
\begin{align}\label{eq:tau-small-y-budget}
    \tau
    :=\inf\left\{t\ge0:
       \frac{Y_t^{\rho,y}}
       {\inf_{0\le s\le t}Y_s^{\rho,y}}
       \ge\frac{\Gamma(h)}{z_\rho(h)}
       \right\}\wedge1\,.
\end{align}
Since $Y_t^{\rho,y}=yY_t^{\rho,1}$, $\tau$ does not depend on $y$.
By continuity of $Y^{\rho,y}$ and $\Gamma(h)/z_\rho(h)>1$, we have $0<\tau\le1$.
For $0\le t<\tau$,
\begin{align}
    Y_t^{\rho,y}<\inf_{0\le s\le t}Y_s^{\rho,y}\frac{\Gamma(h)}{z_\rho(h)}\le z_\rho(H_t^{\rho,y,h})\frac{\Gamma(H_t^{\rho,y,h})}{z_\rho(H_t^{\rho,y,h})}\,.
\end{align}
Here we used \eqref{eqn:3-38} and the fact that $\zeta_\rho$
is nonincreasing by Proposition~\ref{prop:discounted-boundary-scale}.
Since $\Gamma(H_t^{\rho,y,h})\le b(H_t^{\rho,y,h})$,
\eqref{eq:c-pointwise-general} implies
$c^*(Y_t^{\rho,y},H_t^{\rho,y,h})=H_t^{\rho,y,h}$ for $0\le t<\tau$.
Thus, by \eqref{eq:J-y-budget}, 
\begin{align}
    -\mathcal J_{\rho,y}(y,h)
    &=\mathbb{E}\left[\int_0^\infty K_t
       c_t^{\rho,y,h}\,d t\right]\ge\mathbb{E}\left[\int_0^\tau K_t
       H_t^{\rho,y,h}\,d t\right]\ge z_\rho^{-1}(y)\mathbb{E}\left[\int_0^\tau K_t\,d t\right]\,.
\end{align}
The expectation $\mathbb E[\int_0^\tau K_t\,dt]$ is positive,
finite, and independent of $y$.
By Proposition~\ref{prop:discounted-boundary-scale}, we have
$\lim_{y\to0}z_\rho^{-1}(y)=\infty$.
This proves the second limit in \eqref{eq:budget-limits-discounted} and completes the proof.
\end{proof}

\section{Some auxiliary results in Section \ref{sec:undis}}\label{sec:B}

Recall that
\begin{align}
    &\vartheta(h)
    =\frac{g'(h)}{D(h)}\,,
    \qquad
    \chi(h)
    =-\frac{D'(h)}{D(h)}\bigl(h-\ell(h)\bigr)\,,\qquad\Gamma(h):=g'(h)u'(g(h))\,,\\
    &\vartheta(h)=1-\ell'(h)-\chi(h),
    \qquad 0<\vartheta(h)<1\,,\qquad \delta_0=1+2r/\kappa^2>1.\label{eqn:6-11}
\end{align}
\begin{lemma}
\label{lemma:zero-primitive-tail-revised}
Suppose Assumptions~\ref{assume:common} and
\ref{assume:zero-general-new} hold. Then

\begin{enumerate}[label=(\roman*)]
    \item\label{B-1} For every $z>0$, we have $u(z)<0$ and $R_u(z)>0$, and the limit
\begin{align}
    \gamma_\infty:=\lim_{z\to\infty}R_u(z)
\end{align}
exists in $(1,\infty]$. 
\item\label{B-2} The functions $\vartheta$ and $g'$ are
nonincreasing and 
\begin{align}
    \lim_{h\to\infty}g(h)=\infty.
    \label{eq:g-infinity-derived-weak}
\end{align}
\item\label{B-3}The function $\Gamma$ is strictly decreasing and satisfies
\begin{equation}\label{eqn:4-1}
\begin{aligned}    &\lim_{h\to\infty}\Gamma(h)=0\,,
    \qquad \lim_{h\to\infty}h\Gamma(h)=0\,,\\
    &-u(g(h))
    =\int_h^\infty\Gamma(a)\,da\,,\qquad\int_h^\infty
       \left(-\frac{\Gamma'(a)}{\Gamma(a)}\right)
       |u(g(a))|d a<\infty,
    \qquad h>0.
\end{aligned}
\end{equation}
\item\label{B-4} For every $h_*>0$, there exists a constant $C(h_*)>0$ such that
\begin{align}
   \lim_{h\to \infty} \bar u(\Gamma(h)z,h)&=0
       \quad\text{for each fixed } z>0,
       \label{eq:scaled-reward-limit-revised}\\
    |\bar u(\Gamma(h)z,h)|&\le C(h_*)(1+z),
       \qquad h\ge h_*,\ z>0.
       \label{eq:scaled-reward-growth-revised}
\end{align}

\end{enumerate}
\end{lemma}

\begin{proof}
First, we prove \ref{B-1}.
Since $u$ is strictly increasing and tends to zero, $u(z)<0$ for $z>0$.
Concavity implies $R_u\ge0$. If $R_u(z_0)=0$ for some $z_0>0$, its monotonicity would imply $u''=0$ on $(0,z_0)$, contrary to strict concavity.
Thus $R_u>0$, and its monotonicity ensures that $\gamma_\infty$ exists.
Suppose $\gamma_\infty\le1$. Then
\begin{align}
    (\log u'(z) )'=-\frac{R_u(z)}{z}\ge -\frac{1}{z}\,.
\end{align}
Integrating from $1$ to $z\ge1$, we have
\begin{align}
    u'(z)\ge\frac{u'(1)}{z}.
\end{align}
Integrating once more, we obtain
\begin{align}
    u(z)\ge u(1)+u'(1)\log z\,.
\end{align}
This contradicts $\lim_{z\to\infty}u(z)=0$. Hence $\gamma_\infty>1$.

Next, we prove \ref{B-2}.
By Assumption~\ref{assume:zero-general-new} and \eqref{eqn:6-11}, $\vartheta$ is nonincreasing.
Both $D$ and $\vartheta$ are positive and nonincreasing, and hence $g'=D\vartheta$ is nonincreasing.
The monotonicity of $g$ ensures that $\bar g:=\lim_{h\to\infty}g(h)\in(0,\infty]$ exists.
Suppose $\bar g<\infty$. Then
\begin{align}
g(h)-g(1)=\int_1^h g'(a)\,da
    \ge(h-1)g'(h),\qquad h\ge1.
\end{align}
It follows that $\lim_{h\to \infty}g'(h)=0$.
Since $g''\le0$ and $|u(g(a))|\ge|u(\bar g)|>0$, for any $h_0>0$ we have 
\begin{align}
    \lim_{h\to \infty} \int_{h_0}^h|u(g(a))|\Big(-\frac{g''(a)}{g'(a)}\Big)\,da&\ge \lim_{h\to\infty}|u(\bar g)|\int_{h_0}^h-\frac{g''(a)}{g'(a)}\,da\\
    &=|u(\bar g)|\big( \log g'(h_0)-\lim_{h\to\infty }\log g'(h))=\infty\,. 
\end{align}
This contradicts \eqref{eq:zero-integrated-curvature-weak}.
Thus \eqref{eq:g-infinity-derived-weak} holds.

Now we prove \ref{B-3}. Observe that
\begin{align}\label{eqn:4-4}
    \frac{\Gamma'(h)}{\Gamma(h)}=\frac{g''(h)}{g'(h)}-R_u(g(h))\frac{g'(h)}{g(h)}\,.
\end{align}
By \ref{B-1} and \ref{B-2}, $\Gamma'/\Gamma<0$, and hence $\Gamma$ is strictly decreasing.
By \ref{B-2} and \eqref{eq:zero-utility-tail-weak}, 
\begin{align}
    \lim_{k\to \infty} \int_h^k \Gamma(a)\,da&=\lim_{k\to \infty}u(g(k))-u(g(h))=-u(g(h))\,,\\
    0=\lim_{h\to\infty}\int_{h/2}^h\Gamma(a)\,da
    &\ge \limsup_{h\to\infty}\frac{h}{2}\Gamma(h)\ge0\,.
\end{align}
This proves the first three equations in \eqref{eqn:4-1}.
By \ref{B-1}, there exists $\underline z>0$ such that, for $z\ge\underline z$,
\begin{align}\label{eqn:4-2}
    R_u(z)>1\,.
\end{align}
The monotonicity of $R_u$ implies that, for $z\ge\underline z$,
\begin{align}\label{eqn:4-3}
    \log u'(z)-\log u'(\underline z)=\int_{\underline z}^z-\frac{R_u(a)}{a}\,da\le -R_u(\underline z)\big(\log\frac{z}{\underline z}\big).
\end{align}
Applying \eqref{eqn:4-3} with lower endpoint
$z\ge\underline z$, we have
\begin{align}
    -u(z)=\int_z^\infty u'(a)\,da\le \frac{u'(z)z}{R_u(z)-1}\,.
\end{align}
Choose $h_0$ sufficiently large that $g(h_0)\ge\underline z$ and
\eqref{eq:zero-integrated-curvature-weak} holds. By \eqref{eqn:4-4},
\begin{align}
    \int_{h_0}^\infty \Big(-\frac{\Gamma'(a)}{\Gamma(a)}\Big)|u(g(a))|\,da&=\int_{h_0}^\infty -\frac{g''(a)}{g'(a)}|u(g(a))|\,da+\int_{h_0}^\infty R_u(g(a))\frac{g'(a)}{g(a)}|u(g(a))|\,da\\
    &\le \int_{h_0}^\infty -\frac{g''(a)}{g'(a)}|u(g(a))|\,da+\int_{h_0}^\infty R_u(g(a))\frac{g'(a)}{g(a)}\frac{u'(g(a))g(a)}{R_u(g(a))-1}\,da\\
    &\le \int_{h_0}^\infty -\frac{g''(a)}{g'(a)}|u(g(a))|\,da+\frac{R_u(g(h_0))}{R_u(g(h_0) )-1}\int_{h_0}^\infty \Gamma(a)\,da<\infty\,.
\end{align}
Here we used the monotonicity of $R_u$ and $g$, and the fact that $z/(z-1)$ is decreasing for $z>1$.
Thus \ref{B-3} holds.

Finally, we prove \ref{B-4}. For $y,h>0$, we have
\begin{align}
    u(g(h))-yh\le \sup_{\ell(h)\le c\le h}\Big( u(D(h)c-L(h))-yc\Big)\le 0\,.
\end{align}
Then we have
\begin{align}
    |\bar u (\Gamma(h)z,h)|\le |u(g(h))|+h\Gamma(h)z\,.    
\end{align}
By \ref{B-2} and \ref{B-3}, both $|u(g(h))|$ and $h\Gamma(h)$ tend to zero.
This proves \ref{B-4} and completes the proof.

\end{proof}

\begin{prop}
\label{prop:dh-ordering-revised}
Suppose Assumptions~\ref{assume:common} and \ref{assume:zero-general-new} hold.
For $h,q>0$, define $d_h(q):=\frac{\bar u_h(\Gamma(h)q,h)}{\Gamma(h)}$.
Then
\begin{align}
    d_h(q)
    =\begin{cases}
       1-q,
       &0<q\le \frac{1}{\vartheta(h)},\\[1mm]
       \displaystyle
       q\left[
          -\chi(h)
          \frac{ I(\vartheta(h)q\,u'(g(h)))}{g(h)}
          -\ell'(h)
       \right],
       &q>\frac{1}{\vartheta(h)}.
      \end{cases}
    \label{eq:dh-explicit-revised}
\end{align}
\begin{enumerate}[label=(\roman*)]
    \item\label{B-2-1} For every fixed $h>0$, $d_h$ is continuous and nonincreasing, and
    \begin{align}
        d_h(q)>0\,,\quad 0<q<1\,,\quad
        d_h(1)=0\,,\quad        d_h(q)\le0\,,\quad q>1,
        \label{eq:dh-sign-revised}
    \end{align}
    with strict inequality on
    $1<q<\frac{1}{\vartheta(h)}$.  Moreover, $0\le-\partial_qd_h(q)\le1$ for almost every $q>0$.

    \item\label{B-2-2} For every fixed $q>0$, $d_h(q)$ is nonincreasing in $h$.
    
    \item\label{B-2-3}
For all $h,q>0$, we have $|d_h(q)|\le1+q$.
For every fixed $q>0$, $d_h(q)$ is locally Lipschitz in $h$.
Moreover, for every compact $K\subset(0,\infty)$, there exists a constant $C_K>0$ such that, for every $q>0$,
\begin{align}
    \left|\frac{\partial}{\partial h}d_h(q)\right|
    \le C_K(1+q)
    \qquad\text{for a.e. }h\in K.
    \label{eq:dh-h-regularity-derived-clean}
\end{align}

    \item\label{B-2-4} For every fixed $q>0$, $\frac{\bar u_h(q,h)}{\Gamma(h)}=d_h\left(\frac{q}{\Gamma(h)}\right)$ is nonincreasing in $h$.
\end{enumerate}
\end{prop}

\begin{proof}
As in the proof of Proposition~\ref{prop:normalized-marginal-order},
we obtain \eqref{eq:dh-explicit-revised}.

First, we prove \ref{B-2-1}. Continuity follows from Proposition~\ref{prop:pointwise}.
For $\frac{1}{\vartheta(h)}< q < \frac{m_0}{u'(g(h))\vartheta(h)}$,
\begin{align}
    \frac{\partial}{\partial q}d_h(q)
    &=\bar u_{hy}(\Gamma(h)q,h)=\bar u_{yh}(\Gamma(h)q,h)=-\left.\frac{\partial c^*(y,h)}{\partial h}\right|_{y=\Gamma(h)q}\\
    &=-\Big(\ell'(h)+\Big(-\frac{D'(h)}{D(h)}\Big)\frac{I(\frac{\Gamma(h)q}{D(h)})}{D(h)}\Big(1-\frac{1}{R_u(I(\frac{\Gamma(h)q}{D(h)}))}\Big)\Big)\,.
\end{align}
For $\frac{1}{\vartheta(h)}<q<
\frac{m_0}{\vartheta(h)u'(g(h))}$, we have
$0<I(\frac{\Gamma(h)q}{D(h)})
=I(\vartheta(h)q\,u'(g(h)))<g(h)$.
Thus \eqref{eq:zero-single-crossing-weak} implies $\frac{\partial}{\partial q}d_h(q)\le0$.
Moreover,
\begin{align}
     -\frac{\partial}{\partial q}d_h(q)&=\ell'(h)+\chi(h)\frac{I(\frac{\Gamma(h)q}{D(h)})}{g(h)}-\chi(h)\frac{I(\frac{\Gamma(h)q}{D(h)})}{g(h)}\frac{1}{R_u(I(\frac{\Gamma(h)q}{D(h)}))}\\
     &\le \ell'(h)+\chi(h)=1-\vartheta(h)\le1\,.
\end{align}
For $0<q<\frac{1}{\vartheta(h)}$, we have $d_h(q)=1-q$ and $ -\frac{\partial}{\partial q}d_h(q)=1$. For $q>\frac{m_0}{u'(g(h))\vartheta(h)}$, $d_h(q)=-q\ell'(h)$ and $0\le-\frac{\partial}{\partial q}d_h(q)=\ell'(h)=1-\vartheta(h)-\chi(h)\le1$.
By continuity, $d_h$ is nonincreasing, and $0\le-\frac{\partial}{\partial q}d_h(q)\le1$ for almost every $q>0$.
The signs in \eqref{eq:dh-sign-revised} follow from \eqref{eq:dh-explicit-revised} and $\frac{1}{\vartheta(h)}>1$.

Next, the argument in the proof of Proposition~\ref{prop:normalized-marginal-order} applies under the present assumptions and proves \ref{B-2-2}.

We now prove \ref{B-2-3}.
Observe that 
\begin{align}
    0\le \frac{I(\vartheta(h)qu'(g(h)))}{g(h)}\le 1\,,\quad 0\le \ell'(h),\chi(h)\,, \quad 0\le \ell'(h)+\chi(h)\le 1\, 
\text{for }q\ge \frac{1}{\vartheta(h)}\,.
\end{align}
Then we have $|d_h(q)|\le 1+q$ for every $h,q>0$. We now prove \eqref{eq:dh-h-regularity-derived-clean}. The bound is immediate when $q<\frac{1}{\vartheta(h)}$ or
$q>\frac{m_0}{(\vartheta(h)u'(g(h)))}$.
It therefore suffices to consider $h\in K$ and
$\frac1{\vartheta(h)}<q<
\frac{m_0}{\vartheta(h)u'(g(h))}$.
Let $\bar I(q,h)=I(\vartheta(h)q\,u'(g(h)))$.
By a simple calculation, we have
\begin{align}
    \frac{\partial}{\partial h}d_h(q)&=q\Big(-\chi'(h)\frac{ \bar I(q,h)}{g(h)}-\chi(h) \frac{\partial}{\partial h}\Big(\frac{ \bar I(q,h)}{g(h)}\Big)-\ell''(h)\Big)\,,\label{eqn4-13}\\
    \frac{\partial}{\partial h}\frac{\bar I(q,h)}{g(h)}&=\frac{\bar I(q,h)}{g(h)}\Big(-\frac{\vartheta'(h)}{\vartheta(h)R_u(\bar I(q,h))}+\Big(\frac{R_u(g(h))}{R_u(\bar I(q,h))}-1 \Big)\frac{g'(h)}{g(h)}\Big)\label{eqn:4-10}\,.
\end{align}
By \eqref{eq:zero-single-crossing-weak}, \begin{align}
    0\le\ell'(h)
      +\left(-\frac{D'(h)}{D(h)}\right)
       \frac{\bar I (q,h)}{D(h)}
       \left(1-\frac1{R_u(\bar I(q,h))}\right)=\ell'(h)+\chi(h)\frac{\bar I(q,h)}{g(h)}-\frac{\chi(h)  \bar I(q,h)}{R_u(\bar I(q,h))g(h)}\,.
\end{align}
This implies that
\begin{align}
    0\le \frac{\chi(h)  \bar I(q,h)}{R_u(\bar I(q,h))g(h)}\le \ell'(h)+\chi(h)\frac{\bar I(q,h)}{g(h)}\le \ell'(h)+\chi(h)<1\label{eqn:4-11}\,.
\end{align}
Here we used $0<\bar I(q,h)<g(h)$.
By \eqref{eqn:4-10} and \eqref{eqn:4-11}, we have
\begin{align}
    \Big|\chi(h)\frac{\partial}{\partial h}\frac{\bar I(q,h)}{g(h)}\Big|\le \Big|\frac{\vartheta'(h)}{\vartheta(h)}\Big|+\Big|R_u(g(h))\frac{g'(h)}{g(h)}\Big|+\Big|\frac{\chi(h) g'(h)}{g(h)}\Big|\,.\label{eqn:4-12}
\end{align}
By \eqref{eqn:4-12} and \eqref{eqn4-13}, we obtain
\eqref{eq:dh-h-regularity-derived-clean}.
Continuity and the piecewise derivative bounds imply local Lipschitz
continuity in $h$.

Finally, we prove \ref{B-2-4}.
By Lemma~\ref{lemma:zero-primitive-tail-revised}, \ref{B-2-1}, and \ref{B-2-2}, for $h_2>h_1$ we have
\begin{align}
    d_{h_2}\Big(\frac{q}{\Gamma(h_2)}\Big)\le d_{h_1}\Big(\frac{q}{\Gamma(h_2)}\Big)\le d_{h_1}\Big(\frac{q}{\Gamma(h_1)}\Big)\,.
\end{align}
This completes the proof.

\end{proof}

\begin{lemma}
\label{lemma:B-finite-zero-new}
Suppose Assumptions~\ref{assume:common} and \ref{assume:zero-general-new} hold.
Then $B_0$ defined by \eqref{eq:B-zero-new} is finite for every $h>0$ and satisfies
\begin{align}
    B_0'(h)
    =\Phi_{0,y}(z_0(h),h)z_0'(h),
    \qquad
    \lim_{h\to\infty}B_0(h)=0.
    \label{eq:B-derivative-zero-new}
\end{align}
\end{lemma}

\begin{proof}
First, we prove that, for every $h>0$,
\begin{align}
    \int_h^\infty |\Phi_{0,y}(z_0(a),a)z_0'(a)|\,da<\infty\,.
\end{align}
Observe that, for $s,h>0$, we have
\begin{align}
    u(g(h))-sh\le \bar u(s,h)\le 0\,.
\end{align}
Then by \eqref{eq:Phi-y-zero-new} and $\delta_0>1$, there exists a constant $C>0$ such that
\begin{align}
    |\Phi_{0,y}(z_0(a),a)|
    &=\Big|\frac{2}{\kappa^2}(z_0(a))^{\delta_0-1}
      \int_{z_0(a)}^\infty s^{-\delta_0-1}\bar u(s,a)\,ds\Big|\\
    &\le C\Big(\frac{-u(g(a))}{z_0(a)}+a\Big)\,.   
\end{align}
Since $z_0=\Gamma\zeta_0$ and $z_0'\le0$, we have
\begin{align}
    &\int_h^\infty |\Phi_{0,y}(z_0(a),a)z'_0(a)|\,da\notag\\
    &\quad\le C\Big(
       \int_h^\infty -u(g(a))\Big(-\frac{\Gamma'(a)}{\Gamma(a)}
       -\frac{\zeta_0'(a)}{\zeta_0(a)}\Big)\,da
       +\int_h^\infty a(-z'_0(a))\,da\Big).\label{eqn:4-15}
\end{align}
By Lemmas~\ref{lemma:zero-primitive-tail-revised} and \ref{lemma:zero-boundary-new}, we obtain
\begin{align}
    &\int_h^\infty-u(g(a))\Big(-\frac{\Gamma'(a)}{\Gamma(a)}
      -\frac{\zeta_0'(a)}{\zeta_0(a)}\Big)\,da\notag\\
    &\quad\le\int_h^\infty-u(g(a))\Big(-\frac{\Gamma'(a)}{\Gamma(a)}\Big)\,da
      +\int_h^\infty-u(g(a))\Big(-\frac{\zeta_0'(a)}{\zeta_0(a)}\Big)\,da\\
    &\quad\le\int_h^\infty-u(g(a))\Big(-\frac{\Gamma'(a)}{\Gamma(a)}\Big)\,da
      -u(g(h))\Big(\log\zeta_0(h)-\lim_{a\to\infty}\log\zeta_0(a)\Big)
      <\infty\,.\label{eqn:4-13}
\end{align}
Since $0<z_0<\Gamma$, Lemma~\ref{lemma:zero-primitive-tail-revised} and integration by parts imply
\begin{align}
    \int_h^\infty a(-z_0'(a))\,da=hz_0(h)-\lim_{\bar h\to \infty }\bar h z_0(\bar h)+\lim_{\bar h\to \infty }\int_h^{\bar h}z_0(a)\,da<\infty\,.\label{eqn:4-14}
\end{align}By \eqref{eqn:4-15}, \eqref{eqn:4-13}, and
\eqref{eqn:4-14}, $B_0$ is finite and
\eqref{eq:B-derivative-zero-new} holds.
This completes the proof.

\end{proof}

\begin{lemma}
\label{lemma:reflected-moments-zero-new}

Suppose Assumptions~\ref{assume:common} and \ref{assume:zero-general-new} hold.
Then, for every $y,h>0$ and $p\in(1,\delta_0)$,
\begin{align}
    \sup_{T\ge0}\mathbb{E}\left[\Big(\frac{Y_T^y}{ z_0(H_T^{y,h})}\Big)^p\right]<\infty\,,\quad \lim_{T\to\infty}H_T^{y,h}=\infty
    ,
    \label{eq:uniform-R-zero-general-revised}
\end{align}
and 
\begin{align}
    \lim_{T\to\infty}
    \mathbb{E}\left[
      \left|V_0(Y_T^y,H_T^{y,h})\right|
    \right]=0.
    \label{eq:candidate-terminal-zero-new}
\end{align}
\end{lemma}

\begin{proof}
First, we prove \eqref{eq:uniform-R-zero-general-revised}.
By \eqref{eq:H-star-zero-general-new}, $\frac{Y_t^y}{z_0(H_t^{y,h})}\ge1$.
Let $\alpha_t=\log\frac{Y_t^y}{z_0(H_t^{y,h})}\ge0$.
Observe that
\begin{align}
    \alpha_t=Z_t+\bar Z_t\,,
\end{align}where $Z_t=\log\frac{Y_t^y}{z_0(h)}=\log\frac{y}{z_0(h)}-(r+\frac{\kappa^2}{2})t-\kappa W_t$
and $\bar Z_t=-\min(0,\inf_{0\le s\le t}Z_s)$.
For every $T>0$, 
\begin{align}
    \int_0^T\mathds{1}_{\alpha_t>0}d\bar Z_t=0\,.\label{eqn:4-24}
\end{align}
Let $f_p(x)=e^{px}-1-px$ for $x\ge0$.
By It\^{o}'s formula, we have
\begin{align}
    f_p(\alpha_T)
    &=f_p(\alpha_0)+\int_0^T
      \Big[p\Big(r+\frac{\kappa^2}{2}\Big)
      -\frac{\kappa^2p}{2}(\delta_0-p)e^{p\alpha_t}\Big]\,dt\\
    &\quad+\int_0^T f_p'(\alpha_t)\,d\bar Z_t
      -p\kappa\int_0^T(e^{p\alpha_t}-1)\,dW_t.
\end{align}
For each fixed $T>0$, we have $\sup_{0\le t\le T}\alpha_t\le C(1+\sup_{0\le t\le T}|W_t|)$ for some $C>0$.
Hence, for every $n\in\mathbb N$,
\begin{align}
    \mathbb{E}[\sup_{0\le t\le T}e^{n\alpha_t}]<\infty\,.\label{eqn:5-1}
\end{align}
By \eqref{eqn:5-1}, the stochastic integral is a martingale.
Using \eqref{eqn:4-24} and $f_p'(0)=0$, we obtain

\begin{align}
    \mathbb E[f_p(\alpha_T)]
    =f_p(\alpha_0)+\int_0^T
      \Big[p\Big(r+\frac{\kappa^2}{2}\Big)
      -\frac{\kappa^2p}{2}(\delta_0-p)\mathbb E[e^{p\alpha_t}]\Big]\,dt\,.
\end{align}
By \eqref{eqn:5-1} and $e^{px}\ge1+f_p(x)$, we have
\begin{align}
    \bar f'(T)\le C_1-C_2\bar f(T),
\end{align}
where $\bar f(T)=\mathbb E[f_p(\alpha_T)]$ and $C_1,C_2>0$.
By a simple calculation, we obtain
\begin{align}
    \mathbb{E}[f_p(\alpha_T)]=\bar f(T)\le \bar f(0)+\frac{C_1}{C_2}\,.
\end{align}
Therefore, we obtain
\begin{align}
    \sup_{T\ge0}\mathbb{E}\Big[\big(\frac{Y_T^y}{z_0(H_T^{y,h})} \big)^p\Big]=\sup_{T\ge 0}\mathbb{E}\big[e^{p\alpha_T}\big]\le e^2+2\bar f(0)+2\frac{C_1}{C_2}\,.
\end{align}
Here we used $e^{px}\le e^2+2f_p(x)$ for $x\ge0$.
This proves the first inequality in \eqref{eq:uniform-R-zero-general-revised}.
Next, $\lim_{T\to \infty}Y_T^y=0$ and $z_0(H_T^{y,h})\le Y_T^y$.
By Lemma~\ref{lemma:zero-boundary-new}, $\lim_{T\to \infty}H_T^{y,h}=\infty$.

We now prove \eqref{eq:candidate-terminal-zero-new}.
By \eqref{eq:V-zero-general-new} and continuity, we have
\begin{align}
    V_0(Y_T^y,H^{y,h}_T)
    =B_0(H_T^{y,h})+\Phi_0(Y_T^y,H^{y,h}_T)-\Phi_0(z_0(H_T^{y,h}),H_T^{y,h})\,.
    \end{align}
    Here we used $Y_T^y\ge z_0(H_T^{y,h})$.
Observe that
\begin{align}
    |B_0(H_T^{y,h})|
    =\Big|\int_{H_T^{y,h}}^\infty
       \Phi_{0,y}(z_0(a),a)z_0'(a)d a\Big|\le \int_h^\infty|\Phi_{0,y}(z_0(a),a)z_0'(a)|\,da.
\end{align}
By Lemma~\ref{lemma:B-finite-zero-new}, \eqref{eq:uniform-R-zero-general-revised}, and the dominated convergence theorem,
\begin{align}
    \lim_{T\to \infty } \mathbb{E}[|B_0(H_T^{y,h})|]=0\,.
\end{align}
Thus, it suffices to prove that
\begin{align}
    \lim_{T\to \infty }\mathbb{E}[|\Phi_0(Y_T^y,H^{y,h}_T)-\Phi_0(z_0(H_T^{y,h}),H_T^{y,h})|]=0\,.
\end{align}
Observe that $|\bar u(y,h)|\le-u(g(h))+hy$.
By \eqref{eq:Phi-y-zero-new} and a simple calculation, for $y\ge z_0(h)$ and $z\ge1$, we have
\begin{align}
    |\Phi_{0,y}(y,h)|&\le \frac{2}{\kappa^2}\Big(\frac{-u(g(h))}{\delta_0 y}+\frac{h}{\delta_0-1}\Big)\,,\label{eqn:4-27}\\
    \left|\Phi_0(z_0(h)z,h)-\Phi_0(z_0(h),h)\right|&\le\int_{z_0(h)}^{z_0(h)z}|\Phi_{0,y}(s,h)|\,ds\\
    &\le\frac{2}{\kappa^2}
       \Big(\frac{-u(g(h))}{\delta_0}\log z
          +\frac{hz_0(h)}{\delta_0-1}(z-1) \Big)\,.
\end{align}
Since $\log z\le z$ and $z-1\le z$ for $z\ge1$, we obtain
\begin{align}
    &\quad\mathbb{E}\Big[\left|\Phi_0(Y_T^y,H_T^{y,h})-\Phi_0(z_0(H_T^{y,h}),H_T^{y,h})\right|\Big]\\
    &=\mathbb{E}\Big[\Big|\Phi_0\Big(z_0(H_T^{y,h})\frac{Y_T^y}{z_0(H_T^{y,h})},H_T^{y,h}\Big)-\Phi_0(z_0(H_T^{y,h}),H_T^{y,h})\Big|\Big]\\
    &\le \mathbb{E}\Big[\frac{2}{\kappa^2}\Big(\frac{-u(g(H_T^{y,h}))}{\delta_0}+\frac{H_T^{y,h}z_0(H_T^{y,h})}{\delta_0-1}\Big)\frac{Y_T^y}{z_0(H_T^{y,h})}\Big]\,.
\end{align}
By Lemmas~\ref{lemma:zero-primitive-tail-revised}
and \ref{lemma:zero-boundary-new} and
\eqref{eq:uniform-R-zero-general-revised}, we have
\begin{align}\label{eqn:4-25}
    &\lim_{T\to \infty}H_T^{y,h}=\infty\,,\quad \lim_{T\to \infty}u(g(H_T^{y,h}))=0\,,\\
     0\le &\lim_{T\to \infty} H_T^{y,h}z_0(H_T^{y,h})\le \lim_{T\to \infty} H_T^{y,h}\Gamma(H_T^{y,h})=0 \,.
\end{align}
By H\"older's inequality,
\begin{align}
   &\quad\mathbb{E}\Big[\left|\Phi_0(Y_T^y,H_T^{y,h})-\Phi_0(z_0(H_T^{y,h}),H_T^{y,h})\right|\Big]\\&\le \frac{2}{\kappa^2}\mathbb{E}[\Big(\frac{-u(g(H_T^{y,h}))}{\delta_0}+\frac{H_T^{y,h}z_0(H_T^{y,h})}{\delta_0-1}\Big)^{p'}\Big]^{\frac{1}{p'}}\mathbb{E}\Big[\Big(\frac{Y_T^y}{z_0(H_T^{y,h})}\Big)^{p}\Big]^{\frac{1}{p}},
\end{align}
where $\frac{1}{p}+\frac{1}{p'}=1$ and $1<p<\delta_0$.
By \eqref{eqn:4-25}, \eqref{eq:uniform-R-zero-general-revised}, and the dominated convergence theorem,
\begin{align}
    \lim_{T\to \infty }\mathbb{E}\Big[\left|\Phi_0(Y_T^y,H_T^{y,h})-\Phi_0(z_0(H_T^{y,h}),H_T^{y,h})\right|\Big]=0\,.
\end{align}
Here we used $\sup_{a\ge h}(-u(g(a))+az_0(a))<\infty$.
This proves \eqref{eq:candidate-terminal-zero-new} and completes the proof.
\end{proof}

\begin{prop}
\label{prop:candidate-payoff-zero-general-new} Suppose Assumptions~\ref{assume:common} and \ref{assume:zero-general-new} hold.
Then, for every $y,h>0$,
\begin{align}
    V_0(y,h)
    =\mathbb{E}\left[\int_0^\infty
      \bar u(Y_t^y,H_t^{y,h})\,dt\right].
    \label{eq:candidate-payoff-zero-general-new}
\end{align}
\end{prop}

\begin{proof}
First, \eqref{eq:V-zero-general-new} implies $V_0(y,H_0^{y,h})=V_0(y,h)$.
By Proposition~\ref{prop:VI-zero-general-new} and It\^{o}'s formula, we have
\begin{align}
    V_0(Y_T^y,H_T^{y,h})
    &=V_0(y,h)
      +\int_0^T
       \left(
          \frac{\kappa^2}{2}(Y_t^y)^2V_{0,yy}(Y_t^y,H_t^{y,h})
          -rY_t^yV_{0,y}(Y_t^y,H_t^{y,h})
       \right)\,dt
       \\
    &\quad+\int_0^T V_{0,h}(Y_t^y,H_t^{y,h})\,dH_t^{y,h}
      -\kappa\int_0^T Y_t^yV_{0,y}(Y_t^y,H_t^{y,h})\,dW_t\,.
    \label{eq:payoff-Ito-before-PDE}
\end{align}
Observe that $Y_t^y\ge z_0(H_t^{y,h})$, and $H^{y,h}$ increases only when
$Y_t^y=z_0(H_t^{y,h})$.
Thus,
\begin{align}
    \int_0^T
       \mathbf 1_{\{Y_t^y>z_0(H_t^{y,h})\}}\,dH^{y,h}_t=0\,.
\end{align}
By \eqref{eqn:4-21} and continuity, $V_{0,h}(z_0(a),a)=0$ for every $a>0$. Hence
\begin{align}
    \int_0^T V_{0,h}(Y_t^y,H_t^{y,h})\,dH^{y,h}_t=0\,.
    \label{eqn:4-26}
\end{align}
By \eqref{eqn:4-27}, we have
\begin{align}
    |Y_t^yV_{0,y}(Y_t^y,H^{y,h}_t)|
    &=|Y_t^y\Phi_{0,y}(Y_t^y,H^{y,h}_t)|\notag\\
    &\le \frac{2}{\kappa^2}\Big(
       \frac{-u(g(H_t^{y,h}))}{\delta_0}
       +\frac{H_t^{y,h}z_0(H^{y,h}_t)}{\delta_0-1}
        \frac{Y_t^y}{z_0(H^{y,h}_t)}\Big)\,.
\end{align}
Since $\sup_{a\ge h}(-u(g(a))+az_0(a))<\infty$, \eqref{eqn:5-1} implies that
$\int_0^T Y_t^yV_{0,y}(Y_t^y,H^{y,h}_t)\,dW_t$ is a martingale.
Taking expectations in \eqref{eq:payoff-Ito-before-PDE}
and using Proposition~\ref{prop:VI-zero-general-new},
\eqref{eqn:4-26}, and continuity, we obtain
\begin{align}
    V_0(y,h)
    =\mathbb E\left[\int_0^T\bar u(Y_t^y,H^{y,h}_t)\,dt\right]
      +\mathbb E[V_0(Y_T^y,H^{y,h}_T)].
    \label{eq:finite-horizon-candidate-identity-detailed}
\end{align}
Since $\bar u\le0$, the monotone convergence theorem and Lemma~\ref{lemma:reflected-moments-zero-new} yield
\begin{align}
    V_0(y,h)
    =\mathbb E\left[\int_0^\infty\bar u(Y_t^y,H^{y,h}_t)\,dt\right]\,.
\end{align}
This completes the proof.
\end{proof}

Define
\begin{align}
    \mathfrak a(h)
    :=-u(g(h))
    =\int_h^\infty\Gamma(s)\,ds.
    \label{eq:a-transform-zero-new}
\end{align}
Since $\mathfrak a$ is strictly decreasing, its inverse $\mathfrak h=\mathfrak a^{-1}$ is defined on $\mathfrak a((0,\infty))$. Define
\begin{align}
    \Psi(y,A):=\bar u(y,\mathfrak h(A)).
    \label{eq:Psi-transform-zero-new}
\end{align}

\begin{lemma}
\label{lemma:Psi-concavity-revised}
Suppose Assumptions~\ref{assume:common} and \ref{assume:zero-general-new} hold.
Then, for every fixed $y>0$, $\Psi(y,\cdot)$ is $C^1$, piecewise $C^2$, and concave.
More precisely,
\begin{align}
    \Psi_A(y,A)
    &=-\frac{\bar u_h(y,h)}{\Gamma(h)}\,,
    \label{eq:Psi-A-zero-revised}\\
    \Psi_{AA}(y,A)
    &=\frac1{\Gamma(h)}
      \frac{\partial}{\partial h}
      \left(\frac{\bar u_h(y,h)}{\Gamma(h)}\right)
    \le0\,,\qquad a.e.,\label{eq:Psi-AA-zero-revised}
\end{align}
where $h=\mathfrak h(A)$.
\end{lemma}

\begin{proof}
First, we prove that $\Psi(y,\cdot)$ is $C^1$
and piecewise $C^2$.
Since $\mathfrak a'(h)=-\Gamma(h)<0$ and
$\Gamma\in C^1((0,\infty))$, $\mathfrak h\in C^2$, with
\begin{align}
    \frac{\partial h}{\partial A}
    =\frac1{\mathfrak a'(h)}
    =-\frac1{\Gamma(h)}\,.
\end{align}
By Proposition~\ref{prop:dh-ordering-revised}, we have
\begin{align}
    \Psi_A(y,A)
    &=\bar u_h(y,h)\frac{\partial h}{\partial A}
      =-\frac{\bar u_h(y,h)}{\Gamma(h)}\,,
    \\
    \Psi_{AA}(y,A)
    &=-\frac{\partial}{\partial h}
       \left(\frac{\bar u_h(y,h)}{\Gamma(h)}\right)
       \frac{\partial h}{\partial A}
      =\frac1{\Gamma(h)}
       \frac{\partial}{\partial h}
       \left(\frac{\bar u_h(y,h)}{\Gamma(h)}\right)\le0\,.
    \label{eq:Psi-second-chain-rule-detailed}
\end{align}
Next, we prove concavity in $A$.
Let $A_1<A_2$ belong to $\mathfrak a((0,\infty))$.
Since $\mathfrak a$ is strictly decreasing, $h_1=\mathfrak h(A_1)>\mathfrak h(A_2)=h_2$.
By Proposition~\ref{prop:dh-ordering-revised},
\begin{align}
    \Psi_A(y,A_1)=-\frac{\bar u_h(y,h_1)}{\Gamma(h_1)}
    \ge-\frac{\bar u_h(y,h_2)}{\Gamma(h_2)}
    =\Psi_A(y,A_2)\,.
\end{align}
Thus $\Psi_A(y,\cdot)$ is nonincreasing, and $\Psi(y,\cdot)$ is concave.
This completes the proof.
\end{proof}

Let
\begin{align}
    w(y,A)
    :=\frac{V_{0,h}(y,\mathfrak h(A))}
           {\Gamma(\mathfrak h(A))}.
    \label{eq:w-general-revised}
\end{align}
Since $V_{0,h}\le0$ and $\Gamma>0$, we have $w(y,A)\le0$.
Moreover, $V_{0,h}(y,h)=0$ for $0<y\le z_0(h)$, and hence $w(y,A)=0$ for $0<y\le z_0(\mathfrak h(A))$.

\begin{lemma}
\label{lemma:adjoint-zero-revised}
Suppose Assumptions~\ref{assume:common} and \ref{assume:zero-general-new} hold.
For $y,h>0$, let $A_t^*:=\mathfrak a(H_t^{y,h})$.
Then
\begin{align}
    dw(Y_t^y,A_t^*)
    =\Psi_A(Y_t^y,A_t^*)\,dt
      -\kappa Y_t^y w_y(Y_t^y,A_t^*)\,dW_t.
    \label{eq:adjoint-dynamics-zero-revised}
\end{align}
Moreover, there exists $C(h)<\infty$ such that, for $t\ge0$,
\begin{align}
    |w(Y_t^y,A_t^*)|
    +|Y_t^yw_y(Y_t^y,A_t^*)|
    \le C(h)\Big(1+\frac{Y_t^y}{z_0(H_t^{y,h})}\Big)\,.
    \label{eq:adjoint-growth-zero-revised}
\end{align}
\end{lemma}
\begin{proof}
First, we prove \eqref{eq:adjoint-dynamics-zero-revised}.
Since $Y_t^y\ge z_0(H_t^{y,h})$ and $A_t^*=\mathfrak a(H_t^{y,h})$, we consider $(y,\mathfrak a(a))$ with $y\ge z_0(a)$.
For $a>0$ and $y>z_0(a)$, \eqref{eq:Phi-yh-zero-new} and \eqref{eqn:4-21} imply
\begin{align}
    w_y(y,\mathfrak a(a))
    &=\frac{2}{\kappa^2\Gamma(a)}y^{\delta_0-1}
       \int_y^\infty s^{-\delta_0-1}\bar u_h(s,a)\,ds\,,\\
    w_{yy}(y,\mathfrak a(a))
    &=\frac{2}{\kappa^2\Gamma(a)}
       \left(
          (\delta_0-1)y^{\delta_0-2}
          \int_y^\infty s^{-\delta_0-1}\bar u_h(s,a)\,ds
          -\frac{\bar u_h(y,a)}{y^2}
       \right).
    \label{eq:adjoint-wy-integral}
\end{align}
By Proposition~\ref{prop:dh-ordering-revised}, $|\bar u_h(s,a)|\le\Gamma(a)+s$.
Thus the integrals in \eqref{eq:adjoint-wy-integral} are finite, and a simple calculation yields
\begin{align}
    \mathcal L_0w(y,A)=\Psi_A(y,A),
    \qquad y>z_0(\mathfrak h(A)).
    \label{eq:w-PDE-detailed}
\end{align}
By \eqref{eq:w-general-revised} and Proposition~\ref{prop:VI-zero-general-new}, we have
$w_y(y,\mathfrak a(a))=\frac{V_{0,yh}(y,a)}{\Gamma(a)}$ and
$w_A(y,\mathfrak a(a))=-\frac{V_{0,hh}(y,a)}{\Gamma(a)^2}
       +\frac{V_{0,h}(y,a)\Gamma'(a)}{\Gamma(a)^3}$.
Since $V_0\in C^2$ and $V_{0,h}(y,a)=0$ for $y\le z_0(a)$,
\begin{align}
    w(z_0(a),\mathfrak a(a))=w_y(z_0(a),\mathfrak a(a))=w_A(z_0(a),\mathfrak a(a))=0\,.
    \label{eqn:3-43}
\end{align}
Moreover, $w$ is $C^1$ in $(y,A)$ and vanishes for $0<y<z_0(\mathfrak h(A))$.
For $y>z_0(a)$, we have
\begin{align}
    |w_{yy}(y,\mathfrak a(a))|
    &\le \frac{2}{\kappa^2}\Big(\frac{2\delta_0-1}{\delta_0y^2}+\frac{2}{\Gamma(a)y}\Big)\notag\\
    &\le\frac{2}{\kappa^2}\Big(\frac{2\delta_0-1}{\delta_0(z_0(a))^2}+\frac{2}{\Gamma(a)z_0(a)}\Big)\,.
    \label{eqn:3-41}
\end{align}
Here we used $|\bar u_h(s,a)|\le\Gamma(a)+s$.
By \cite[Chapter~IV, Corollary~1 to Theorem~70]{protter2012stochastic},
\begin{align}
    \kappa^2\int_0^T\mathbf 1_{\{P_t=0\}}\,dt
    &=\int_0^T\mathbf 1_{\{P_t=0\}}\,d\langle P\rangle_t\notag\\
    &=\int_{\mathbb R}\mathbf 1_{\{0\}}(\xi)L_T^\xi(P)\,d\xi
      =0
      \quad\text{a.s.}
    \label{eq:boundary-zero-Lebesgue-time-adjoint-detailed}
\end{align}
where $P_t=\log\frac{Y_t^y}{z_0(H_t^{y,h})}$
and $L_T^\xi(P)$ denotes its local time.
Thus,
\begin{align}
    \int_0^T\mathds{1}_{Y_t^y=z_0(H_t^{y,h})}\,dt=0\,.
    \label{eqn:4-40}
\end{align}
Since $z_0$ and $\mathfrak h$ are strictly decreasing,
$A\mapsto z_0(\mathfrak h(A))$ is increasing.
By \eqref{eqn:3-41}, \eqref{eqn:4-40}, and
\cite[Theorem~2.1 and Remark~2.2]{cai2023change}, we have
\begin{align}
    w(Y_T^y,A_T^*)
    =w(Y_0^y,A_0^*)+\int_0^T w_y(Y_t^y,A_t^*)\,dY_t^y+\int_0^T w_A(Y_t^y,A_t^*)\,dA_t^*
      +\frac12\int_0^T w_{yy}(Y_t^y,A_t^*)\,d\langle Y^y\rangle_t.
\end{align}
Observe that $dA_t^*=-\Gamma(H_t^{y,h})\,dH_t^{y,h}$ and $H^{y,h}$ increases only when $Y_t^y=z_0(H_t^{y,h})$.
By \eqref{eq:w-PDE-detailed}, \eqref{eqn:3-43}, and \eqref{eqn:4-40}, we have
\begin{align}
    w(Y_T^y,A_T^*)
    =w(Y_0^y,A_0^*)+\int_0^T\Psi_A(Y_t^y,A_t^*)\,dt-\int_0^T\kappa Y_t^yw_y(Y_t^y,A_t^*)\,dW_t\,.
\end{align}

Next, we prove \eqref{eq:adjoint-growth-zero-revised}.
By \eqref{eqn:4-21} and \eqref{eq:Phi-yh-zero-new}, we have
\begin{align}
    w(y,\mathfrak a(a))
    =\frac{2}{\kappa^2\Gamma(a)}
       \int_{z_0(a)}^y q^{\delta_0-1}
       \left(
          \int_q^\infty
             s^{-\delta_0-1}\bar u_h(s,a)\,ds
       \right)dq,
    \qquad y\ge z_0(a).
    \label{eq:growth-w-before-scaling-explained}
\end{align}
For $z\ge\zeta_0(a)$, changing variables by $y=\Gamma(a)z$, $q=\Gamma(a)v$, and $s=\Gamma(a)\xi$, we obtain
\begin{align}
    |w(\Gamma(a)z,\mathfrak a(a))|
    &=\Big|\frac{2}{\kappa^2}
       \int_{\zeta_0(a)}^z v^{\delta_0-1}
       \left(
          \int_v^\infty
             \xi^{-\delta_0-1}d_a(\xi)\,d\xi
       \right)dv\Big|\label{eqn:4-45}\\
    &\le\frac{2}{\kappa^2}
       \left(\frac1{\delta_0}\log\frac{z}{\zeta_0(a)}
       +\frac{z-\zeta_0(a)}{\delta_0-1}\right)\,.
    \label{eq:w-scaled-family-revised}
\end{align}
Here we used $|d_a(\xi)|\le1+\xi$ from Proposition~\ref{prop:dh-ordering-revised}.
By Lemma~\ref{lemma:zero-boundary-new}, $\zeta_0(a)\ge\underline\zeta>0$, and hence
\begin{align}
    |w(Y_t^y,\mathfrak a(H_t^{y,h}))|
    \le C\Big(1+\frac{Y_t^y}{\Gamma(H_t^{y,h})}\Big)
    \le C\Big(1+\frac{Y_t^y}{z_0(H_t^{y,h})}\Big),
\end{align}
for some $C>0$.
By a change of variables in \eqref{eq:adjoint-wy-integral}, we have
\begin{align}
    \Gamma(a)w_y(\Gamma(a)z,\mathfrak a(a))
    =\frac{2}{\kappa^2}z^{\delta_0-1}
       \int_z^\infty\xi^{-\delta_0-1}d_a(\xi)\,d\xi\,.
\end{align}
Multiplying by $z$ and using $|d_a(\xi)|\le1+\xi$, we obtain
\begin{align}
    |yw_y(y,\mathfrak a(a))|
    =\Big|\frac{2}{\kappa^2}z^{\delta_0}
       \int_z^\infty\xi^{-\delta_0-1}d_a(\xi)\,d\xi\Big|
    \le C(1+z)\,,
\end{align}
for some $C>0$. Thus \eqref{eq:adjoint-growth-zero-revised} holds.
This completes the proof.
\end{proof}

\begin{lemma}
\label{lemma:competitor-terminal-zero-new}
Suppose Assumptions~\ref{assume:common} and
\ref{assume:zero-general-new} hold.
For $y,h>0$, let $H\in\mathcal H(h)$ satisfy
\begin{align}
    \mathbb{E}\left[\int_0^\infty
      \bar u(Y_t^y,H_t)\,dt\right]>-\infty.
    \label{eq:finite-payoff-H-zero-new}
\end{align}
Define $A_t:=\mathfrak a(H_t)$ and $A_t^*:=\mathfrak a(H_t^{y,h})$. Then
\begin{align}
    \lim_{T\to\infty}A_T=0,
    \quad
    \lim_{T\to\infty}A_T^*=0,
    \quad
    \lim_{T\to\infty}\mathbb{E}\left[|A_T-A_T^*|\,|w(Y_T^y,A_T^*)|\right]=0.
    \label{eq:terminal-product-zero-new}
\end{align}
\end{lemma}

\begin{proof}
First, we prove
\begin{align}
    \lim_{T\to\infty}A_T=0,
    \qquad
    \lim_{T\to\infty}A_T^*=0.
\end{align}
By Lemma~\ref{lemma:reflected-moments-zero-new} and
\eqref{eq:a-transform-zero-new}, we have
\begin{align}
    \lim_{T\to\infty}A_T^*=0.
\end{align}
Since $H$ is nondecreasing, $H_\infty:=\lim_{T\to\infty}H_T$ exists.
Let $E:=\{H_\infty<\infty\}$.
On $E$, $u(g(H_t))\le u(g(H_\infty))<0$, and
\eqref{eq:bar-u-definition} implies
\begin{align}
    \int_0^\infty\bar u(Y_t^y,H_t)\,dt\le 
    \int_0^\infty u(g(H_t))\,dt
    \le\int_0^\infty u(g(H_\infty))\,dt
    =-\infty.
\end{align}
Thus \eqref{eq:finite-payoff-H-zero-new} implies $\mathbb P(E)=0$.
By \eqref{eq:a-transform-zero-new}, $\lim_{T\to\infty}A_T=0$.

Now we prove the last equation in \eqref{eq:terminal-product-zero-new}.
Observe that $0\le A_T,A_T^*\le\mathfrak a(h)$. For $p\in(1,\delta_0)$, H\"older's inequality, dominated convergence,
and Lemmas~\ref{lemma:reflected-moments-zero-new} and
\ref{lemma:adjoint-zero-revised} imply
\begin{align}
\limsup_{T\to\infty}\mathbb E\left[|A_T-A_T^*|\,|w(Y_T^y,A_T^*)|\right]
    &\le
    \limsup_{T\to \infty}\left(\mathbb E\left[|A_T-A_T^*|^{p'}\right]\right)^{ \frac{1}{p'} }\limsup_{T\to \infty}
    \left(\mathbb E\left[|w(Y_T^y,A_T^*)|^p\right]\right)^{ \frac{1}{p} }
    =0,
\end{align}
where $\frac{1}{p'}+\frac{1}{p}=1$.
This completes the proof.
\end{proof}

\section{Counterexample to \cite[Lemma A.1]{guasoni2020shortfall}}\label{sec:benchmark-comparison}
We construct a consumption process $c$ and its running maximum $H$
that satisfy the budget condition of \cite[Lemma A.1]{guasoni2020shortfall}
but strictly reverse its asserted inequality. Choose
\begin{align}
  u(\xi)&=-\frac{1}{\xi},\qquad D(h)=h^{-\frac{1}{2}},\qquad L(h)=0,
  \label{eq:ce-preferences}\\
  \rho&=0,\qquad r=\frac{1}{2},\qquad \mu=\frac{3}{2},
  \qquad \sigma=1,\qquad h=1.\label{eq:ce-parameters}
\end{align}
Then $\kappa=\frac{\mu-r}{\sigma}=1$. Denoting the parameter $\mu$ in \cite{guasoni2020shortfall}
by $\mu^G$, we have
\begin{align}
  \alpha=\frac{1}{2},\qquad \gamma=2,\qquad
  \mu^G=\mu-r=1,\qquad r=\frac{1}{2},\qquad\sigma=1.
\end{align}The state-price density and its running minimum are
\begin{align}
  K_t=e^{-t-W_t},
  \qquad
  \underline K_t:=\inf_{0\le s\le t}K_s.
  \label{eq:ce-K}
\end{align}
This is the process denoted by $M$ in \cite{guasoni2020shortfall}.
Fix the multiplier at $y=1$. For $0<k<1$, define
\begin{align}
  H_t^k&:=1\vee\left(\frac{k}{\underline K_t}\right)^{\frac{2}{3}},
       \label{eq:ce-Hk}\\
  z_t^k&:=K_t(H_t^k)^{\frac{3}{2}},\label{eq:ce-zk}\\
  c_t^k&:=\min\left\{H_t^k,(H_t^k)^{\frac{1}{4}}K_t^{-\frac{1}{2}}\right\}
       =H_t^k\min\{1,(z_t^k)^{-\frac{1}{2}}\}.
       \label{eq:ce-ck}
\end{align}
Equation \cite[(A.3)]{guasoni2020shortfall}, with
$y=1$, $h=1$, $\alpha=\frac{1}{2}$, and $\gamma=2$, gives
\begin{align}
  \widehat H_t^G(1)
     =1\vee\left(\frac{\frac{1}{2}}{\underline K_t}\right)^{\frac{2}{3}}
     =H_t^{\frac{1}{2}}.
\end{align}
Consequently, $k=\frac{1}{2}$ reproduces the consumption and habit processes
used in their dual construction. We shall use the different, fixed
choice $k=\frac{1}{4}$.
The following lemma verifies that $H^k$ is the running
maximum generated by $c^k$.
\begin{lemma}\label{lem:ce-running-maximum}
For $0<k<1$, the processes $H^k$ and $c^k$ are continuous and adapted,
$z_t^k\ge k$, and
\begin{align}
  H_t^k=1\vee\sup_{0\le s\le t}c_s^k.
  \label{eq:ce-actual-habit}
\end{align}
Moreover, $H^k$ can increase only when $z_t^k=k$.
\end{lemma}

\begin{proof}
Continuity and adaptedness follow from those of $K$ and $\underline K$. If $H_t^k=1$, then
$z_t^k=K_t\ge\underline K_t\ge k$. If $H_t^k>1$, then
$z_t^k=\frac{kK_t}{\underline K_t}\ge k$. The process $\underline K$ can decrease
only when $K_t=\underline K_t$, so an increase of $H^k$ can occur only
at $z_t^k=k$.

We next prove \eqref{eq:ce-actual-habit}.
Since $c_s^k\le H_s^k\le H_t^k$ for $s\le t$, we have $H_t^k\ge1\vee\sup_{0\le s\le t}c_s^k$. If $H_t^k>1$, continuity of $K$ ensures that
$K_s=\underline K_t$ for some $s\in[0,t]$.
Then $H_s^k=H_t^k$ and $z_s^k=k<1$, so $c_s^k=H_s^k=H_t^k$. Thus, $H_t^k\le 1\vee\sup_{0\le s\le t}c_s^k$.
If $H_t^k=1$, the reverse inequality is immediate. This proves \eqref{eq:ce-actual-habit}.
\end{proof}

Define
\begin{align}
  A_k&:=\frac{5}{6}k^2-4k+\log k-3,
  \qquad B_k:=-\frac{5}{3}k^2+4k,\\
  q_k(z)&:=
  \begin{cases}
    \displaystyle A_k+\frac{z^2}{6}-2z-\log z,&k\le z\le1,\\[2mm]
    \displaystyle A_k+\frac{7}{2}-\frac{16}{3}\sqrt z,&z\ge1,
  \end{cases}\label{eq:ce-qk}\\
  b_k(z)&:=
  \begin{cases}
    \displaystyle B_k-\frac{z^2}{3}+2z,&k\le z\le1,\\[2mm]
    \displaystyle B_k-1+\frac{8}{3}\sqrt z,&z\ge1.
  \end{cases}\label{eq:ce-bk}
\end{align}
The functions in \eqref{eq:ce-qk}--\eqref{eq:ce-bk}
are used to evaluate the cost and expected utility of $c^k$.

\begin{lemma}\label{lem:ce-expectations}
For every $0<k<1$,
\begin{align}
  \mathbb{E}\left[\int_0^\infty K_tc_t^k d t\right]&=b_k(1),
       \label{eq:ce-cost-identity}\\
  \mathbb{E}\left[\int_0^\infty
    \left({u\bigl(D(H_t^k)c_t^k\bigr)}-K_tc_t^k\right)d t
    \right]&=q_k(1),\label{eq:ce-lagrangian-identity}\\
  \mathbb{E}\left[\int_0^\infty
      {u\bigl(D(H_t^k)c_t^k\bigr)}d t\right]
      &=q_k(1)+b_k(1).\label{eq:ce-utility-identity}
\end{align}
\end{lemma}

\begin{proof}
Define
\begin{align}
  j(z)&:=
    \begin{cases}-1-z,&k\le z\le1,\\-2\sqrt z,&z\ge1,\end{cases}
  \qquad
  p(z):=
    \begin{cases}z,&k\le z\le1,\\\sqrt z,&z\ge1.\end{cases}
\end{align}
Substituting \eqref{eq:ce-preferences} and
\eqref{eq:ce-ck} gives
\begin{align}
  {u\bigl(D(H_t^k)c_t^k\bigr)}-K_tc_t^k
       &=(H_t^k)^{-\frac{1}{2}}j(z_t^k),\\
  K_tc_t^k&=(H_t^k)^{-\frac{1}{2}}p(z_t^k).
  \label{eq:ce-integrands}
\end{align}
The values and first two derivatives of $q_k$ and $b_k$
agree at $z=1$, so both functions are $C^2$ on $[k,\infty)$.For $\varphi=q_k$ or $b_k$, It\^o's formula and
Lemma~\ref{lem:ce-running-maximum} yield
\begin{align}
  d\bigl((H_t^k)^{-\frac{1}{2}}\varphi(z_t^k)\bigr)
  &=(H_t^k)^{-\frac{1}{2}}\bar{\mathcal L}\varphi(z_t^k)d t
      -(H_t^k)^{-\frac{1}{2}}z_t^k\varphi'(z_t^k)dW_t\notag\\
  &\quad +(H_t^k)^{-\frac{3}{2}}
       \left(-\frac{1}{2}\varphi(k)+\frac{3}{2}k\varphi'(k)\right)d H_t^k,
  \label{eq:ce-ito}
\end{align}
where $\bar{\mathcal L} \varphi(z):=\frac{1}{2}z^2\varphi''(z)-\frac{1}{2}z\varphi'(z)$.
Direct differentiation gives, for $k<z<1$,
\begin{align*}
  q_k'(z)=\frac{z}{3}-2-\frac{1}{z},\quad
  q_k''(z)=\frac{1}{3}+\frac{1}{z^2},\quad
  b_k'(z)=2-\frac{2z}{3},\quad b_k''(z)=-\frac{2}{3},
\end{align*}
whereas on $z>1$,
\begin{align*}
  q_k'(z)=-\frac{8}{3\sqrt z},\quad q_k''(z)=\frac{4}{3z^{\frac{3}{2}}},\quad
  b_k'(z)=\frac{4}{3\sqrt z},\quad b_k''(z)=-\frac{2}{3z^{\frac{3}{2}}}.
\end{align*}
Consequently,
\begin{align}
  \bar{\mathcal L} q_k=-j,\qquad \bar{\mathcal L} b_k=-p,\label{eq:ce-poisson}\\
  -\frac{1}{2}q_k(k)+\frac{3}{2}kq_k'(k)=0,\qquad
  -\frac{1}{2}b_k(k)+\frac{3}{2}kb_k'(k)=0.\label{eq:ce-reflection}
\end{align}
For every finite $T$, the explicit derivatives and \eqref{eq:ce-K}
give $\mathbb E[\int_0^T|(H_t^k)^{-\frac{1}{2}}z_t^k\varphi'(z_t^k)|^2d t]<\infty$,
since $\mathbb E[e^{a\sup_{0\le t\le T}|W_t|}]<\infty$ for every $a>0$.
Thus the stochastic integral in \eqref{eq:ce-ito} is a square-integrable
martingale. Using \eqref{eq:ce-poisson}--\eqref{eq:ce-reflection} and
$H_0^k=z_0^k=1$, we have
\begin{align}
  q_k(1)&=\mathbb E\left[\int_0^T(H_t^k)^{-\frac{1}{2}}j(z_t^k)d t\right]
          +\mathbb E\left[(H_T^k)^{-\frac{1}{2}}q_k(z_T^k)\right],
          \label{eq:ce-q-finite}\\
  b_k(1)&=\mathbb E\left[\int_0^T(H_t^k)^{-\frac{1}{2}}p(z_t^k)d t\right]
          +\mathbb E\left[(H_T^k)^{-\frac{1}{2}}b_k(z_T^k)\right].
          \label{eq:ce-b-finite}
\end{align}
The formulas \eqref{eq:ce-qk}--\eqref{eq:ce-bk} imply
\begin{align}
  |q_k(z)|+|b_k(z)|\le C_k(1+\sqrt z),\qquad z\ge k,
  \label{eq:ce-growth}
\end{align}
for some constant $C_k>0$.
Since $\lim_{T\to\infty}K_T=0$ almost surely,
$\lim_{T\to\infty}H_T^k=\infty$ almost surely. As
$0<(H_T^k)^{-\frac{1}{2}}\le1$, dominated convergence gives
\begin{align}
  \lim_{T\to\infty}\mathbb E[(H_T^k)^{-\frac{1}{2}}]=0.
  \label{eq:ce-H-limit}
\end{align}
By \eqref{eq:ce-Hk} and a direct calculation, we have
\begin{align}
  0\le\mathbb E\left[(H_T^k)^{-\frac{1}{2}}\sqrt{z_T^k}\right]
  &=\mathbb E\left[K_T^{\frac{1}{2}}(H_T^k)^{\frac{1}{4}}\right]\notag\\
  &\le e^{-\frac{3T}{8}}
      +k^{\frac{1}{6}}\mathbb E\left[
        e^{-\frac{1}{2}(T+W_T)+\frac{1}{6}\sup_{s\le T}(s+W_s)}\right]\notag\\
  &\le e^{-\frac{3T}{8}}
      +k^{\frac{1}{6}}e^{-\frac{T}{3}}\mathbb E\left[e^{-\frac{W_T}{2}+\frac{1}{6}\sup_{s\le T}W_s}\right]\notag\\
  &\le e^{-\frac{3T}{8}}
      +k^{\frac{1}{6}}e^{-\frac{T}{3}}
        \bigl(\mathbb E[e^{-W_T}]\bigr)^{\frac{1}{2}}
        \bigl(\mathbb E[e^{\frac{1}{3}\sup_{s\le T}W_s}]\bigr)^{\frac{1}{2}}\notag\\
  &\le e^{-\frac{3T}{8}}
      +k^{\frac{1}{6}}e^{-\frac{T}{3}}e^{\frac{T}{4}}(2e^{\frac{T}{18}})^{\frac{1}{2}}\notag\\
  &=e^{-\frac{3T}{8}}+\sqrt2\,k^{\frac{1}{6}}e^{-\frac{T}{18}}.
  \label{eq:ce-mixed-bound}
\end{align}
Letting $T\to\infty$ in \eqref{eq:ce-mixed-bound} , we obtain
\begin{align}
  \lim_{T\to\infty}\mathbb E\left[(H_T^k)^{-\frac{1}{2}}\sqrt{z_T^k}\right]=0.
  \label{eq:ce-mixed-limit}
\end{align}
Combining \eqref{eq:ce-growth}, \eqref{eq:ce-H-limit},
and \eqref{eq:ce-mixed-limit} yields
\begin{align}
  \lim_{T\to\infty}\mathbb E\left[(H_T^k)^{-\frac{1}{2}}
        \bigl(|q_k(z_T^k)|+|b_k(z_T^k)|\bigr)\right]=0.
  \label{eq:ce-terminal-limit}
\end{align}
Since $-j\ge0$ and $p\ge0$, by the monotone convergence theorem
and \eqref{eq:ce-terminal-limit}, we have
\eqref{eq:ce-cost-identity} and \eqref{eq:ce-lagrangian-identity}.
Equation \eqref{eq:ce-utility-identity} follows immediately.
This completes the proof.
\end{proof}

We now construct a counterexample to
\cite[Lemma A.1]{guasoni2020shortfall}. Set $k=\frac{1}{4}$ and choose
\begin{align}
  H_t^{\frac{1}{4}}&=1\vee\left(\frac{1}{4\underline K_t}\right)^{\frac{2}{3}},\\
  c_t^{\frac{1}{4}}&=\min\left\{H_t^{\frac{1}{4}},
                            (H_t^{\frac{1}{4}})^{\frac{1}{4}}K_t^{-\frac{1}{2}}\right\}.
  \label{eq:ce-counterexample-policy}
\end{align}
Lemma~\ref{lem:ce-running-maximum} shows that $H^{\frac{1}{4}}$ is exactly
the habit generated by $c^{\frac{1}{4}}$.
At $k=\frac{1}{4}$, the auxiliary functions satisfy
\begin{align}
  b_{\frac{1}{4}}(1)&=\frac{41}{16},\qquad
  q_{\frac{1}{4}}(1)=-\frac{185}{32}-2\log2.
\end{align}
For $x=\frac{41}{16}$, \eqref{eq:ce-cost-identity} gives
\begin{align}
  \mathbb E\left[\int_0^\infty K_tc_t^{\frac{1}{4}}d t\right]
      =b_{\frac{1}{4}}(1)=\frac{41}{16}=x.
  \label{eq:ce-budget-counterexample}
\end{align}
Thus $c^{\frac{1}{4}}$ satisfies the budget condition in
\cite[Lemma A.1]{guasoni2020shortfall}. Equation
\eqref{eq:ce-utility-identity} gives
\begin{align}
  \mathbb E\left[\int_0^\infty u(D(H_t^{\frac{1}{4}})c_t^{\frac{1}{4}})d t\right]
      =-\frac{103}{32}-2\log2.
  \label{eq:ce-utility-counterexample}
\end{align}
At $y=1$, \eqref{eq:ce-lagrangian-identity} gives
\begin{align}
  &\mathbb E\left[\int_0^\infty
       \widetilde U^G(K_t,H_t^{\frac{1}{2}})d t\right]+\frac{41}{16}\notag\\
  &\quad=\mathbb E\left[\int_0^\infty
     \bigl(u(D(H_t^{\frac{1}{2}})c_t^{\frac{1}{2}})-K_tc_t^{\frac{1}{2}}\bigr)d t\right]
       +\frac{41}{16}\notag\\
  &\quad=q_{\frac{1}{2}}(1)+\frac{41}{16}
       =-\frac{65}{16}-\log2,
  \label{eq:ce-bound-counterexample}
\end{align}
where
\begin{align}
  \widetilde U^G(a,h)
  &:=\sup_{c>0}\{u(D(h\vee c)c)-ac\}\notag\\
  &=\begin{cases}
      \displaystyle-3\,2^{-\frac{2}{3}}a^{\frac{1}{3}},&0<a\le\frac{1}{2}h^{-\frac{3}{2}},\\[1mm]
      \displaystyle-h^{-\frac{1}{2}}-ah,&\frac{1}{2}h^{-\frac{3}{2}}<a\le h^{-\frac{3}{2}},\\[1mm]
      \displaystyle-2a^{\frac{1}{2}}h^{\frac{1}{4}},&a>h^{-\frac{3}{2}}.
     \end{cases}\label{eq:ce-static-G}
\end{align}
The function in \eqref{eq:ce-static-G} is the conjugate in Equation
 \cite[(A.1)]{guasoni2020shortfall}. Since $z_t^{\frac{1}{2}}\ge\frac{1}{2}$,
$c_t^{\frac{1}{2}}$ attains the supremum in \eqref{eq:ce-static-G} at
$(a,h)=(K_t,H_t^{\frac{1}{2}})$. Thus, the first equality in \eqref{eq:ce-bound-counterexample} holds.
By \eqref{eq:ce-utility-counterexample} and
\eqref{eq:ce-bound-counterexample}, we have
$-\frac{103}{32}-2\log2-(-\frac{65}{16}-\log2)
=\frac{27}{32}-\log2>0$, and hence
\begin{align}
  \mathbb E\left[\int_0^\infty u(D(H_t^{\frac{1}{4}})c_t^{\frac{1}{4}})d t\right]
  >\mathbb E\left[\int_0^\infty
        {\widetilde U^G(K_t,H_t^{\frac{1}{2}})}d t\right]
       +\frac{41}{16}.
  \label{eq:ce-reverse-inequality}
\end{align}
Thus, Lemma A.1 of \cite[Lemma A.1]{guasoni2020shortfall} does not hold at $x=\frac{41}{16}$ and $y=1$ for
$u(\xi)=-\frac{1}{\xi}$, $D(h)=h^{-\frac{1}{2}}$, $L(h)=0$,
$\rho=0$, $r=\frac{1}{2}$, $\mu=\frac{3}{2}$, $\sigma=1$,
and $h=1$.

\end{document}